\documentclass{statsoc}
\usepackage[hmargin=1.0in,vmargin=1.0in]{geometry}
\usepackage{amsmath,amssymb,graphicx,setspace,natbib}
\usepackage{colortbl}
\newcolumntype{L}[1]{>{\raggedright\let\newline\\\arraybackslash\hspace{0pt}}m{#1}}
\newcolumntype{C}[1]{>{\centering\let\newline\\\arraybackslash\hspace{0pt}}m{#1}}
\newcolumntype{R}[1]{>{\raggedleft\let\newline\\\arraybackslash\hspace{0pt}}m{#1}}
\usepackage[ruled]{algorithm}
\usepackage{algorithmicx}
\usepackage{algpseudocode}
\usepackage{tikz}
\usepackage{paralist}
\usepackage{enumerate}
\usepackage{hyperref}
\hypersetup{letterpaper=true,colorlinks=true,linkcolor=red,citecolor=blue}
\usepackage{ulem}

\def\tog{\mathrm{tog}}

\newtheorem{theorem}{Theorem}[section]
\newtheorem{lemma}[theorem]{Lemma}

\newtheorem{remark}[theorem]{Remark}

\algnewcommand{\algorithmicgoto}{\textbf{go to}}
\algnewcommand{\Goto}[1]{\algorithmicgoto~\ref{#1}}

\newcommand{\Exp}{{\mathbb E}}

\newcommand{\bgamma}{ {\boldsymbol \gamma} }

\newcommand{\blambda}{ {\boldsymbol \lambda} }
\newcommand{\bLambda}{ {\boldsymbol \Lambda} }
\newcommand{\bmu}{ {\boldsymbol \mu} }

\newcommand{\brho}{ {\boldsymbol \rho} }

\newcommand{\bSigma}{ {\boldsymbol \Sigma} }
\newcommand{\btheta}{ {\boldsymbol \theta} }

\newcommand{\bxi}{ {\boldsymbol \xi} }

\DeclareMathOperator*{\argmin}{\mathop{\mathrm{argmin}}}

\newcommand{\bzero}{ {\boldsymbol 0} }

\newcommand{\bA}{ {\boldsymbol A} }

\newcommand{\bC}{ {\boldsymbol C} }

\newcommand{\bD}{ {\boldsymbol D} }

\newcommand{\boldf}{ {\boldsymbol f} }

\newcommand{\bI}{ {\boldsymbol I} }

\newcommand{\bK}{ {\boldsymbol K} }

\newcommand{\bR}{ {\boldsymbol R} }
\newcommand{\bs}{ {\boldsymbol s} }

\newcommand{\bv}{ {\boldsymbol v} }

\newcommand{\bx}{ {\boldsymbol x} }

\newcommand{\by}{ {\boldsymbol y} }

\newcommand{\norm}[1]{\left\lVert#1\right\rVert}

\newcommand{\mt}[1]{\tilde{#1}}
\newcommand{\wt}[1]{\widetilde{#1}}

\newcommand{\scX}{\mathcal{X}}
\newcommand{\cse}{C^{\rm \scriptsize SE}}

\title[Additive Gaussian Process Regression]{Additive Gaussian Process Regression}
\author[S. Qamar and S. T. Tokdar]{Shaan Qamar}
\address{Department of Statistical Science, Duke University, Durham, NC, USA.}
\email{shaan@stat.duke.edu}
\author[S. Qamar and S. T. Tokdar]{Surya T. Tokdar}
\address{Department of Statistical Science, Duke University, Durham, NC, USA.}

\begin{document}

\begin{abstract}
Additive-interactive regression has recently been shown to offer attractive minimax error rates over traditional nonparametric multivariate regression in a wide variety of settings, including cases where the predictor count is much larger than the sample size and many of the predictors have important effects on the response, potentially through complex interactions. 
We present a Bayesian implementation of additive-interactive regression using an additive Gaussian process (AGP) prior and develop an efficient Markov chain sampler that extends stochastic search variable selection in this setting. Careful prior and hyper-parameter specification are developed in light of performance and computational considerations, and key innovations address difficulties in exploring a joint posterior distribution over multiple subsets of high dimensional predictor inclusion vectors. The method offers state-of-the-art support and interaction recovery while improving dramatically over competitors in terms of prediction accuracy on a diverse set of simulated and real data. Results from real data studies provide strong evidence that the additive-interactive framework is an attractive modeling platform for high-dimensional nonparametric regression.
\end{abstract}


\section{Introduction} \label{sec:intro}

Much of the high dimensional regression literature focuses on parametric linear models regularized with sparsity and shrinkage \citep{tibshirani1996regression, dantzig, esl}. 
Restricted by their linearity and additivity assumptions, these methods often fail to adequately model many naturally occurring predictor-response relations. By requiring the parametric relation to hold globally, these methods remain prone to introducing bias when quantifying intervention effect from nonrandomized studies \citep{hill2011bayesian}. In addition, these methods also run the risk of over-parametrization when attempting to adjust for non-linearity or to uncover predictor interaction.

Several smoothing based nonparametric regression methods \citep{rodeo, bertin, o1978curve, gp-paper} accommodate a wider range of predictor-response relations and come with mathematical guarantees of delivering good performance in various settings \citep{rodeo, vdV, vandervaart&vanzanten09, bhattacharya2011, tokdar2012}. Unfortunately, the computational demands of these methods scale poorly with predictor dimension due to costly likelihood and score function evaluations. More importantly, with a traditional sparse nonparametric regression model in which $f$ is assumed to depend on $d$ of the original $p$ predictors, statistical estimation greatly suffers from the curse of dimensionality in the high dimensional setting. Under this framework, the minimax $L_2$ estimation risk $r_n$ based on $n$ observations is of the order $r_n^2 \asymp n^{-2\alpha/(2\alpha + d)} + d/n \log (p/d)$, where $\alpha$ denotes the degree of smoothness of $f$ \citep{yytokdar}. The second term is the penalty paid for variable selection, and remains small even in {\it large $p$ small $n$} situations, i.e., when $p$ is as large as $\exp(n^\beta), ~\beta \in (0,1)$. For $p$ of this order, the first term gives the standard fixed-dimension minimax risk for a $d$-variate $\alpha$-smooth function \citep{stone82}, and is small only when $d = o(\log n) = o(\log\log p)$, i.e., $f$ is extremely sparse in the observed predictors.

Popular nonparametric methods based on the theory of ensemble weak learners \citep{mars,randomForest,bart} use simple additive structures and sparsity, and their computational costs scale reasonably well with dimension. However, these methods come with few theoretical guarantees, do not address the large $p$ small $n$ curse of dimensionality, may greatly underperform when the actual predictor-response relation is smooth or near parametric, and often behave as black-box forecasting machines offering little in terms of inference on predictor importance and interaction. Additive structures are also assumed in the widely used generalized additive model (GAM) regression of \cite{hastie1986generalized} and its high dimensional extensions \citep{ravikumar2007spam}. These approaches consider only univariate components and ignore predictor interaction, however, and the approach by \cite{suzuki2012pac} considers interaction only within predetermined subgroups of predictors. 

We pursue a novel multivariate regression model and method where the unknown regression function $f$ is theorized to decompose as $f = f_1 + \cdots + f_k$ for some $k \ge 1$, with each component function $f_s$ depending nonparametrically on a small number of predictors. Here, predictors are treated exchangeably and learning their interaction patterns is considered a primary objective.
Although such an additive-interactive regression resembles the theoretical framework behind ensemble methods \citep{freund-schapire}, the emphases are entirely distinct. 
Ensemble methods add many ``weak learners'' to boost their overall efficiency while avoiding the risk of overfitting. In contrast, the additive-interactive framework is a modeling assumption which postulates that a high dimensional function $f$ can be divided into low dimensional pieces added together, enabling a statistical method to estimate each piece with an efficient low-dimensional learner. Seminal works by \cite{vdV,vandervaart&vanzanten09} on Gaussian process (GP) regression indicate that a GP prior specification on the component functions could be ideally suited for this task, motivating the additive GP method pursued here.

In a recent work, \cite{yytokdar} establish attractive minimax theory for such additive-interactive regression, with a focus on the high dimensional setting. They show that the additive-interactive regression breaks away from the extreme sparsity assumption, provided the number of predictors included in any single component is bounded. In particular, if every component function is $\alpha$-smooth and includes $d$ predictors, the minimax risk remains small even with $k = o(n^\xi), ~ \xi \in (0,1)$ components. This corresponds to the inclusion of $dn^\xi \asymp (\log p)^{\xi/\beta}$ predictors. By restricting the size of each component, which restricts the maximum order of interaction, one can therefore learn the effect of many more predictors by increasing the number of additive components. 

We develop a fully Bayesian implementation of additive-interactive regression using an additive Gaussian process (AGP) prior. \cite{yytokdar} show that a theoretical abstraction of our method offers posterior contraction rates that adaptively match the minimax error rate across a wide range of regression settings. The significance of our contribution is threefold: First, we present in Section \ref{sec:sparse_agp_priors} details of how to set the global and component specific model hyper-parameters, allowing the method to adapt to varying degrees of sparsity, additivity and complexity of interaction patterns. Second, 
in high dimensions, computational challenges arise in efficient sampling of the component inclusion vectors while maintaining a reversible Markov chain sampler. Section \ref{sec:backfitting_agp} develops an efficient and reproducible MCMC scheme, adapting concepts from stochastic neighborhood search \citep{sss} and generalizing the multiple-try Metropolis \citep{mtm} algorithm. Additional strategies are proposed in Sections \ref{sec:icm_sampling}, \ref{sec:component_adaptation} and \ref{sec:inclusion_propensity_adapt} to improve mixing and efficiency. Finally, we demonstrate AGP's dramatic improvement over competitors in terms of prediction accuracy, as well as its state-of-the-art support and interaction recovery on a diverse set of simulated and real data studies in Sections \ref{sec:simulation} and \ref{sec:real_data_analysis}. 
Our results provide a strong evidence that the additive-interactive framework is a practically attractive modeling platform for high-dimensional nonparametric regression, and also validate the proposed Markov chain sampler as an effective computational learning tool. Section \ref{sec:discussion} concludes with a discussion of extensions and future work.

\section{Additive-interactive regression with Gaussian processes}
\label{sec:agp-regression}

\subsection{The additive GP model}



Recall that a stochastic process $w = (w(x) : x \in \scX)$ on a Euclidian domain $\scX$ is called a Gaussian process (GP) if every finite collection of elements $(w(x_1), \ldots, w(x_n))$, $n \ge 1$, $\{x_1, \ldots, x_n\} \subset \scX$, has a joint Gaussian distribution. Such a process is completely determined by its mean and covariance functions $\mu(x) = \mathbb{E}[w(x)]$, $C(x, \mt{x}) = \mathbb{E}[(w(x) - \mu(x))(w(\mt{x}) - \mu(\mt{x}))]$, with the latter being non-negative definite over $\scX \times \scX$. For any real function $\mu(x)$ and non-negative definite function $C(x, \mt{x})$, there exists a GP with these functions as its mean and covariance. We refer to the probability law of such a process as $\mathrm{GP}(\mu, C)$. When $\mu$ and $C$ are continuous, $w \sim \mathrm{GP}(\mu, C)$ may be viewed as a random element of $\mathcal{C}(\scX)$, the space of continuous functions over $\scX$ equipped with the supremum norm. The smoothness of function valued realizations of such a random element depends on the smoothness of $C$. For $C = \cse(x, \mt{x}) \mathrel{\mathop:} = \exp(-\norm{x - \mt{x}}^2)$, the squared exponential covariance function, realizations of $w$ are infinitely smooth elements of $\mathcal{C}(X)$.

For paired predictor-response observations $(\bx_i, y_i) \in \Re^p \times \Re$, $i = 1, \ldots, n$, an additive GP (AGP) model facilitating additive-interactive regression can be written as:
\begin{align} \label{eq:agp-model}
\begin{split}
& y_i = f(\bx_i) + \epsilon_i, ~\epsilon_i \overset{\mathrm{iid}}{\sim} \mathrm{N}(0,\sigma^2) \\
& f(\bx_i) = f_1(\bx_i) + \cdots + f_k(\bx_i) \\
& f_l | (\rho_l, \lambda_l, \gamma_l, \sigma) \sim \mathrm{GP}\big(\mu(\cdot), \: \sigma^2 \rho_l^2 \cse(\cdot,\cdot | \gamma_l, \lambda_l)\big) \\
& \rho_l \overset{\mathrm{iid}}{\sim} \pi_\rho, ~ \lambda_l \overset{\mathrm{iid}}{\sim} \pi_\lambda, ~ \gamma_l \overset{\mathrm{iid}}{\sim} \pi_\gamma, ~\sigma^2 \sim \pi_\sigma.
\end{split}
\end{align}
The proposed model uses a sum of $k$ GP components to model an unknown regression function $f_0$ having $k_0$ additive components, namely $f_0(\bx) \approx \mathbb{E}(y | \bx) = \sum_{l = 1}^k f_l(\bx)$. Let $k_\mathrm{max}$ denote a fixed large number that serves as an upper bound on $k$. Since $k_0$ is unknown, components are allowed to be empty to allow the {\em effective} number of components be random within this bound. 
Additional parameters $\gamma_l$ and $\lambda_l$ inserted in each GP component's covariance function facilitate selective predictor inclusion and adaptive local smoothing. In addition, parameter $\rho_l$ captures the signal-to-noise ratio for each component. Appropriately chosen prior distributions $\pi_\rho, \pi_\lambda, \pi_\gamma,$ and $\pi_\sigma$ induce sparsity by imposing regularity on individual GP components, allowing AGP to adapt to the degree of additivity in the data, while ensuring that the number of selected predictors is small relative to $p$ (see Section \ref{sec:sparse_agp_priors}).

\subsection{Regularity and prior specification}
\label{sec:sparse_agp_priors}

Response vector $\by = (y_1, \dots, y_n)$ is assumed centered and scaled having mean 0 and variance 1. AGP model \eqref{eq:agp-model} embeds the notion of controlled interaction, with inclusion vector $\gamma$ modeling a $d$-order predictor interaction with common scaling $\rho,\lambda$. Each GP component is initialized to have mean $\mu = 0$ and covariance function $\cse(\bx, \mt{\bx}) = \rho^2 \exp(-\lambda^2 \norm{\bx_{\gamma} - \mt{\bx}_{\gamma}}^2)$. Here, $\bx_\gamma = \{x_j \gamma_j, j = 1, \dots, p\}$ denotes the sub-vector of $\bx$ corresponding to the nonzero entries of inclusion vector $\gamma = (\gamma_1, \dots, \gamma_p) \in \{0,1\}^p$. The noise variance is modeled under conjugate prior $\sigma^2 \sim \mathrm{IG}(a,b)$ with suitably chosen hyper-parameters (see Section \ref{sec:prior_initialization}).

\subsubsection{Sparse priors for inclusion vectors} \label{sec:prior_incusion_vector}

Sparse priors placed on component inclusion vectors regularize model complexity in terms of the number of active components, in addition to limiting the order of interaction in each. Predictors are treated exchangeably, each having prior inclusion probability $\tau \in (0,1)$, with the prior probability for inclusion vector $\gamma = (\gamma_1, \dots, \gamma_p)$ given by $\pi(\gamma | \tau) = \prod_{j=1}^p \tau^{\gamma_j} (1-\tau)^{1-\gamma_j}.$
\begin{figure}[h]
\centering
\setlength{\tabcolsep}{1pt}
\begin{tabular}{C{0.32\columnwidth} C{0.32\columnwidth} C{0.32\columnwidth}}
\includegraphics[width=\linewidth]{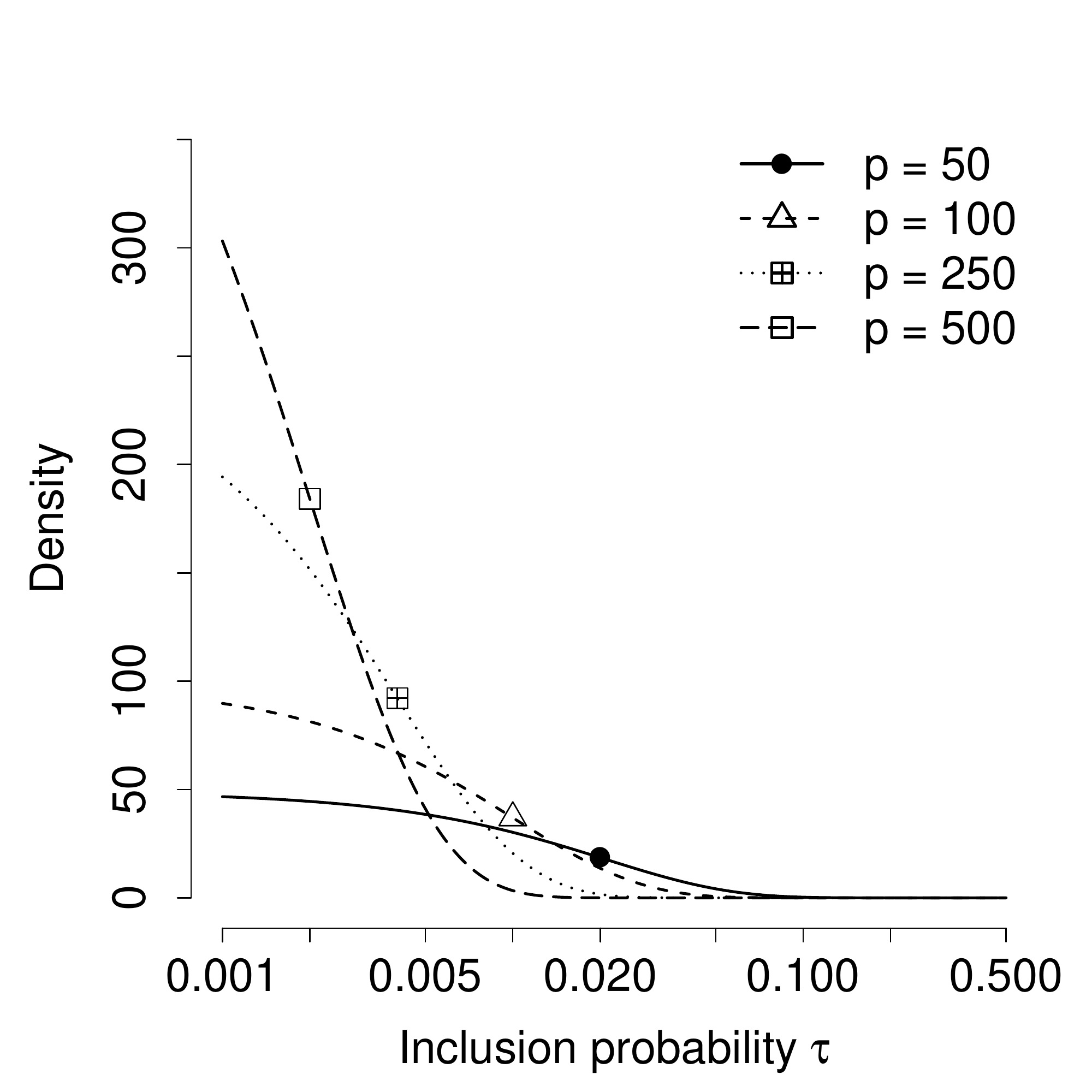} & 
\includegraphics[width=\linewidth]{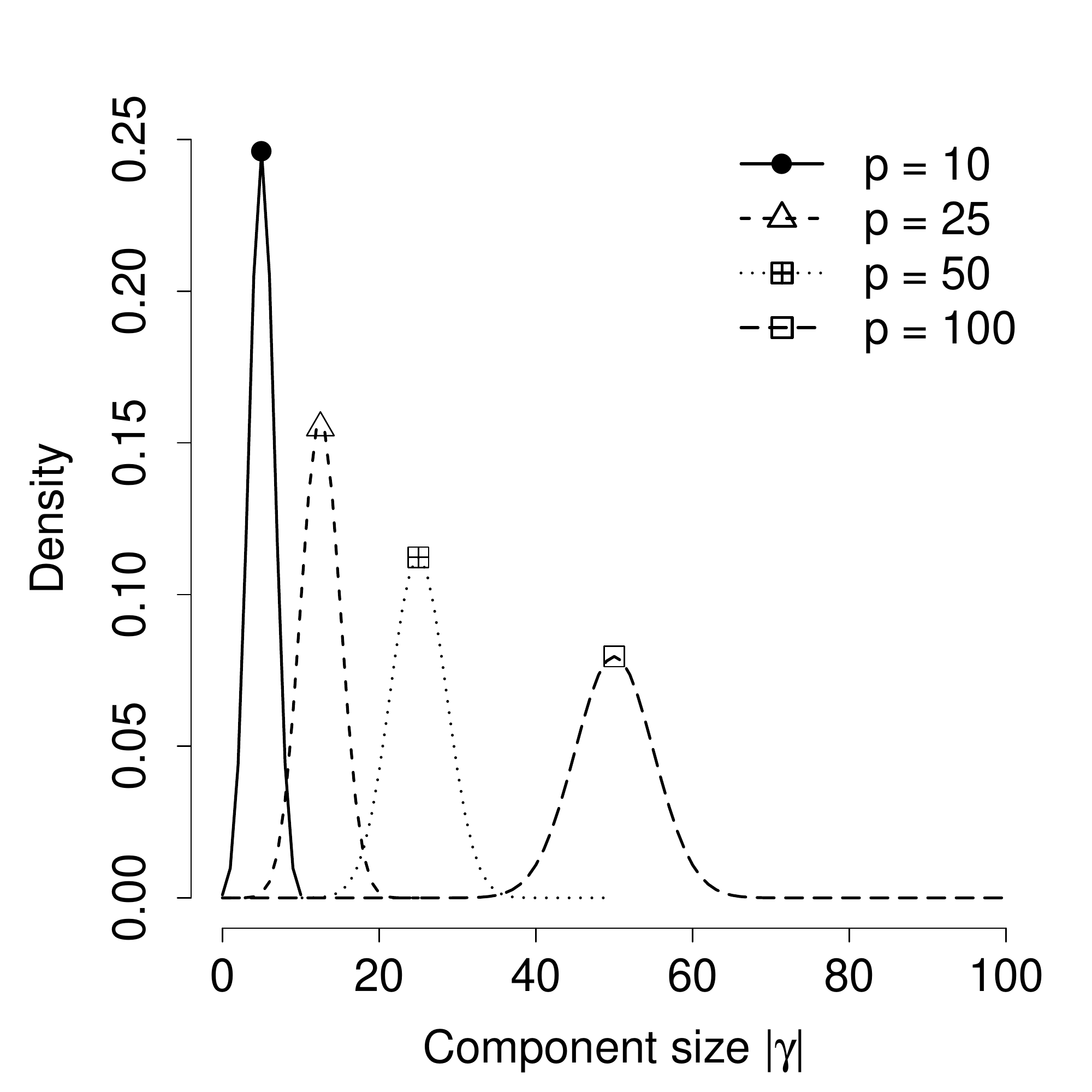} &
\includegraphics[width=\linewidth]{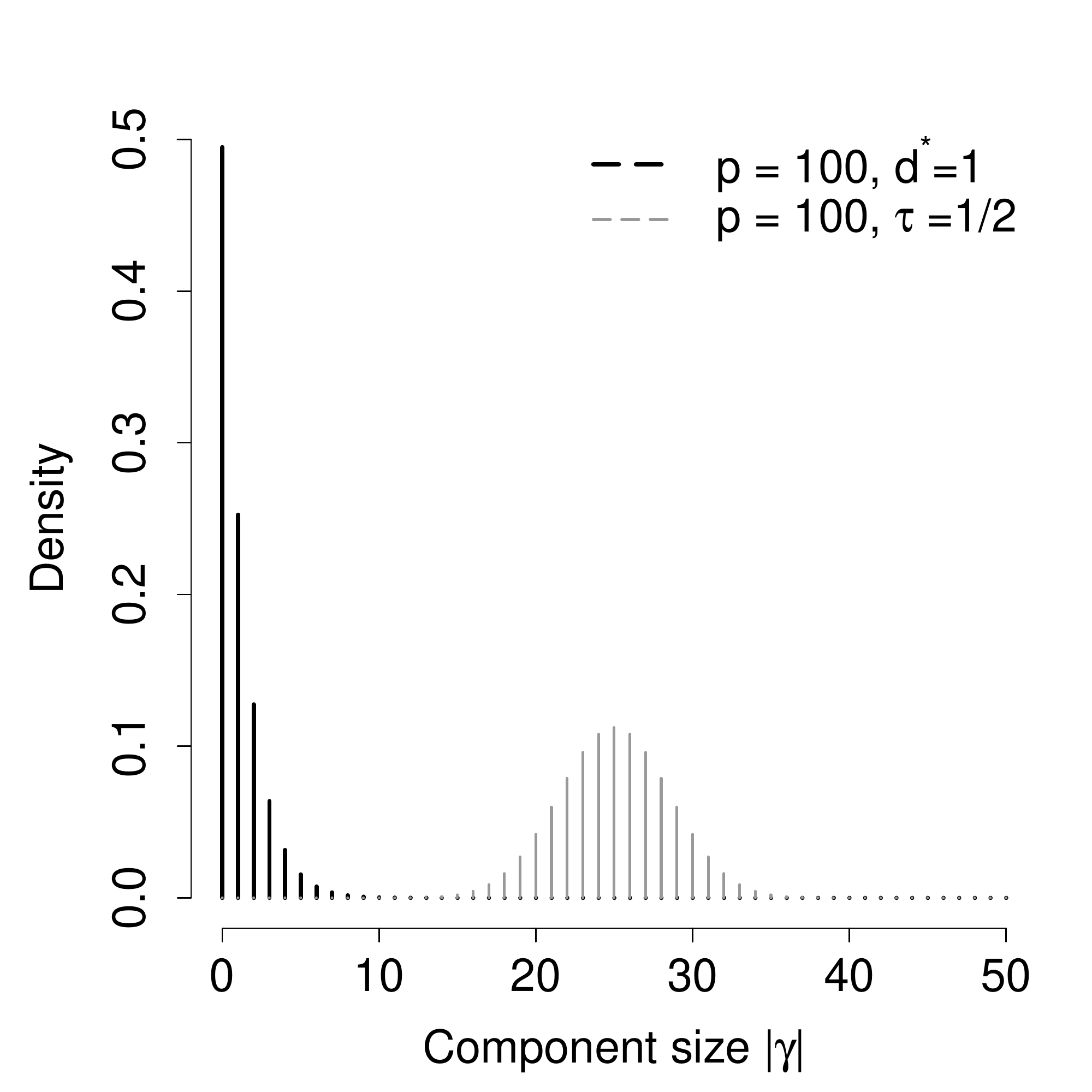} \\
{(a) $\pi(\tau | d^* = 1)$} & {(b) $\pi(\gamma | \tau = 1/2)$} & {(c) $\pi(\gamma | d^* = 1)$ vs. $\pi(\gamma | \tau = 1/2)$}
\end{tabular}
\caption{{\em Left}: prior for predictor inclusion probability $\tau$ with default hyper-parameters; {\em Middle}: prior probability over component inclusion size for varying predictor dimensions $p$ and fixed inclusion probability $\tau = 1/2$; {\em Right}: for $p = 100$, the latter is compared to $\pi(\gamma | d^* = 1)$, the induced marginal prior over component inclusion size.}
\label{fig:scale_priors}
\end{figure}

When $p$ is small, fixing $\tau = 1/2$ leads to a distribution over model sizes depicted in Figure \ref{fig:scale_priors}(b). Having an average model size of $p/2$, this prior clearly does not favor sparse configurations as $p$ grows. Similarly, the ``uniform indifference'' prior $\pi(\gamma) = 2^{-p}, ~\gamma \in \{0,1\}^p$ fails to adequately penalize large models. 
The Beta-binomial prior provides the desired degree of regularity, with $\tau \sim \mathrm{Beta}(\omega, \nu)$ parametrized by its prior mean and ``sample size.''\footnote{The standard Beta parameterization is given by $\alpha = \omega\nu, ~\beta = (1-\omega)\nu$.} Setting $\omega = d^* / p$ and $\nu = p$, the prior expected component size is $d^*$ independent of the predictor dimension, and $\mathrm{var}(\tau) \approx 1/p^2$ when $p$ is large. In such a case, $\tau$ concentrates near its prior mean as shown in Figure \ref{fig:scale_priors}(a). Marginal distribution $\pi(\gamma | d^*)$ shown in Figure \ref{fig:scale_priors}(c) places most of its mass on small models, with alternate choices of $d^*$ allowing for additional flexibility.

\subsubsection{Discrete priors for covariance scale parameters} \label{sec:greedy-gibbs}

Scale parameters of the GP covariance function play a key role in the ability to learn lower-dimensional interaction effects in the additive-interactive framework. In particular, grouping of predictors within components is encouraged through component-specific scaling. However, coupling between the $n$-vector of GP realizations and its hyper-parameters can lead to poor mixing \citep{murray} even in low dimensions. Variable selection introduces an additional layer of complexity, where component inclusion vector $\gamma$ and scale parameters $\rho, \lambda$ are intricately linked within the GP covariance function (see Section \ref{sec:sparse_agp_priors}). 
AGP's performance depends crucially on good mixing of the component inclusion vectors. Consequently, poor mixing of the scale parameters can significantly hamper MCMC efficiency if updates to $\gamma$ are rejected because the correct scaling was not proposed. 

To address this challenge, we propose a griddy-Gibbs \citep{ritter1992facilitating} discretization of the support for scale parameters $\rho, \lambda$. Though simple, this strategy is essential for robust and efficient MCMC over the AGP model space and is highly effective. In particular, this avoids a Metropolis step over $\{\rho_l, \lambda_l : l = 1, \dots, k\}$ by allowing posterior weights over the grid to be calculated, and enables marginalization over scale parameters when sampling component inclusion vectors (see Section \ref{sec:cond_gp_sampling}). Define $S_\rho$ as the support grid for signal-to-noise ratio $\rho$, with prior weight vector $W_\rho = \{w_i : \rho_i \in S_\rho\}$, and correspondingly $S_\lambda$ for inverse length-scale $\lambda$, with prior weight vector $W_\lambda = \{w_i : \lambda_i \in S_\lambda\}$. Discrete priors for component scale parameters $\rho, \lambda$ are specified as
\begin{align} \label{eq:gp_covar_priors}
\begin{aligned}
\pi(\rho) &= \sum_{\rho_i \in S_\rho} w_i \delta(\rho - \rho_i), &w_i \propto& (1 + \rho_i)^{-\alpha}, &\alpha &\ge 0 \\
\pi(\lambda) &= \sum_{\lambda_i \in S_\lambda} w_i \delta(\lambda - \lambda_i), &w_i \propto& \exp(-\beta \lambda_i), &\beta &\ge 0.
\end{aligned}
\end{align}
Here, if $X \sim \delta(x)$, then $X = 0$ w.p. 1 (i.e., $\delta(\cdot)$ is the Dirac delta distribution). 

\subsubsection{Prior hyper-parameter initialization}
\label{sec:prior_initialization}

The response and predictors are standardized to elicit meaningful grid points and default specification for prior hyper-parameters. By default, we set $a = b = 1$ for the prior on noise variance $\sigma^2$, and $\alpha = \beta = 0$ for \eqref{eq:gp_covar_priors}. In experiments, sensitivity to the initialization of these hyper-parameters is negligible, though a careful choice for grid support $S_\rho$ and $S_\lambda$ is necessary to allow AGP to adapt to additive structures having varying degrees smoothness and signal strength. 

Correlation between nearby points is characterized by inverse length-scale $\lambda$. Components with larger length-scales capture smoother patterns in the regression surface, while those with smaller length-scales characterize finer shapes and features. Restricting to higher values of correlation offers regularization of GP smoothness. 
Grid $S_\lambda$ is chosen so that the GP correlation at a distance of 0.10 given by $\exp(-0.01 \lambda_l^2)$ ranges over $\{0.70, 0.80, 0.88, 0.94, 0.99\}$\footnote{$\lambda \approx 1$ characterizes a near-linear relationship for nearby points, while larger values characterize higher degrees of non-linearity.}.
The covariance scaling for a GP component is parameterized as the product of noise variance $\sigma^2$ with signal-to-noise ratio $\rho^2$. Components with smaller values of this parameter concentrate increasingly on their prior mean, whereas components with larger values capture important variation in the response. Holding all other components fixed, the ratio of explained variance to total variance for component $l$ is $R_l^2 = \rho_l^2 / (1 + \rho_l^2)$. Grid values for $S_\rho$ are chosen so that $R^2$ ranges over $\{0, 0.25, 0.50, 0.70, 0.85, 0.99\}$. 

In addition, we define $k_a = \sum_{l = 1}^{k_\mathrm{max}} I(\rho_l > 0 \text{ or } d_l > 0)$ as the number of {\it active} components, and $\mathcal{I}_A$ denotes the active component index set. For $d^* \ll p$, prior component size $d_l$ is approximately $\mathrm{Poisson}(d^*)$, and $\Exp k_a \approx k_\mathrm{max} (1 - w_0 \exp(-d^*))$. With $\alpha = 0$ and $d^* = 1$, the prior expected number of active components is approximately $0.95 k_\mathrm{max}$. The minimum and maximum number of active components are set as $k_\mathrm{min} = \lfloor \log(p) \rfloor$ and $k_\mathrm{max} = \lceil p^{1/2} \rceil$, respectively. This allows multiple attempts at identifying important predictors across a larger number of active components initially. Thereafter, the number of active components adapts in a data dependent (see Section \ref{sec:component_adaptation}).

\section{Posterior computation for AGP regression} \label{sec:backfitting_agp}

\subsection{A basic Bayesian back-fitting algorithm}
\label{sec:cond_gp_sampling}
We propose a fast, efficient, and reproducible Metropolis-within-Gibbs sampler for posterior computation in AGP model \eqref{eq:agp-model}. Posterior computation is performed over the joint space of component parameters $\bxi = \{(\rho_l, \lambda_l, \gamma_l) : l = 1, \dots, k\}$, and sampling proceeds by sequentially updating component-specific parameters via the ``back-fitting'' algorithm \citep{ppr}.

In the regression setting, the $n$-vector of GP realizations $ (f_l(x_1), \dots, f_l(x_n))'$, $1 \le l \le k$, over the observed predictors is conjugate and may be analytically integrated out. In addition, the covariance function for AGP priors have scaling which allows marginalization over the variance of the error process. Doing so, one obtains a marginal likelihood for the data which depends only on $\bxi = \{\brho, \blambda, \bgamma\}$, namely $\by | \bxi \sim t_n(\bzero, (b/a) \bSigma)$. Here, $\bSigma = \bI_n + \bK$, $\bK = \sum_{l=1}^k \rho_l^2 \bC_l$, and $\bC_l$, $1 \le l \le k$, denote component covariance matrices defined in Section \ref{sec:sparse_agp_priors}. In addition, $t_\nu(\bmu, \bSigma)$ denotes a multivariate-$t$ distribution with $\nu$ degrees of freedom, mean $\bmu$ and 	covariance $\bSigma$. For $l = 1, \dots, k$, back-fitting for AGP proceeds to:
\begin{enumerate}[(1)]
\item Sample component inclusion vector $\gamma_l \sim \pi( \cdot | \tau, \bxi_{(l)}, \by)$ according to
\begin{align} \label{eq:gp_gamma_update}
\pi(\gamma_l | \tau, \bxi_{(l)}, \by) \propto \pi(\gamma_l | \tau) \sum_{(\rho_l, \lambda_l) \in \mathcal{G}} \pi(\rho_l, \lambda_l) \: p(\by | (\rho_l, \lambda_l, \gamma_l), \bxi_{(l)}),
\end{align}
where $\mathcal{G} = S_\rho \times S_\lambda$ is the discrete grid for covariance scale parameters discussed in Section \ref{sec:greedy-gibbs}. Enumeration of $\gamma_l \in \{0,1\}^p$ is intractable when $p$ is large, so sampling here proceeds via a Metropolis step (see Section \ref{sec:gamma-sampling}). 
\item Sample scale parameters $(\rho_l, \lambda_l) \in \mathcal{G}$ as a block according to
\begin{align} \label{eq:gp_scale_param_update}
\pi( \rho_l, \lambda_l | \gamma_l, \bxi_{(l)}, \by) = \frac{\pi(\rho_l, \lambda_l) \: p(\by | (\rho_l, \lambda_l, \gamma_l), \bxi_{(l)})}{\sum_{(\mt{\rho_l}, \mt{\lambda_l}) \in \mathcal{G}} \pi(\mt{\rho_l},\mt{\lambda_l}) \: p(\by | (\mt{\rho_l}, \mt{\lambda_l}, \gamma_l), \bxi_{(l)})}. 
\end{align}
\item (Optional) Auxiliary posterior draws for $\sigma^2$ and $\boldf_l = (f_l(x_1), \dots, f_l(x_n))'$ are available in closed-form as
\begin{align} \label{eq:gp_update}
\begin{split}
&\pi(\sigma^2 | \bxi, \by) = {\rm IG}(a+n / 2, b + \by' \bSigma^{-1} \by / 2) \\
&\pi(\boldf_l | \sigma^2, (\rho_l,\lambda_l, \gamma_l), \bxi_{(l)},\by) = {\rm N}\big( \bSigma_l \bSigma_{(l)}^{-1}\by, \sigma^2 \bSigma_l\big),
\end{split}
\end{align}
where $\bSigma_{(l)} = \bI_n + \bK_{(l)}$, $\bK_{(l)} = \sum_{s \ne l} \rho_s^2 \bC_s$ and $\bSigma_l = \big(\bSigma_{(l)}^{-1} + (\rho_l^2 \bC_l)^{-1}\big)^{-1}$. Posterior draws for  variance parameter $\sigma^2$ may serve as a model fit diagnostic and a measure of MCMC mixing and stability.
\end{enumerate}

The complete AGP sampling scheme is given in Algorithm \ref{alg1}. There, a posterior draw for the predictor inclusion probability precedes component updates, i.e., $\tau \sim {\rm Beta}(\mu', \nu')$, $\nu' = (1 + k_a)p$ and $\mu' = (d^* + \sum_{l \in \mathcal{I}_A} d_l) / \nu'$. 

\subsection{Updating variable inclusion vectors}
\label{sec:gamma-sampling}

\subsubsection{Reversible neighborhood sampling}
\label{sec:neighborhood_search}

On $\{0,1\}^p$, the space of $p$ dimensional inclusion vectors, a random walk that chooses a variable index at random and proposes to flip its inclusion status, quickly becomes inefficient as $p$ grows as it takes on an average $p$ moves before any index gets a ``look-at.'' When the assumed prior concentrates on sparse inclusion vectors, such random walks are heavily biased toward adding additional predictors rather than removing unwanted ones.
To overcome these issues, ``Shotgun Stochastic Search'' (SSS) \citep{sss} uses a neighborhood search procedure in high dimensional spaces to quickly identify inclusions vectors with large posterior mass, and is demonstrated to perform well in linear regression and graphical modeling. Additional MCMC schemes have been developed in the context of exploring high dimensional discrete spaces in applications to statistical physics and genomics \citep{hamze2013,evolLinReg,strens2003evolutionary,mansinghka2009exact}.


It is relatively straightforward to adapt shotgun stochastic search to an MCMC setting for the additive-interactive AGP model. Consider the following single-predictor changes to component inclusion vector $\gamma$, namely 
\begin{enumerate}[(1)]
\item Add neighbors: $N_A(\gamma) = \{\gamma' : \gamma = \gamma + 1_j$, $j \in [\gamma]^c \}$
\item Remove neighbors: $N_R(\gamma) = \{\gamma' : \gamma = \gamma - 1_j$, $j \in \gamma \}$
\item Swap neighbors: $N_S(\gamma) = \{\gamma' : \gamma - 1_j + 1_k$, $(j,k) \in \gamma \times [\gamma]^c\}$.
 \end{enumerate}
Swap moves are equivalent to adding and then removing a (different) predictor, but does so in a single step. A proposal distribution 
\begin{align}
p(\gamma' | \gamma) &= \frac{\pi(\gamma' | - )}{\sum_{\mt{\gamma} \in N(\gamma)} \pi(\mt{\gamma} | -) }, \quad \gamma' \in N(\gamma) = N_A(\gamma) \cup N_R(\gamma) \cup N_S(\gamma), \label{eq:sss_proposal}
\end{align}
defined over these ``one-away neighbors'' may be used in a Markov chain sampler for component inclusion vectors in the AGP model. We call the resulting Markov chain sampler a reversible neighborhood sampler (RNS). RNS allows every predictor to be considered in the context of the current inclusion vector $\gamma$, and is biased towards moves with larger posterior probability. 
A Metropolis step with acceptance probability
\begin{align}
\alpha(\gamma, \gamma') &= \min\left\{1, \frac{\sum_{\mt{\gamma} \in N(\gamma)} \pi(\mt{\gamma} | - )}{\sum_{\mt{\gamma} \in N(\gamma')} \pi(\mt{\gamma} | -)}\right\}\label{eq:sss_mh_alpha}
\end{align}
ensures that updates satisfy detailed balance and the Markov chain maintains the desired stationary distribution in \eqref{eq:gp_gamma_update}. 


The space of inclusion vectors grows exponentially in the predictor dimension, posing potentially serious scalability and mixing issues for RNS when $p$ is large. For an inclusion vector containing $d$ predictors, proposal distribution \eqref{eq:sss_proposal} requires evaluating the likelihood score for $(p-d)$ add, $d$ remove, and $d(p-d)$ swap neighbors. Hence, the per-iteration complexity for sampling component inclusion vectors via back-fitting (see  Section \ref{sec:backfitting_agp}) scales linearly in the predictor dimension. Although likelihood scores may be evaluated in parallel across separate processors, scalable Markov chain sampling of inclusion vectors within the additive-interactive framework requires strategies that address the following:
\begin{inparaenum}[(Problem 1)]
\item
RNS often identifies good configurations, but is unable to quickly transition to them because of the reversibility constraint. This occurs when good inclusion vectors have even better neighborhoods\footnote{Consider modeling $f(x) = g_1(x_1, x_2) + g_2(x_3,x_4,x_5)$ with a two component AGP model, and assume $\gamma_1 = \{1, 2, 5, 6\}$ and $\gamma_2 = \{3\}$. The neighborhoods for a remove proposal $\gamma'_1 = \{1,2,5\}$ or an add proposal $\gamma'_2 = \{3, 4\}$ contains a better inclusion vector than any contained in the current neighborhood, causing the RNS acceptance probability to be small.}; and
\item many neighbors offer little improvement to model fit in high dimensions, and a trade-off exists between the number of additive components used and the neighborhood search size (see Section \ref{sec:component_adaptation}). 
\end{inparaenum}

The first of these issues is addressed by leveraging the {\em disjoint} structure of the defined one-away neighborhood $N(\gamma)$ in terms of its constituent add, remove and swap configurations. In particular, if $\gamma' \in N_A(\gamma)$ then $\gamma \in N_R(\gamma')$; if $\gamma' \in N_R(\gamma)$ then $\gamma \in N_A(\gamma')$; and otherwise $\gamma' \in N_S(\gamma)$ and $\gamma \in N_S(\gamma')$. Using this fact, we propose a paired-moved reversible neighborhood sampler (PRNS) with ``Add-remove'', ``Remove-add'', and ``Swap-swap'' forward-reverse paired neighborhoods. By allowing different moves to be proposed separately and in quick succession, PRNS can dramatically improve mixing in the space of single predictor changes to $\gamma$ (see Section \ref{sec:paired_move_neighborhoods}).


\subsubsection{A paired-move neighborhood sampler}
\label{sec:paired_move_neighborhoods}

The PRNS proposal distribution is given by
\begin{align} \label{eq:paired_move_proposal}
p(\gamma' | \gamma) &= w_A \:p_A(\gamma'|\gamma) + w_R \:p_R(\gamma' | \gamma) + w_S \:p_S(\gamma'|\gamma),
\end{align}
with $w_A, w_R, w_S \ge 0$ and $w_A + w_R + w_S = 1$. These probabilities are allowed to vary with component size to encourage additions to smaller components, removal from larger components, and swaps for intermediately sized ones. An empty component has size $|\gamma| = 0$, so set $w_A = 1$. Otherwise, set $w_A \propto h_A(|\gamma|; \lambda_A)$ and $w_S \propto h_S(|\gamma|; \lambda_S)$, where $h_A$ is monotonically decreasing in $|\gamma|$, and $h_S$ is unimodal at interaction size $\bar{d} = \argmin_\delta P(|\gamma| > \delta | d^* = 1) \le 0.01$.
\begin{figure}[h]
\centering
\includegraphics[width=0.70\linewidth]{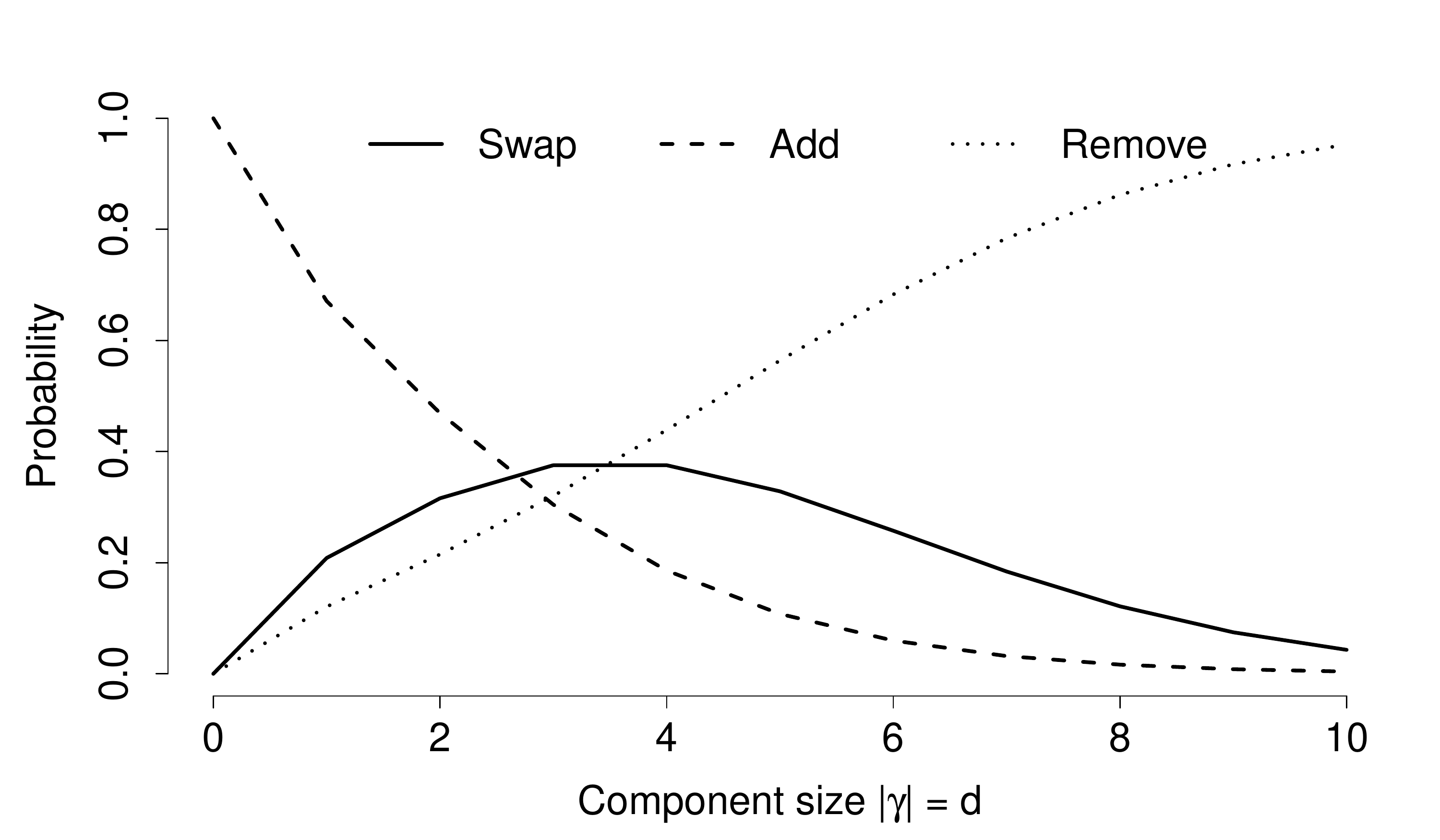}
\caption{Probability of Add/Remove/Swap moves as a function of component-size $|\gamma| = d$. By default, $h_S$ is the Poisson density with $\lambda_S = 4$, $h_A$ is the exponential density with $\lambda_A = d^* = 1$, and $\bar{d} = 4$.}
\label{fig:ARS_prob}
\end{figure}

Paired-moves reduce the number of likelihood evaluations per iteration, though $O(p)$ likelihood evaluations in a GP regression setting remains computationally prohibitive for single computer implementations when $p$ is large. The PRNS transition rule for sampling component inclusion vectors proceeds as follows: 
\begin{enumerate}[(1)]
\item Select move $m \in \{A, R, S\}$ with probability $w_A, w_R,$ and $w_S$, respectively
\item For move $m \in \{A, R, S\}$, construct forward one-away neighborhood $N_m(\gamma)$
\item Propose $\gamma' \sim N_m(\gamma)$ sampled according to PRNS proposal distribution \eqref{eq:paired_move_proposal} restricted to this set 
\item For the corresponding reverse move $m' \in \{R, A, S\}$, construct reverse neighborhood $N_{m'}(\gamma')$. By construction $\gamma \in N_{m'}(\gamma')$
\item Accept proposal $\gamma'$ with probability 
\begin{align}\label{eq:prns_mh_alpha}
\alpha_m(\gamma, \gamma') = \min\left\{1, \frac{ w_{m'}(|\gamma'|) \sum_{\mt{\gamma} \in N_m(\gamma)} \pi(\mt{\gamma} | -)}{w_{m}(|\gamma|) \sum_{\mt{\gamma} \in N_{m'}(\gamma')} \pi(\mt{\gamma} | - )} \right\}.
\end{align} 
\end{enumerate}

\begin{lemma} \label{lem:prns_stationarity}
The paired-move neighborhood sampler (PRNS) satisfies detailed balance and MCMC samples converge to the desired target posterior $\pi(\gamma | -)$.
\end{lemma}
\noindent A proof of Lemma \ref{lem:prns_stationarity} is provided in Section \ref{sec:detailed_balance_pfs} of the Appendix. 

To further reduce the complexity for sampling component inclusion vectors, one might propose random thinning of paired neighborhoods, where a fixed ``budget'' of size $M$ defines the maximum allowable neighborhood size (likelihood evaluations) per component per iteration. Intuitively, $M \asymp o(p)$ seems necessary to be able to explore sufficiently large regions of the inclusion space while being computationally tractable. In particular, consider a randomly thinned paired-move Add transition $\gamma \rightarrow \gamma' = \gamma + 1_j$ having budget $M$. Here, one selects $M$ of $p-|\gamma|$ possible add configurations in the forward set and all $|\gamma|+1 < M$ remove neighbors for the paired reverse set. In the opposite direction direction, transitioning from $\gamma' \rightarrow \gamma = \gamma' - 1_j$ does not require selection of predictor $j$ beforehand as all $|\gamma|+1$ remove neighbors are considered in the forward set, hence the paired reverse (Add) set need only select $M-1$ of $p - |\gamma| - 1$ predictors. This introduces a factor of $M / (p - |\gamma|)$ in the expression for detailed balance which is non-negligible in desired $M \ll p$ limit. Hence, although random thinning of one-away neighborhoods appears to have good empirical performance,  it compromises reversibility of the paired-move Markov chain sampler.

We resolve the issue of reversibility for random neighborhoods by considering a probabilistic formulation for generating Add/Remove/Swap neighbors of a predictor inclusion vector. 
In particular, we adapt the multiple-try Metropolis (MTM) algorithm \citep{mtm} to the discrete setting where transitions are confined to the set of one-away neighbors of component inclusion vector $\gamma$. The approach is general, allowing probabilistic control over neighborhood sizes and adaptive importance scores for predictors comprising these configurations. This addresses problem \#2 raised in Section \ref{sec:neighborhood_search}.

\subsection{Generalized multiple-try Metropolis on discrete spaces}
For a real valued random variable $x \in \mathcal{X} \subseteq \Re$, let $T : \mathcal{X} \times \mathcal{X} \rightarrow [0,1]$ be a transition kernel such that $T(x, x') > 0 \Longleftrightarrow T(x',x) > 0$. In addition, let $\lambda(x, x')$ be some symmetric function and define weight function $\omega(x, x') = \pi(x) T(x, x')\allowbreak\lambda(x, x')$. The MTM algorithm produces draws from target distribution $\pi(x)$ by (1) sampling reference set $x_1^*, \dots, x_M^* \sim T(x, \cdot)$; (2) calculating weights $\omega(x_i^*, x), ~i=1, \dots, M$; (3) drawing $x' \in \{x_1^*, \dots, x_M^*\}$ with probability proportional to $\omega(x',x)$; (4) sampling reverse set $\mt{x}_1, \dots \mt{x}_{M-1} \sim T(x', \cdot)$ having weights $w(\mt{x}_i, x'), \: i = 1, \dots, M-1$ and $\mt{x}_M = x$; and (5) accepting proposal $x'$ w.p. $\alpha(x,x') = \min\Big\{1, \sum_{i=1}^M w(x_i^*,x) / \sum_{i = 1}^M w(\mt{x}_i, x')\Big\}$.

\subsubsection{Random neighborhoods via ``toggled'' predictor inclusion}
\label{sec:toggle_mixed_mtm}

To adapt the MTM algorithm for sampling predictor inclusion vector $\gamma \in \{0,1\}^p$, define $\eta_j \sim \mathrm{Bern}(\omega_j)$ with $\omega : \{0,1\} \times \Re^+ \to [0,1]$ being a weight function taking input $\gamma_j = I(j \in \gamma)$ and nonnegative predictor importance score $v_j$, $j = 1, \ldots, p$. Here, predictor $j$ is ``toggled'' on(off) corresponding to an Add(Remove) whenever $\gamma_j = 0(\gamma_j = 1)$ and $\eta_j = 1$. The unique set of selected predictor ``toggles'' ensures no duplicates among the constructed Add/Remove/Swap neighbors. We initialize $\gamma = 0$ and predictors are treated exchangeably a priori by setting importance $v_j = 1$, $j = 1, \ldots, p$, for example. This frameworks lends itself naturally to adaptive schemes which are useful in sampling high dimensional inclusion vectors. In particular, effective strategies will enhance the chance of selecting predictors with large $v_j$ when $j \not\in \gamma$ and suppress them when $j \in \gamma$.

Let $\pi(\gamma)$ stand for the conditional posterior distribution in \eqref{eq:gp_gamma_update} and define operator $\tog(j, \gamma) : \gamma_j \leftarrow 1-\gamma_j$. Below, we restrict ourselves to Add/Remove one-away neighbors for simplicity but note that Swaps are trivially incorporated by considering proposals $\mt{\gamma} = \tog((r,a), \gamma)$ for predictor pairs $(r,a) \in \{(j,k) : \gamma_j = 1, \gamma_k = 0\}$ subject to $\eta_r = \eta_a = 1$. This case is handled in detail in the section that follows. An algorithm for the proposed generalized MTM over a $p$-variate discrete space (DMTM) is:
\begin{enumerate}[(1)]
\item For $j = 1, \dots, p$, draw $\eta_j \sim \mathrm{Bern}(\omega(\gamma_j, v_j))$ independently
\item For $k \in \{j : \eta_j = 1\}$, $\mt{\gamma} = \tog(k, \gamma)$ defines the forward set of mixed Add and Remove neighbors 
\item Among the constructed forward neighbors, select $\gamma' = \tog(k^*, \gamma)$ with probability $\pi(\gamma') / \sum_{j: \eta_j = 1} \pi(\tog(j, \gamma))$
\item For $j \ne k^*$, draw $\eta_j' \sim \mathrm{Bern}(\omega(\gamma'_j, v_j))$ independently and set $\eta_{k^*}' = 1$. 
For $k \in \{j : \eta_j' = 1\}$, $\mt{\gamma} = \tog(k, \gamma')$ defines the reverse set of mixed Add and Remove neighbors
\item Accept proposal $\gamma'$ with probability 
\begin{align} \label{eq:discrete_mixed_mtm_alpha}
\alpha(\gamma, \gamma') = \min\left\{1, \frac{\omega(\gamma_{k^*}', v_{k^*}) \sum_{j: \eta_{j} = 1} \pi(\tog(j, \gamma))}{\:\omega(\gamma_{k^*}, v_{k^*}) \sum_{j: \eta'_{j} = 1} \pi(\tog(j, \gamma'))}\right\}.
\end{align}
\end{enumerate}

\begin{lemma}\label{lem:discrete_mixed_mtm}
The DMTM algorithm satisfies the detailed balance condition, and MCMC samples converge to the desired target posterior $\pi(\gamma)$. 
\end{lemma}
\noindent The proof of Lemma \ref{lem:discrete_mixed_mtm} follows from the proof of Theorem \ref{thm:paired_move_wmtm}. See Section \ref{sec:detailed_balance_pfs} of the Appendix for details.

\subsubsection{A paired-move reversible MTM sampler}
\label{sec:paired_move_wmtm}
A probabilistic version of the paired-move neighborhood sampler (PRNS) introduced in Section \ref{sec:paired_move_neighborhoods} is obtained as a special case of the DMTM algorithm for a specific choice of weight function $\omega$, namely
\begin{align}\label{eq:weight_fn_wmtm}
\omega(\gamma_j, v_j; m) = (1-\gamma_j) f(v_j) I(m = A) + \gamma_j I(m = R), \quad \gamma_j = I(j \in \gamma).
\end{align}
The DMTM algorithm is modified to incorporate paired-moves as follows:
\begin{enumerate}[(1)]
\item Select move $m \in \{ A, R, S\}$ with probabilities $w_A(|\gamma|), w_R(|\gamma|),$ and $w_S(|\gamma|)$, respectively. 
\item
\begin{enumerate}[(a)]
\item If move $m \in \{A,R\}:$ for $j = 1, \dots, p$, draw $\eta_j \sim \mathrm{Bern}(\omega(\gamma_j, v_j; m))$. For $k \in \{j : \eta_j = 1\}$, $\mt{\gamma} = \tog(k, \gamma)$ defines the set of Add/Remove neighbors 
\item If move $m = S:$ define $(\eta_r, \eta_a) \in \{0, 1\}^2$ and consider predictor pairs $(r,a) \in \{(j,k) : \gamma_j = 1, \gamma_k = 0\}$. For $r \in \gamma$, draw $\eta_r \sim \mathrm{Bern}(\omega(\gamma_r, v_r, m = R))$, and independently for $a \in [\gamma]^c$ draw $\eta_a \sim \mathrm{Bern}(\omega(\gamma_a, v_a, m = A))$. For $(r,a) : \eta_r = \eta_a = 1$, $\mt{\gamma} = \tog((r,a), \gamma)$ defines the set of Swap neighbors
\end{enumerate}
\item Among the constructed forward neighbors, select $\gamma' \in N_m(\gamma)$ with probability $\pi(\gamma') / \sum_{\mt{\gamma} \in N_m(\gamma)} \pi(\mt{\gamma})$. If $m \in \{A, R\}$, denote $\gamma' = \tog(k^*,\gamma)$; otherwise denote $\gamma' = \tog((r^*, a^*), \gamma)$ for $m = S$
\item
\begin{enumerate}[(a)]
\item If move $m = A:$ for $j \ne k^*$, draw $\eta_j' \sim \mathrm{Bern}(\omega(\gamma_j', v_j, m' = R))$ and set $\eta_{k^*}' = 1$. Using the defined weight function, 
note that for $j \not\in \gamma'$, $\omega_j = 0 \Longrightarrow \eta_j' = 0$. For $k \in \{j : \eta_j' = 1\}$, $\mt{\gamma} = \tog(j, \gamma')$ defines the set of reverse (remove) neighbors 
\item If move $m = R:$ for $j \ne k^*$, draw $\eta_j' \sim \mathrm{Bern}(\omega(\gamma_j', v_j, m' = A))$ and set $\eta_{k^*}' = 1$. Using the defined weight function, 
note that for $j \in \gamma'$, $\omega_j = 0 \Longrightarrow \eta_j' = 0$. For $k \in \{j : \eta_j' = 1\}$, $\mt{\gamma} = \tog(j, \gamma')$ defines the set of reverse (add) neighbors 
\item If move $m = S:$ define $(\eta_r', \eta_a') \in \{0, 1\}^2$ as before, with predictor pairs $(r,a) \in \mathcal{S}(
\gamma') = \{(j,k) : \gamma_j' = 1, \gamma_k' = 0\}$. For $(r,a) \in \mathcal{S}(\gamma') : (r,a) \ne (a^*, r^*)$, draw $\eta_r' \sim \mathrm{Bern}(\omega(\gamma_r', v_r, m' = R))$ and independently draw $\eta_a' \sim \mathrm{Bern}(\omega(\gamma_a', v_a, m' = A))$, and set $(\eta_{a^*}', \eta_{r^*}') = (1,1)$. For $(r,a) : \eta_r' = \eta_a' = 1$, $\mt{\gamma} = \tog((r,a), \gamma')$ defines the set of reverse (swap) neighbors
\end{enumerate}
\item For $m \in \{A, R, S\}$, the corresponding reverse paired neighborhood is $m' \in \{R, A, S\}$. Let $s^* = k^*$ if $m \in \{A,R\}$; otherwise $s^* = (r^*, a^*)$ for $m = S$. With forward and reverse paired neighborhoods $N_m$ and $N_{m'}$ as defined above, accept proposal $\gamma' = \tog(s^*, \gamma)$ with probability 
\begin{align} \label{eq:paired_move_wmtm_alpha}
\alpha_m(\gamma, \gamma') = \min\left\{1, \frac{w_{m'}(|\gamma'|)}{w_{m}(|\gamma|)} \frac{\omega(s^*, \bv; m') \sum_{\mt{\gamma} \in N_{m}(\gamma)} \pi(\mt{\gamma})}{\:\omega(s^*, \bv; m) \sum_{\mt{\gamma'} \in N_{m'}(\gamma')} \pi(\mt{\gamma})}\right\}.
\end{align}
 \end{enumerate}

\begin{theorem} \label{thm:paired_move_wmtm}
The paired-move discrete MTM algorithm satisfies the detailed balance condition, and samples therefore converge to the desired target posterior $\pi(\gamma)$. 
\end{theorem}
\noindent A proof of Theorem \ref{thm:paired_move_wmtm} is provided in Section \ref{sec:detailed_balance_pfs} of the Appendix. 

This is a non-trivial generalization of the MTM algorithm which extends MCMC for sampling high dimensional inclusion vectors and enables scalable variable selection outside of the linear regression setting. The framework is general and versatile, allowing sparsity and other model assumptions (e.g., additivity, smoothness etc.) to be incorporated in the sampling of inclusion vectors as is done for the AGP model. Adaptive MCMC strategies within this framework are discussed in Section \ref{sec:inclusion_propensity_adapt}.

\subsection{Additional ensemble-wide strategies}
\subsubsection{Additive inter-component moves}
\label{sec:icm_sampling}

To enable predictors to be rearranged or exchanged between components rapidly, we propose inter-component moves (ICM). These moves are complementary to the sequential updating of component inclusion vectors using paired-moves in the back-fitting MCMC scheme of Section \ref{sec:cond_gp_sampling}. While paired Add/Remove/Swap employ a ``divide-and-learn'' approach to exploring the space of predictor inclusion, ICM moves propose updates to inclusion state $\bxi = (\bgamma, \btheta) = \{(\gamma_l, \theta_l) : l = 1, \dots, k\}$ via transitions $((\gamma_m, \theta_m), (\gamma_n, \theta_n)) \rightarrow ((\gamma_m', \theta_m'), (\gamma_n', \theta_n'))$
between active components $(m,n) \in \mathcal{I}_A$. Here, $\theta_l = (\rho_l, \lambda_l) \in \mathcal{G}$ denotes the scaling for GP component $l$. The following transitions are considered: 

\begin{enumerate}[(1)]
\item \label{itm:cross_donate} Cross-component donate: choose an active component at random, say $\gamma_n$. For predictors $j \in \gamma_n$, define $\mt{\bgamma}_{jm} = \{\dots, \gamma_m + 1_j, \gamma_n - 1_j, \dots\}$ for every other active component $m \ne n$. Set $N_{\rm CD}(\bgamma; n) = \{\mt{\bgamma}_{jm}: j \in \gamma_n, m \ne n\}$ 
\item \label{itm:pair_donate} Paired donate: choose indices of two active components $(m,n)$ at random, with $\gamma_n$ taken to be the non-empty {\it donor} component by convention. For predictors $j \in \gamma_n$, define $\mt{\bgamma}_{jm} = \{\dots, \gamma_m + 1_j, \gamma_n - 1_j, \dots\}$ and set $N_{\rm PD}(\bgamma; (m,n)) = \{\mt{\bgamma}_{jm} : j \in \gamma_n\}$ 
\item \label{itm:pair_swap} Paired swap: choose indices of two active non-empty components $(m,n)$ at random. For predictor pair $(j,k) \in \mathcal{S}(\gamma_m, \gamma_n) = \{(p,q) \in \gamma_m \times \gamma_n$ : $ p \ne q\}$, define $\mt{\bgamma}_{jk} = \{\dots, \gamma_m -1_j + 1_k, \gamma_n - 1_k + 1_j, \dots\}$ and set $N_{\rm PS}(\bgamma; (m,n)) = \{\mt{\bgamma}_{jk} : (j,k) \in \mathcal{S}(\gamma_m, \gamma_n)\}$.
\end{enumerate}

Cross-component moves in \eqref{itm:cross_donate} facilitate a rapid exchange of predictors from a selected active component to any other active component holding fixed all component scalings (i.e. transitions for $\theta$ occur via the identity map). In contrast, the proposal neighborhood for \eqref{itm:pair_donate} and \eqref{itm:pair_swap} is defined over $N_{[\cdot]}(\bgamma; (m,n)) \times \mathcal{G}^2$, where the acceptance probability for paired Donate/Swap proposals between an active component pair is maximized via optimal scaling. Sampling for ICM proceeds as follows:
\begin{enumerate}[(1)]
\item Select move $m \in \{\rm CD, PD, PS\}$ with probability $w_{\rm CD}$, $w_{\rm PD}$, and $w_{\rm PS}$, respectively
\item
\begin{enumerate}[(a)]
\item If $m = {\rm CD}:$ let $k$ be the index for the randomly selected non-empty active component. Then select $(\bgamma', \btheta' = \btheta) \in A_m(\bgamma) = N_m(\bgamma; k)$ with probability $\pi(\bgamma' | -) / \allowbreak \sum_{\mt{\bgamma} \in A_m} \pi(\mt{\bgamma} | -)$
\item If $m \in \{\rm PD, PS\}:$ let $(k,l)$ denote indices of the two randomly selected active components. Then select $(\bgamma', \btheta') \in A_m(\bgamma) = N_m(\bgamma; (k,l)) \times \mathcal{G}^2$ with probability $\pi(\bgamma', \btheta' | -) / \sum_{(\mt{\bgamma}, \mt{\btheta}) \in A_m} \pi(\mt{\bgamma}, \mt{\btheta} | -)$
\end{enumerate}
\item Accept transition to state $\bxi' = (\bgamma', \btheta')$ under the defined proposal distribution for move $m$ with probability 
\begin{align} \label{eq:icm_acceptance_ratio}
\alpha_m(\bxi, \bxi') = \min\left\{1, \frac{\sum_{\mt{\bxi} \in A_m(\bxi)} \pi(\mt{\bxi} | -)}{\sum_{\mt{\bxi} \in A_m(\bxi')} \pi(\mt{\bxi} | -)}\right\}.
\end{align}
\item \label{itm:icm_par_gibbs_update} If $m = {\rm CD}:$ update scale parameters $\btheta = \{\rho_l, \lambda_l : l \in \mathcal{I}_A\}$ via Gibbs transition kernel $T(\btheta, \btheta' | \bgamma, \by) = \prod_{l \in \mathcal{I}_A} \pi(\theta_l | \btheta'_{j < l}, \btheta_{j > l}, \bgamma, \by)$
\end{enumerate}

ICM moves are particularly useful in low signal and high dimensional predictor settings. Across random test/train replicate runs on real data illustrations in Section \ref{sec:real_data_analysis}, variance of the predictive mean-squared-error is consistently lower when MCMC uses ICM moves (see Table \ref{tab:variance_icm}). 
 
\begin{lemma} \label{lem:icm_stationarity}
For state $\bxi = (\bgamma, \btheta)$, $\xi_l = (\gamma_l, \rho_l, \lambda_l)$ and $l \in \mathcal{I}_A$, the ICM transition kernel preserves stationary distribution $\pi(\bxi | \by)$.
\end{lemma}
\noindent A proof of Lemma \ref{lem:icm_stationarity} provided in Section \ref{sec:detailed_balance_pfs} of the Appendix.

\begin{remark}
\label{lem:Tstationary}
Transition kernel $T(\btheta, \btheta' | \bgamma, \by)= \prod_{l \in \mathcal{I}_A} \pi(\theta_l | \btheta'_{j < l}, \btheta_{j > l}, \bgamma, \by)$ has stationary distribution $\pi(\btheta | \bgamma, \by)$. This follows from the fact that $T$ is a Gibbs transition kernel. ICM moves used in conjunction with such a transition kernel for component scaling parameters also preserves stationarity.
\end{remark}

\subsubsection{Updating empty components}
\label{sec:component_adaptation}
From Section \ref{sec:prior_initialization}, recall that $k_a = \sum_{l = 1}^{k_\mathrm{max}} I(\rho_l > 0 \text{ or } d_l > 0)$ defines the number of active components in the AGP model. Algorithm \ref{alg1} is initialized with fixed lower and upper bounds on the number of components, while allowing empty components with no predictors included (i.e., $d_l = 0$). Non-empty components explaining little variation in the response will often have signal-to-noise ratio $\rho = 0$, but remain active to allow AGP to discover additional additive structure in subsequent iterations.

Let $B$ denote a fixed computational budget (likelihood evaluations) per iteration. A trade-off exists between a per-component budget $M$ and the total number of active components $k_a$, namely $B = M k_a$. This suggest the possibility of including only as many components as needed,  and having more detailed neighborhood scans when updating individual inclusion vectors. Component inclusion vectors are updated with probability $\varrho(\rho_l, \gamma_l; \theta_0)$. Active set $\mathcal{I}_A$ is updated by Algorithm \ref{alg2} in each MCMC iteration, with baseline probability $\theta_0$ initialized so that, on average, one inactive component is ``switched on.''
\begin{algorithm}[h]
\begin{algorithmic}[1]
\caption{Updating active component index set} \label{alg2}
\Require 
\begin{inparaenum}[(i)] \item Active index set $\mathcal{I}_A$; \item State vector $\xi_l = (\gamma_l, \rho_l, \lambda_l)$, $l \in \mathcal{I}_A$; \item Fixed target budget $B$ (likelihood evaluations per iteration)
\end{inparaenum}
\Ensure 
\begin{inparaenum}[(i)] \item Updated active index set $\mathcal{I}_A$; \item Target neighborhood budget $M$ \end{inparaenum}
\Function{UPDATE.ACTIVESET}{$\bxi$, $\mathcal{I}_A$, $B$}
\State Initialize $\theta_0 = 0$
\State Compute $k_a = \sum_{l = 1}^{k_{\max}} I(\rho_l > 0 \text{ or } d_l > 0)$
\If{$k_a < k_{\max}$} 
\State set $\theta_0 = (k_{\max} - k_a)^{-1}$
\EndIf

\\
\For{$l = 1 : k_{\max}$}
\State $\varrho_l = \theta_0 + (1 - \theta_0) I(l \in \mathcal{I}_A)$
\If{($l > k - k_{\min}$, $|\mathcal{I}_A| < k_{\min}) \text{ or Unif(0,1)} < \varrho_l$}
	\State $\mathcal{I}_A \leftarrow \mathcal{I}_A \cup \{l\}$
\Else
	\State $\mathcal{I}_A \leftarrow \mathcal{I}_A \setminus \{l\}$
\EndIf
\EndFor

\State Set $M = B / |\mathcal{I}_A|$
\EndFunction
\end{algorithmic}
\end{algorithm}

\subsection{Adaptive predictor importance}
\label{sec:inclusion_propensity_adapt}



We propose an adaptive version of the Markov chain sampler where a predictor's chance of being included in an Add move, for example, is proportional to its importance score. Let $T$ denote the length of the MCMC chain, and define $b_0 = \max\{100, \lfloor T / 10 \rfloor \}$ as a burn-in period. The importance score of a predictor is initialized as $v_j = 1$, $j = 1, \dots, p$, and subsequently updated as
\begin{align} \label{eq:propensity_update}
v_j(t+1) = v_j(t) + \frac{\sum_{l \in \mathcal{I}_A} I(\rho_l > 0, j \in \gamma_l)}{k_a^\zeta}\left( \frac{I(t \le b_0)\: t}{b_0} + \frac{I(t > b_0)}{(t - b_0)^\zeta} \right),
\end{align}
where $\zeta \in (1/2,1]$. Following convention for ``optimal learning rates'' in stochastic gradient descent algorithms, we fix $\zeta = 2/3$. Here, importance scores are increased for predictors frequently included in the past, and adaptations designed to be modest, typically converging to an equilibrium where $\max(v_j) \approx10$ (see Figure \ref{fig:real_posterior_plots}).  Stationarity is maintained subject to diminishing adaptation of predictor importance scores used in the paired-move DMTM sampler \citep{roberts2007coupling}.

The ``toggle'' mechanism introduced in Section \ref{sec:toggle_mixed_mtm} for selecting a subset of predictors from which to form one-away neighborhoods of an inclusion vector is extremely flexible. The weight function \eqref{eq:weight_fn_wmtm} defined in the paired-move DMTM algorithm introduced in Section \ref{sec:paired_move_wmtm} lends itself naturally to adaptive MCMC strategies. Crucially, learning of predictor importance scores borrows information across components, where predictors deemed important in one component may subsequently be promoted via Add/Swap moves in others. This is useful in sampling high dimensional inclusion vectors, allowing probabilistic neighborhood subset selection, with the weight function controlling the size and richness of one-away neighborhoods.

Let $f : \Re^+ \rightarrow [0,1]$ for weight function $\omega(\gamma_j, v_j; m)$ defined in \eqref{eq:weight_fn_wmtm}. In particular, we consider 
\begin{equation}\label{eq:predictor_inclusion_pr}
f(v_j; M) = \frac{M v_j^\alpha}{M v_j^\alpha + p}, ~\alpha \ge 0, 
\end{equation}
where $M$ is the target neighborhood budget set by Algorithm \ref{alg2}. At initialization, $v_j = 1$ for all predictors, so the expected size of an Add neighborhood is $\sum_{j \not\in \gamma} f(v_j) = M\big(\frac{p - |\gamma|}{p + M}\big) \approx M$. Under the sparsity assumption, the vast majority of predictors retain scores $v_j \approx 1$, and so the probabilistic control on neighborhood sizes is maintained. For paired Swap moves, letting $\tilde f(v) = f(v) / |\gamma|$ maintains the same limit. As importance scores are updated, predictors with large $v_j$ are promoted within proposal neighborhoods (note: by our choice of weight function, all remove configurations are considered in paired forward/reverse neighborhoods). Parameter $\alpha$ controls the degree of inclusion bias, and we recommend $\alpha \in [1, 1.5]$. As an example: assume the setup for simulated experiments in Section \ref{sec:simulation} with $\alpha = 1.5$, $p = 1000$, $k_{\max} = \lceil \sqrt{p}\rceil$, $B = 10 k_{\max}$ and assume an active set of size $|\mathcal{I}_A| = 10$. Then, predictors with $v_j = 5(10)$ have inclusion probability 0.26(0.50).

\section{Numerical experiments} \label{sec:simulation}

The AGP method is compared to state-of-the-art competitors in terms of predictive root-mean-squared error (RMSE) and support recovery across a variety of simulated truths. A 1000 iteration MCMC chain is run for AGP using Algorithm \ref{alg1} with default initializations. Averaging across independent replicates and test functions, a non-optimized AGP implementation in R completes in $121_{18}$ minutes for $p = 1000$\footnote{Runtimes are averaged over 5 independent replicates for each test function considered in Section \ref{sec:varSelectionInteractionRecovery}. Simulations were run on an x86$\times$64 Intel(R) Core(TM) i7-3770.}. Results discard the first 200 samples and thin subsequent samples by selecting every fourth draw.

Synthetic data are generated by drawing $x_{ij} \sim \mathrm{Unif}(0,1)$ and $y_i \sim \mathrm{N}(f(x_i), 1)$, $1 \le i \le n = 100, ~ 1\le j \le p$. When the predictor dimension is small, a single GP prior with an ARD squared exponential kernel can identify important predictors but offers no characterization of lower dimensional interaction. Tree models can accomplish both tasks, but often suffer in terms predictive accuracy for nonlinear functions. We select two popular ensemble methods, BART \citep{bart} and Random Forests (RF) \citep{randomForest}. The Lasso \citep{tibshirani1996regression} is selected as a final competitor. The NULL model reports prediction using a na\"{\i}ve average over training outcomes.


\subsection{Variable selection and interaction recovery}
\label{sec:varSelectionInteractionRecovery}
\begin{figure}[!ht]
\caption{Simulated test functions 1-4 for Section \ref{sec:varSelectionInteractionRecovery} with $p = 1000$. {\em Left}: number of active vs. non-empty (utilized) components, with median sizes appearing as vertical dashed lines; {\em Middle}: marginal inclusion probabilities for the AGP model (index on log-scale); {\em Right}: an interaction graph among the most important predictors. Vertex sizes are proportional to predictor importance, and edges between predictor pairs are drawn in proportion to their co-appearance across GP components.}
\label{fig:var_all}
\centering
\arrayrulecolor{gray}
\setlength{\tabcolsep}{2pt}
\begin{tabular}{m{0.16\textwidth} | C{0.26\textwidth} C{0.26\textwidth} C{0.27\textwidth}}
& Components & Marginal inclusion & Interaction graph \\
\hline 
\parbox[t]{\linewidth}{Friedman \\ $n = 100$ \\ $p = 1000$}
& \includegraphics[width=\linewidth]{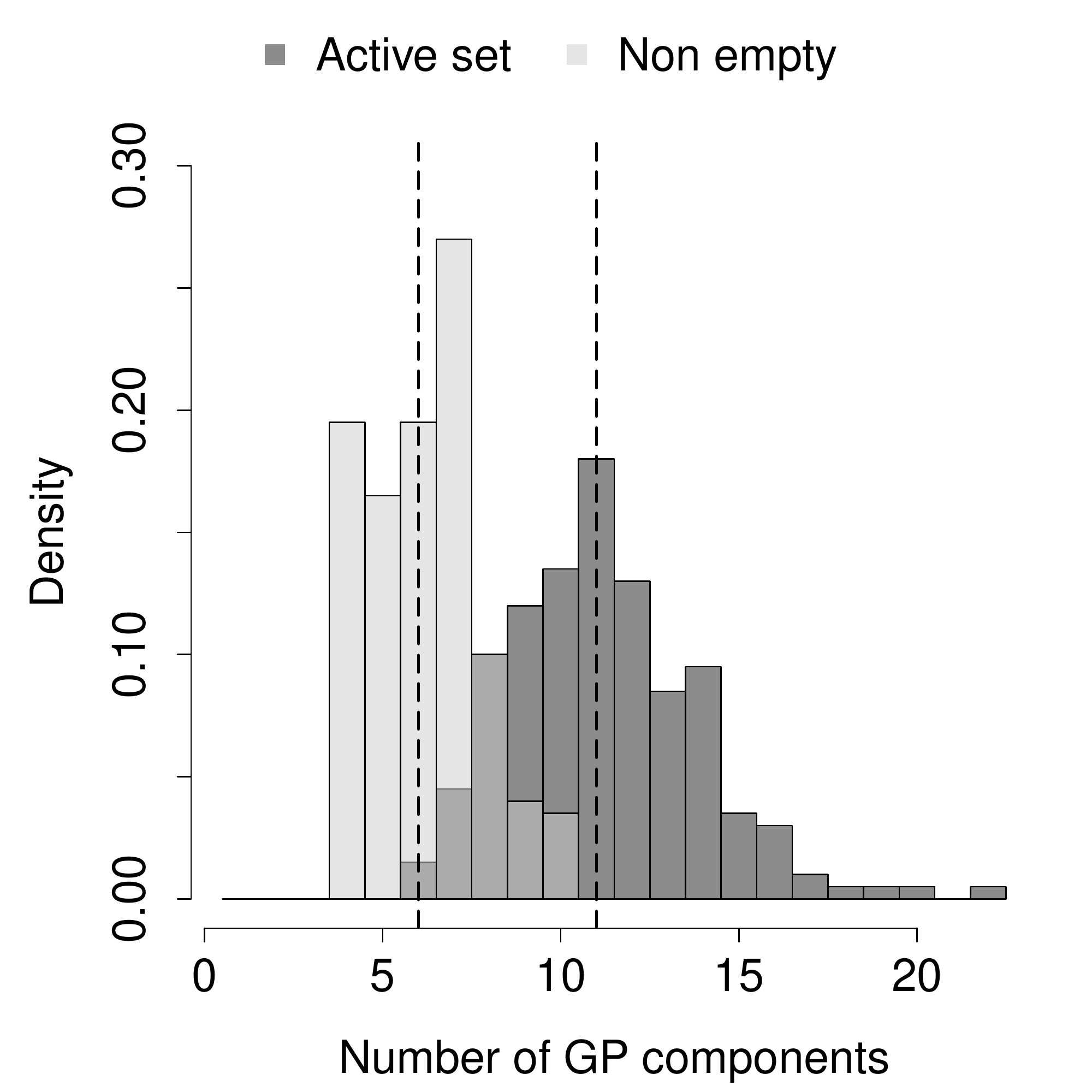} 
& \includegraphics[width=\linewidth]{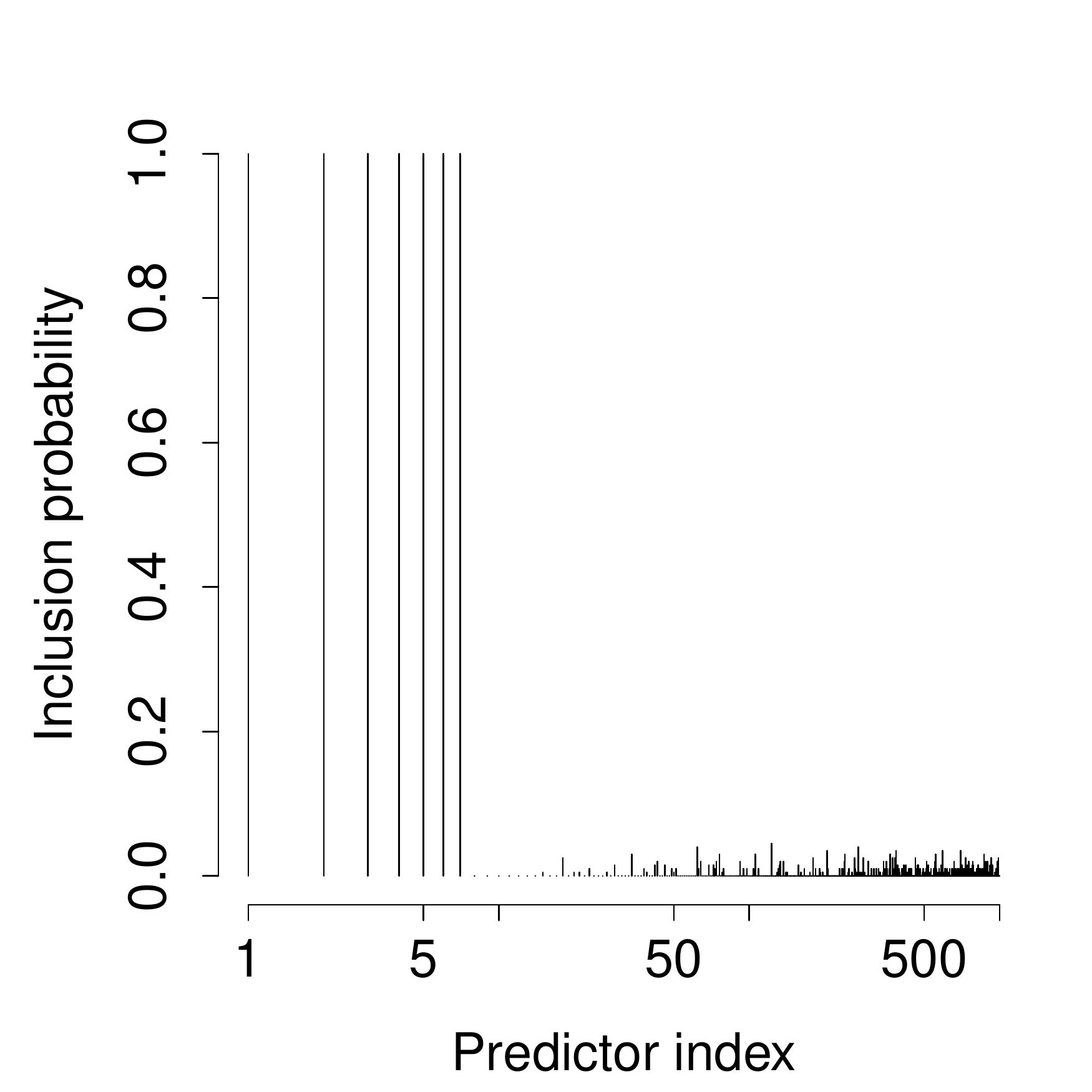}
& \includegraphics[width=\linewidth]{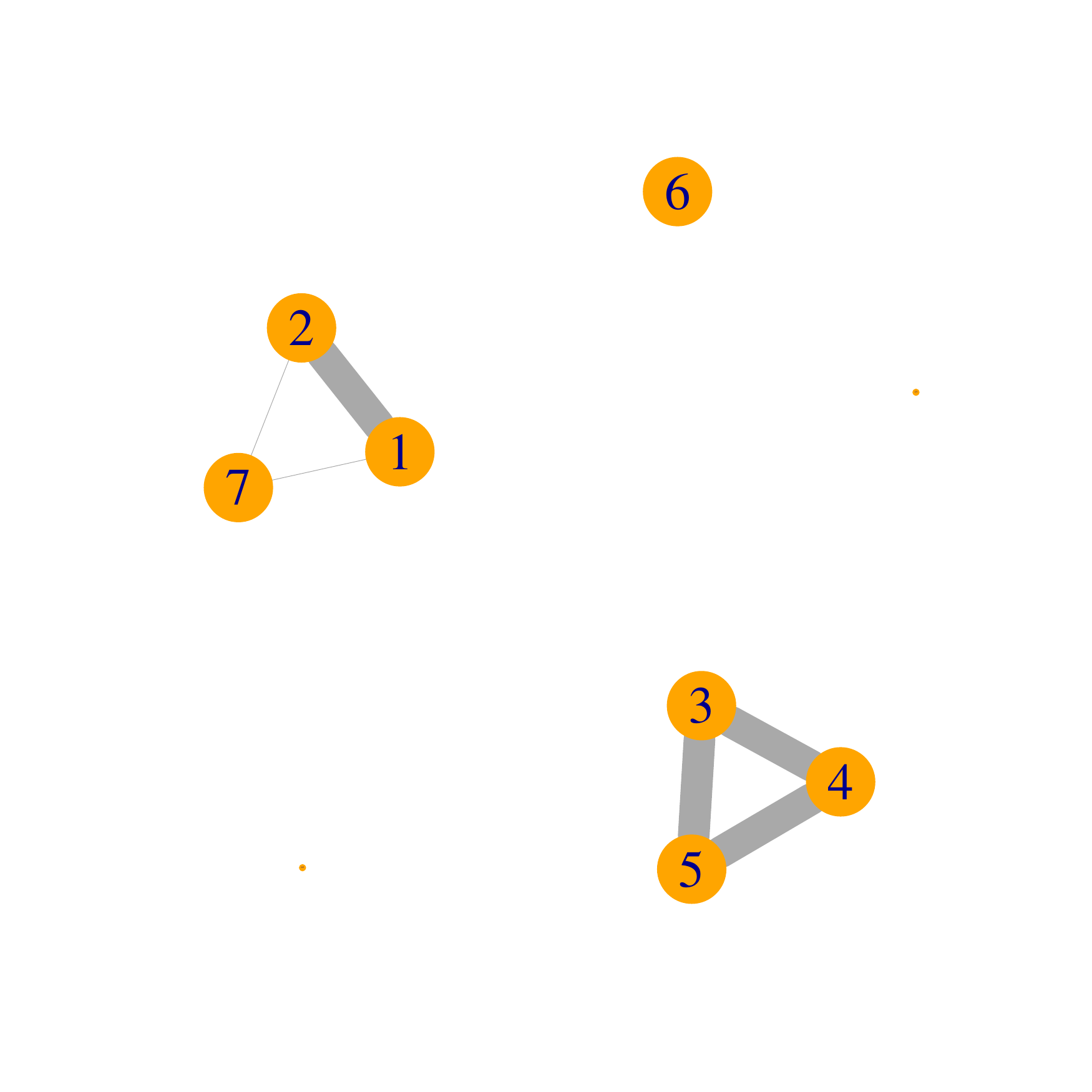} \\
\hline
\parbox[t]{\linewidth}{Confounded \\ $n = 100$ \\ $p = 1000$}
& \includegraphics[width=\linewidth]{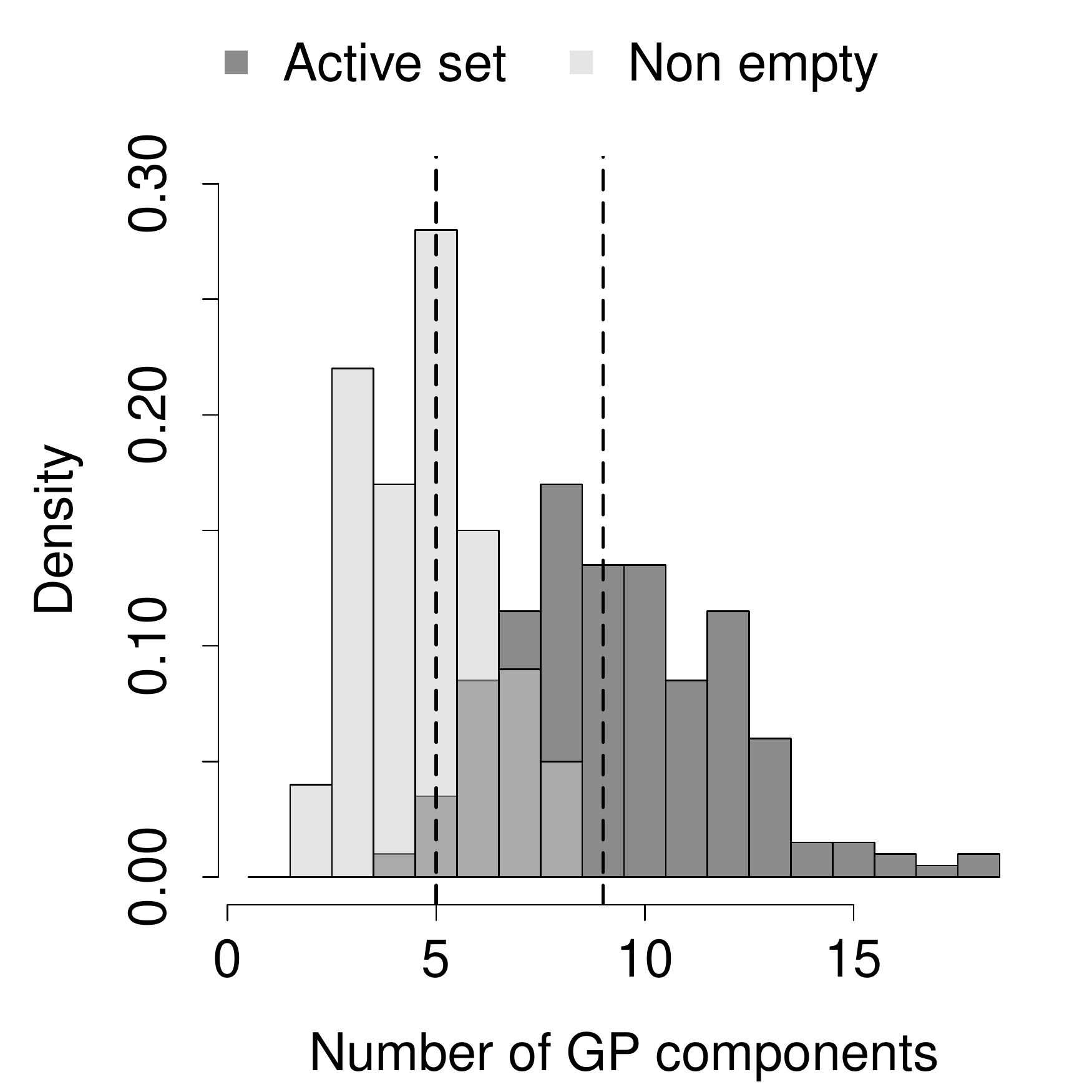} 
& \includegraphics[width=\linewidth]{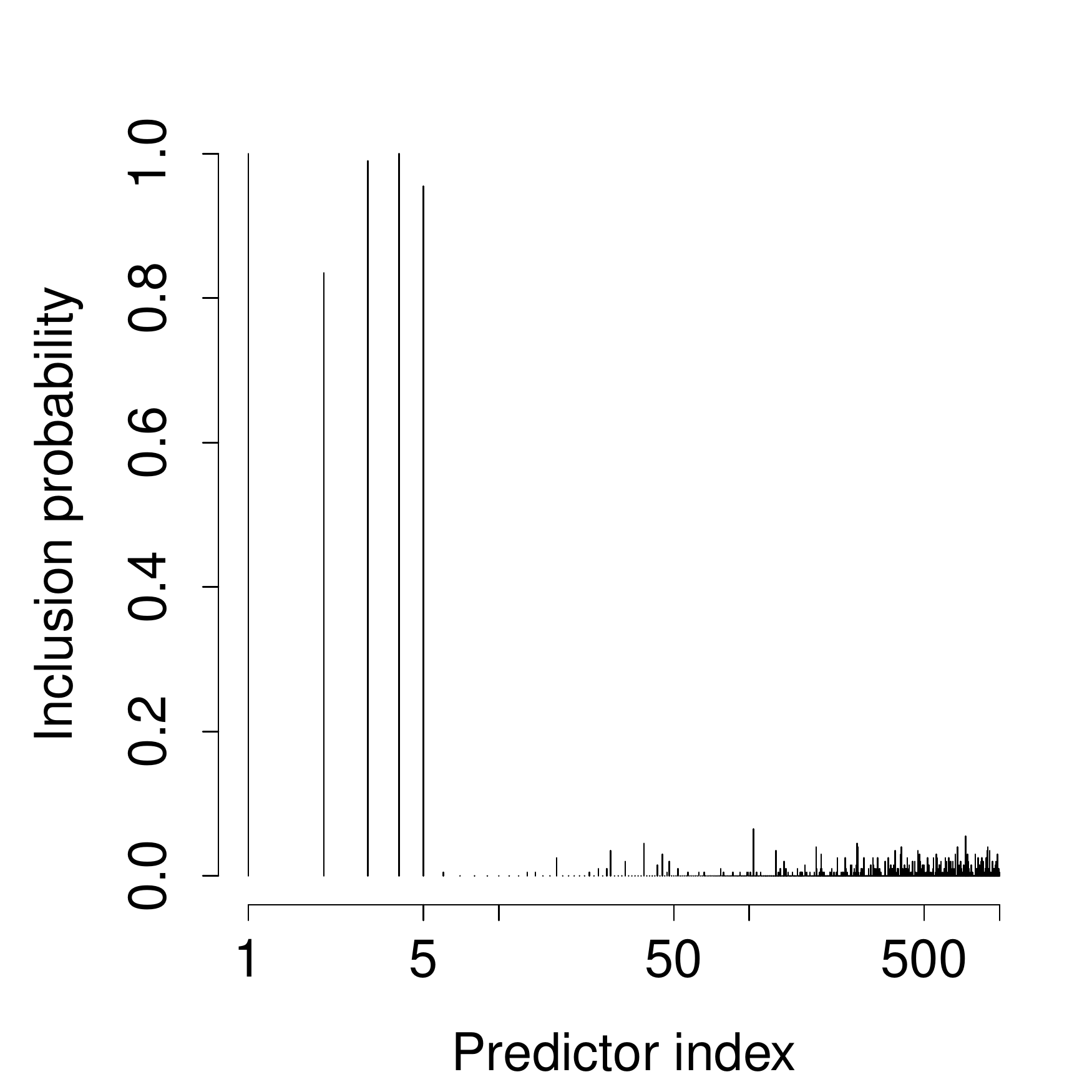}
& \includegraphics[width=\linewidth]{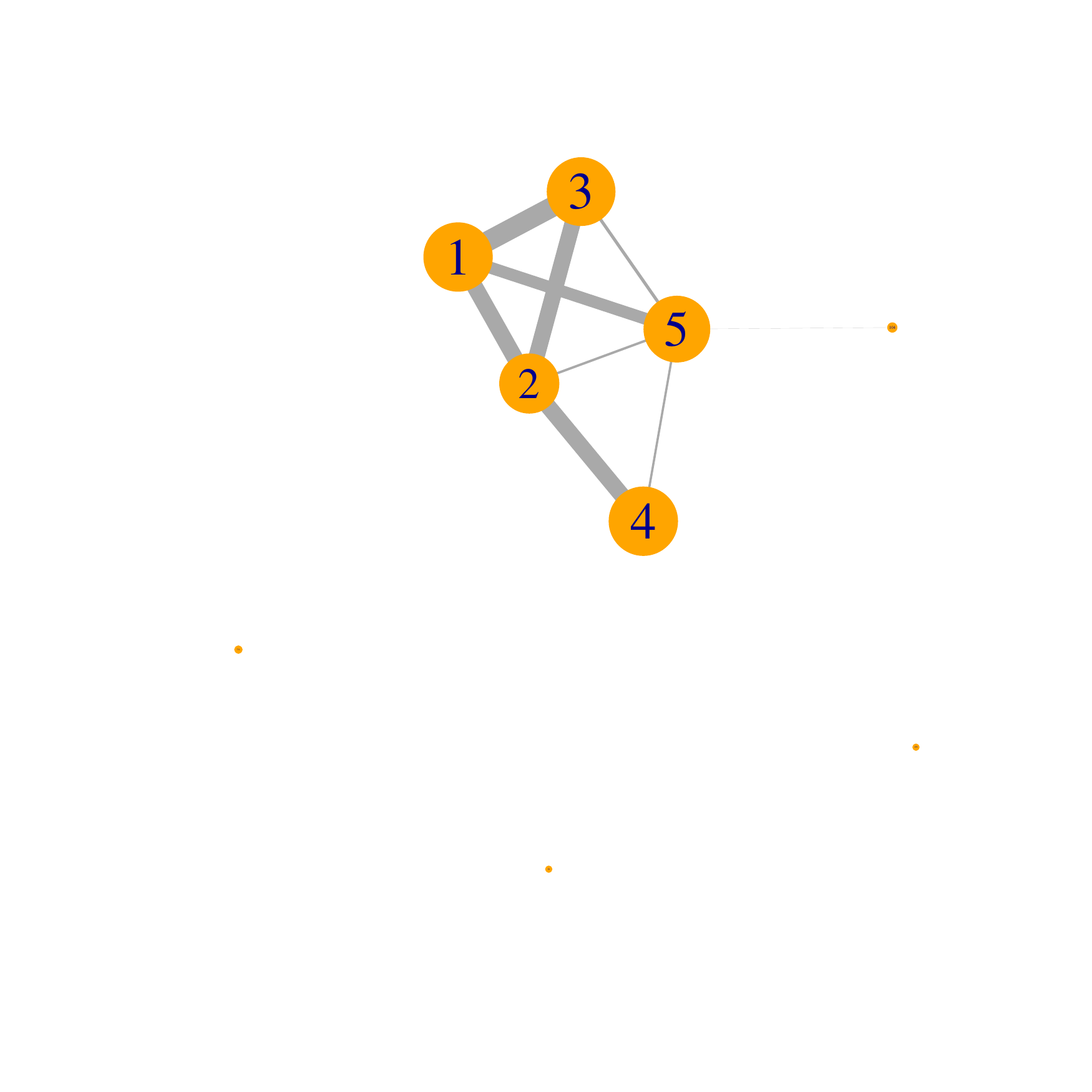} \\
\hline
\parbox[t]{\linewidth}{Linear \\ $n = 100$ \\ $p = 1000$}
& \includegraphics[width=\linewidth]{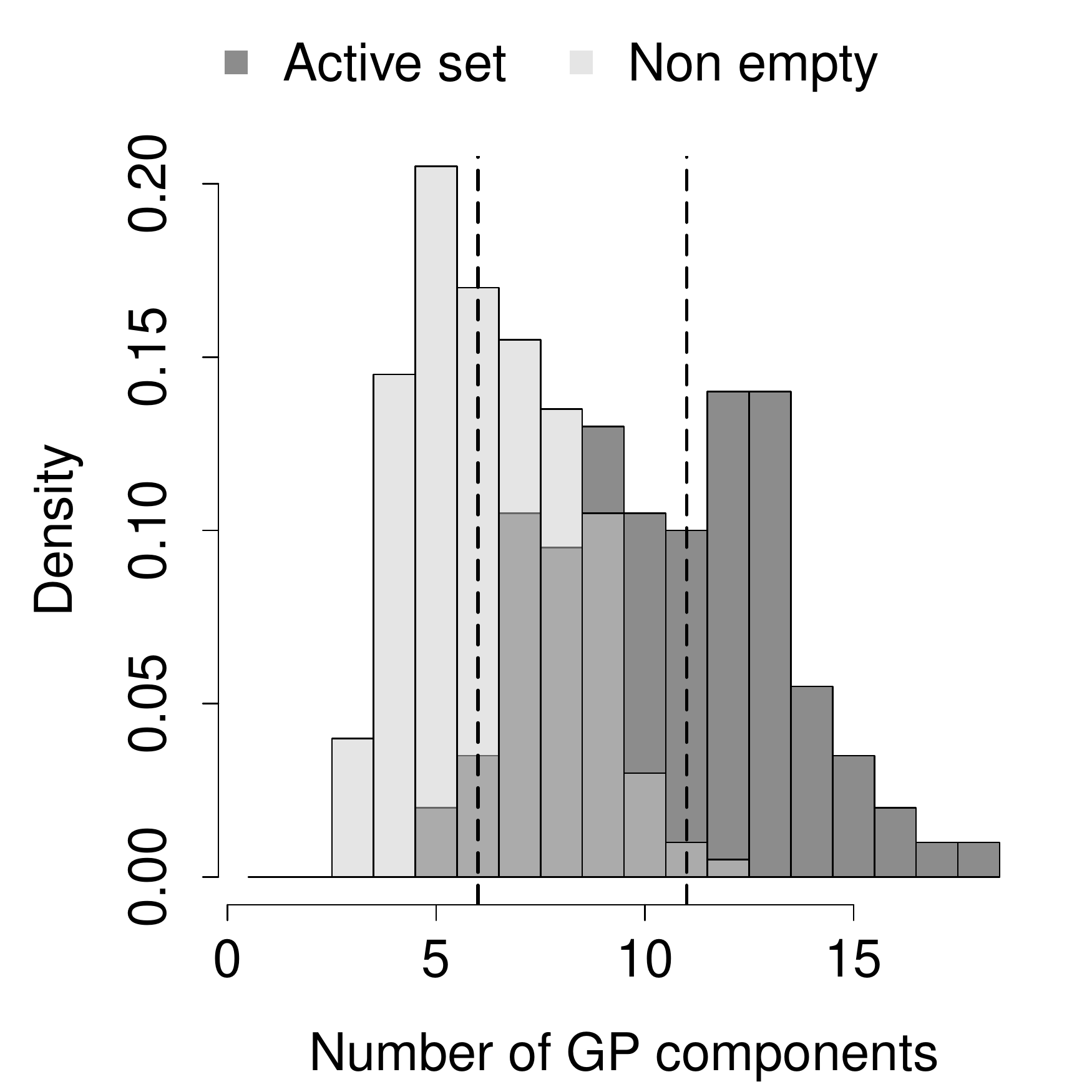} 
& \includegraphics[width=\linewidth]{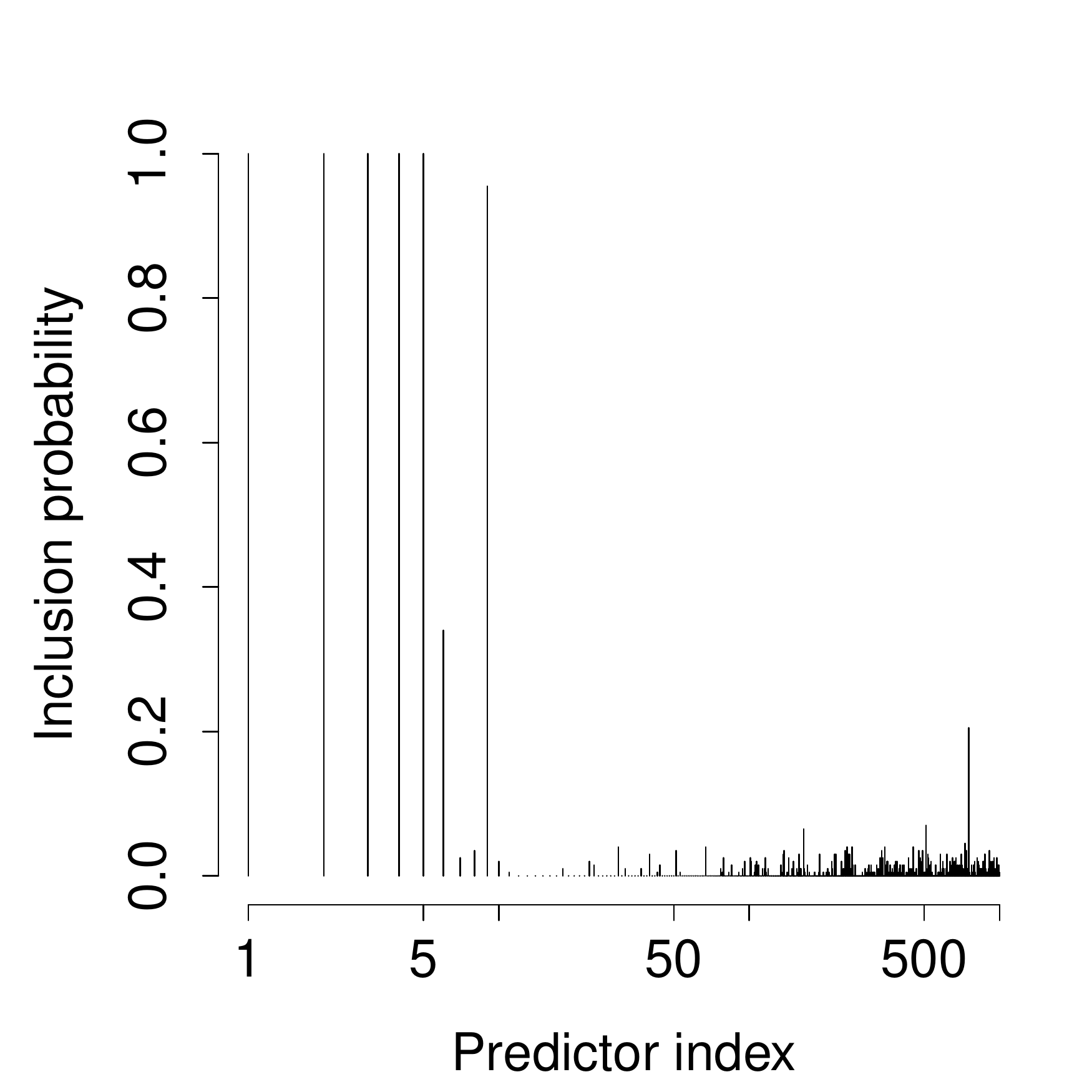}
& \includegraphics[width=\linewidth]{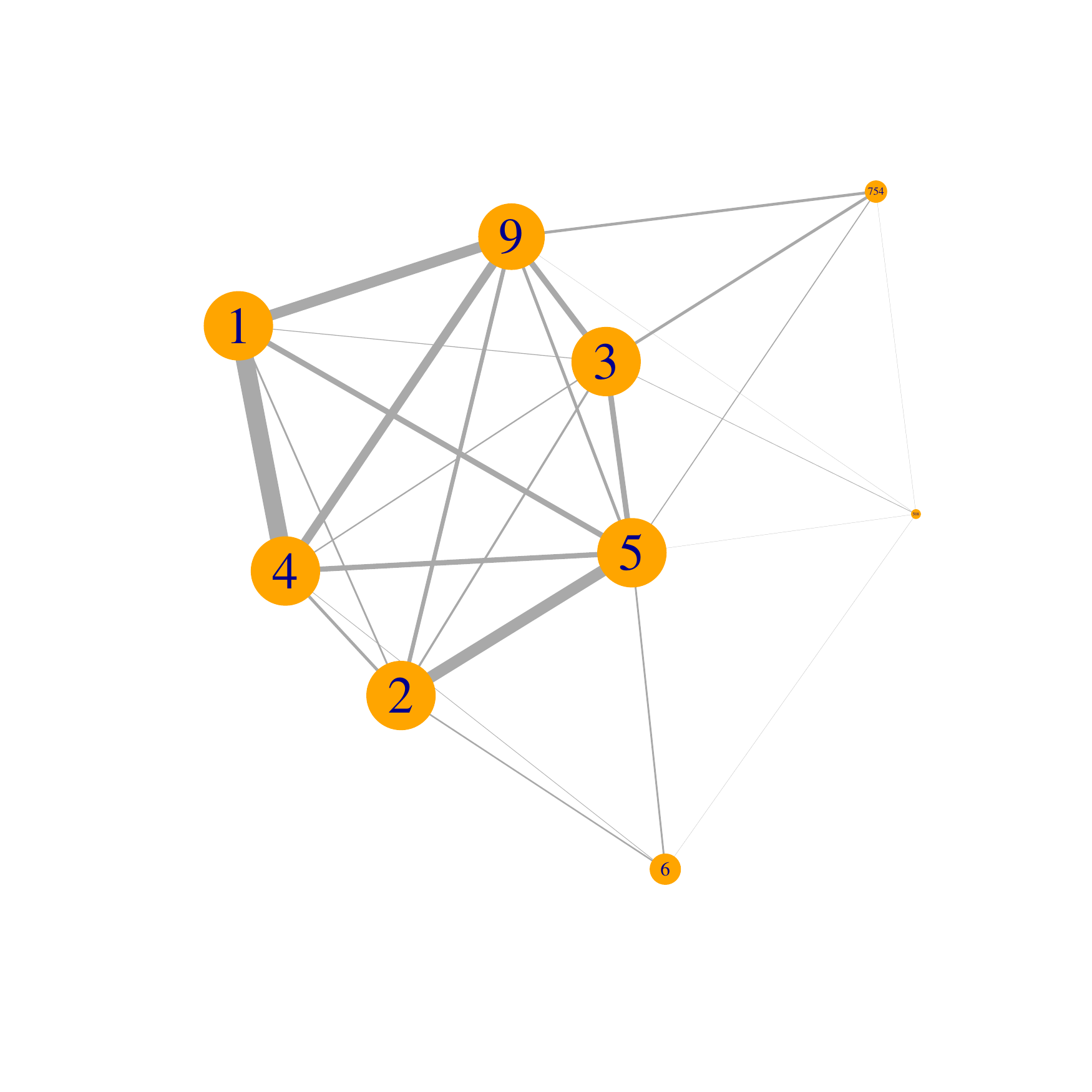} \\
\hline
\parbox[t]{\linewidth}{Single \\ $n = 100$ \\ $p = 1000$}
& \includegraphics[width=\linewidth]{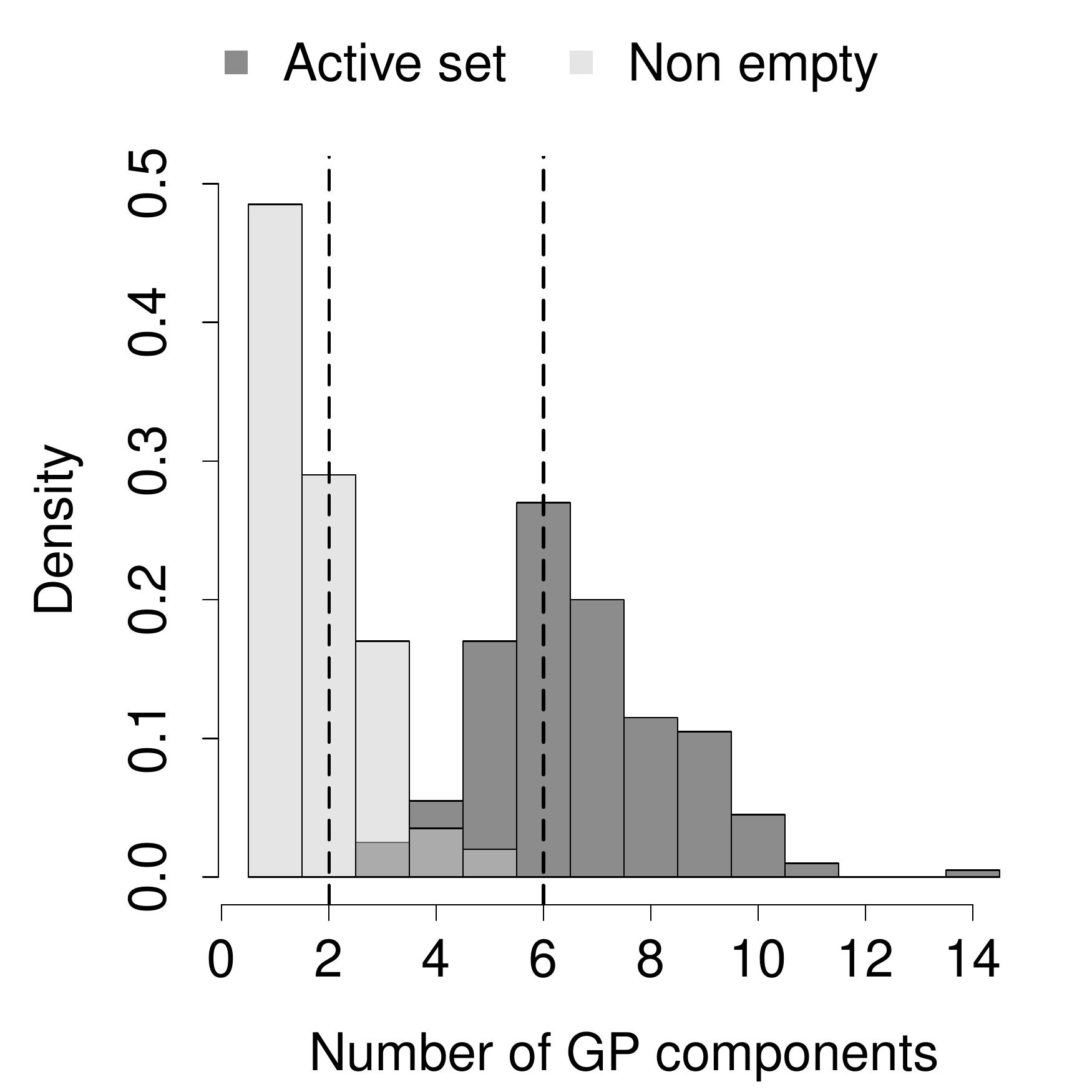} 
& \includegraphics[width=\linewidth]{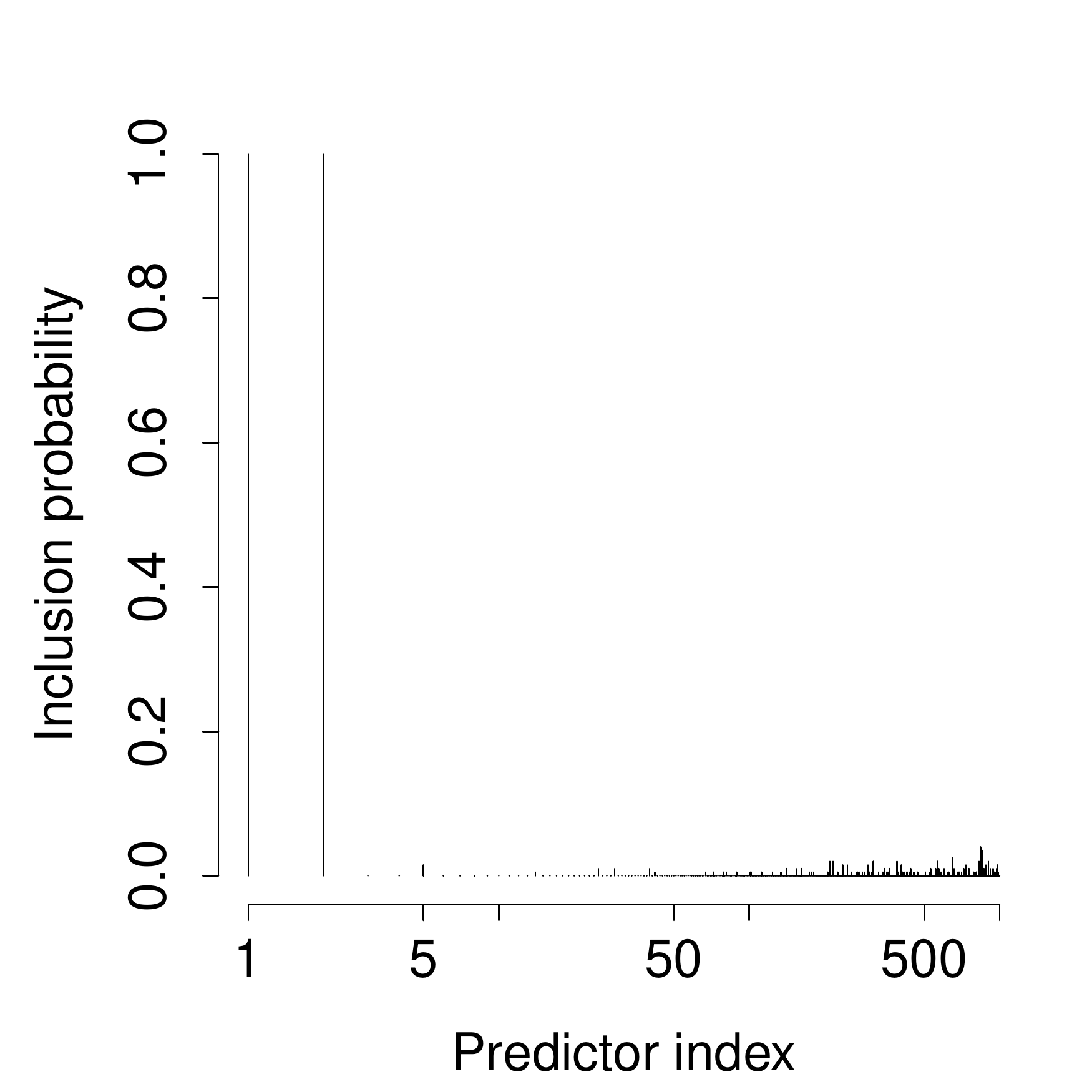}
& \includegraphics[width=\linewidth]{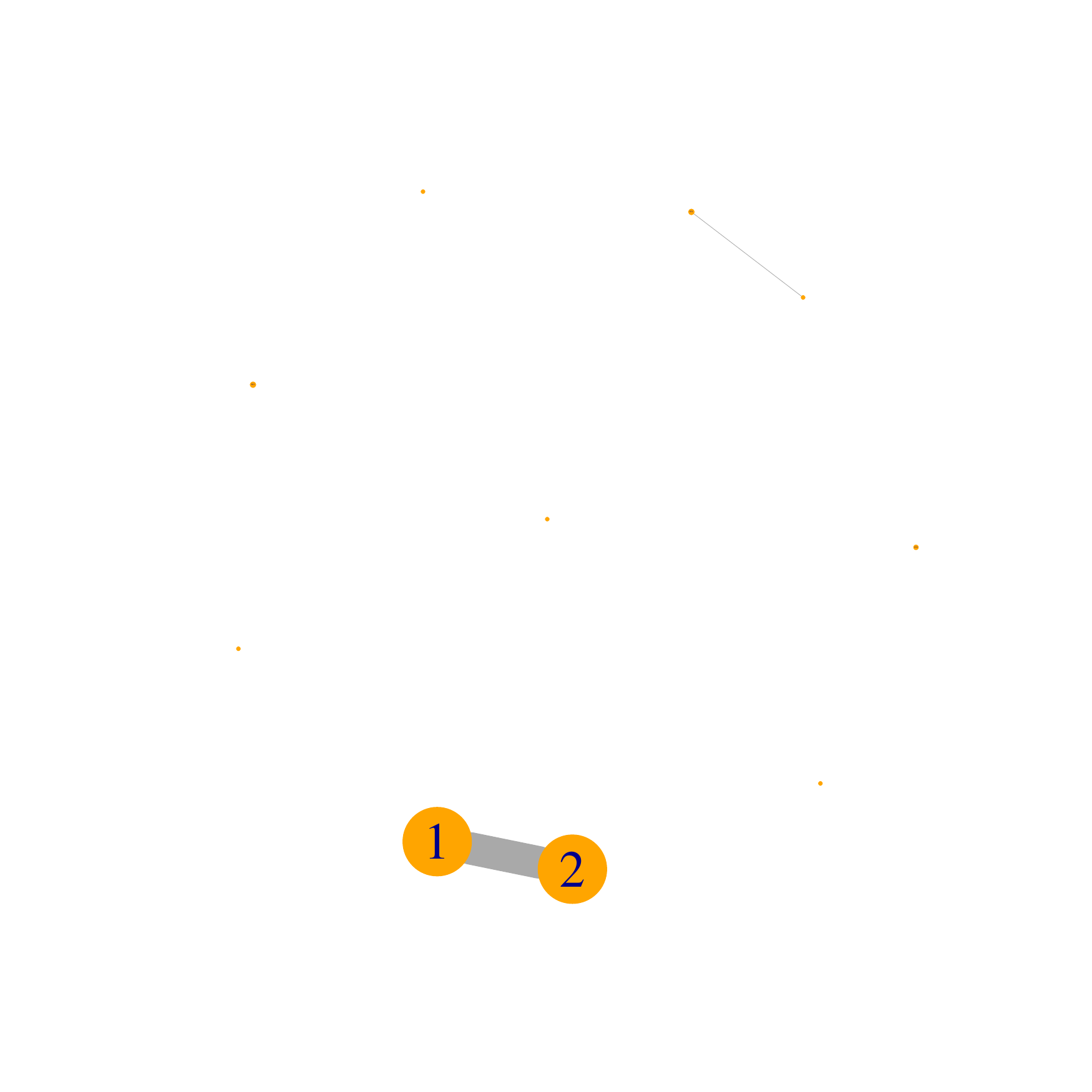}
\end{tabular}
\end{figure}

\begin{figure}[h]
\centering
\includegraphics[width=0.70\linewidth]{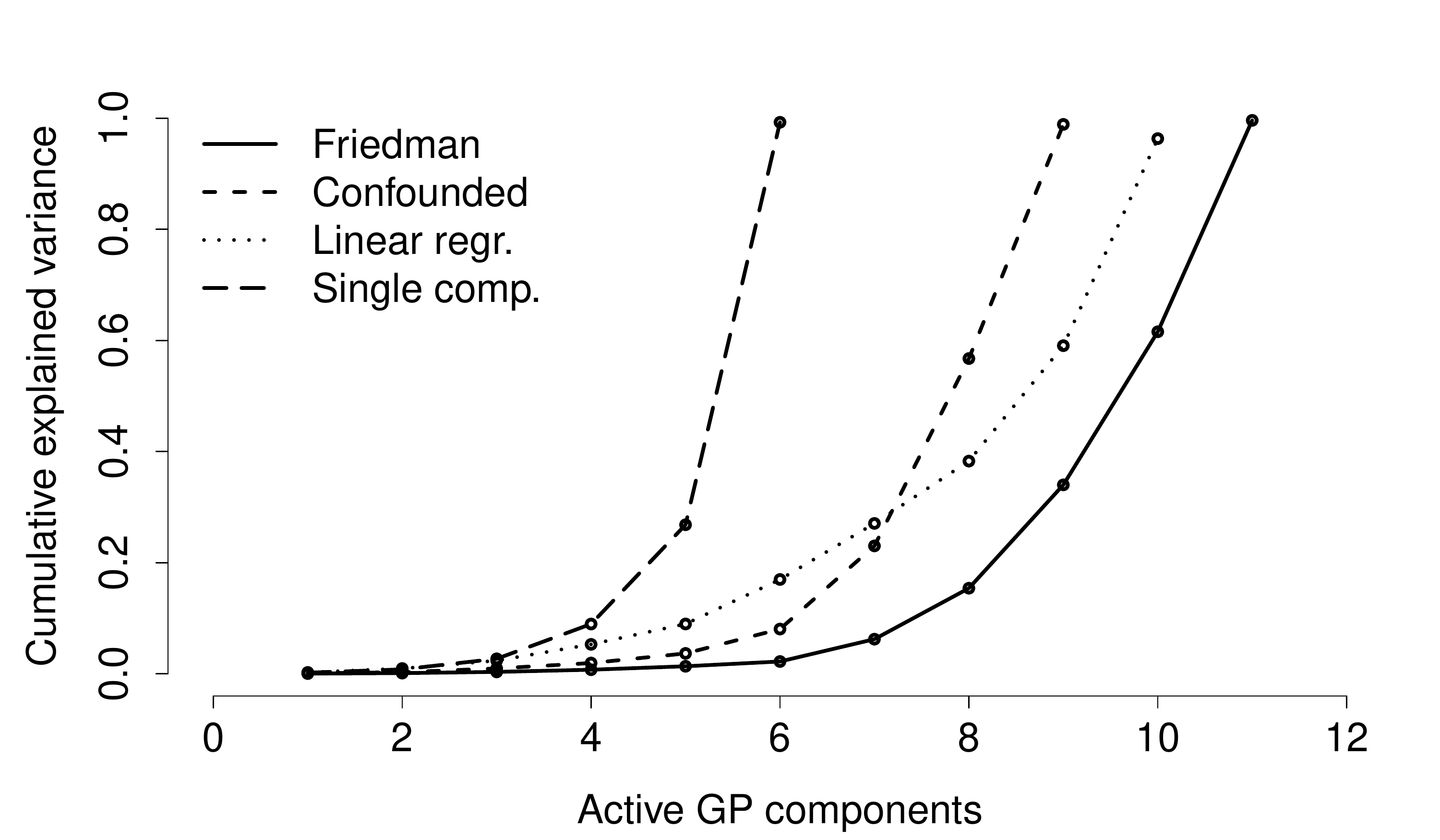} 
\caption{A plot of the cumulative variance explained as a function of the median number of active components (sorted by increasing importance). The fraction of the marginal variance explained by an active GP component is $\rho_l^2 / (1+ \sum_{s \in \mathcal{I}_A} \rho_s^2)$, $l \in \mathcal{I}_A$. Median active sizes appear as vertical dashed lines in the histograms shown in Figure \ref{fig:var_all}.}
\label{fig:explain_var_by_active_comp}
\end{figure}

Simulated experiments in this section concern test functions in Table \ref{tab:sim_functions}.
\begin{table}
\caption{\label{tab:sim_functions} Simulated test functions. All experiments fix $n = 100$ and $p = 1000$.}
\centering
\begin{tabular}{l | l}
\hline
Friedman$^*$ & $f = 10 \sin (\pi x_1 x_2) + 10 \cos(\pi(x_3 x_4 + x_5)) + 20(x_6 - 0.5)^2 + 10x_7$ \\
Confounded & $f = 10 \cos(\pi (x_1 + x_2 + x_3)) + 10 \sin(\pi (x_2 + x_4)) + 10x_5(x_1 + x_2)$ \\
Linear regression & $f = 5(x_1 + \ldots + x_5) + 2(x_6 + \ldots + x_{10})$ \\
Single component & $f = 10 \cos(\pi (x_1 + 5x_2))$
\end{tabular}
\end{table}
The degree of nonlinearity and additivity varies substantially across these functions, but all are sparse considering the assumed data generating process and predictor dimension. The number of active AGP components is adaptively tuned using Algorithm \ref{alg2}, and Figure \ref{fig:var_all} displays a histograms for the number of active (and non-empty) components post burn-in. Figure \ref{fig:explain_var_by_active_comp} plots the cumulative variance explained as a function of the number of active components (sorted by increasing importance). The marginal variance explained by an active AGP component is given by $\rho_l^2 / (1+ \sum_{s \in \mathcal{I}_A} \rho_s^2), ~l \in \mathcal{I}_A.$ Here, one observes that learning is most evenly distributed across active components in the linear regression example, whereas most of the learning for the ``single component'' example is captured by a single AGP component.

An interaction graph summarizes co-appearances of a predictor pairs across any of AGP's active components. Here, marginal inclusion probabilities are thresholded at the $(1- q / p)^{\rm th}$ quantile, focussing on relationships between predictors deemed to be most important (by default, we set $q = 10$). A measure for the marginal importance of the $j$-th predictor is calculated as:
\begin{inparaenum}[(1)]
\item LASSO: $|\hat{\beta}_j|$;
\item BART: $\tfrac{1}{T} \sum_{t=1}^T (n_{jt} - \check{n}_t)/(\hat{n}_t - \check{n}_t) $, with $\hat{n}_t = \max_{j}(n_{jt})$, $\check{n}_t = \min_j(n_{jt})$, where $n_{jt}$ denotes the number of unique occurrences of the $j$-th predictor across the tree ensemble at MCMC iteration $t$;
\item RF: using `{\tt importance}' in the {\tt randomForest} R package; and
\item AGP: $\tfrac{1}{S} \sum_{t > T-S} \max\{I( j \in \gamma_l(t)),\: l \in \mathcal{I}_A\}$, namely the fraction of post burn-in iterations containing the $j$-th predictor in any component.
\end{inparaenum}

In terms of support recovery, AGP is the clear winner for all test functions, identifying the most important predictors with the least number of false positives. In contrast, support recovery using inclusion probabilities from BART and RF can be challenging even when $p$ is small (comparing Figures \ref{fig:var_all} and \ref{fig:simu_competitor_inclusion}). In addition, AGP improves dramatically over all competitors in terms of predictive mean-squared-error (see Section \ref{sec:pred_perf}). 

\paragraph{Friedman$^*$}
Additive effects are well separated for the modified Friedman function \citep{mars}. As shown in Figure \ref{fig:var_all}, the median number of non-empty components used by AGP to recover this structure is 5, two of which are used to isolate the two- and three-way interaction effects, while another two components are used to capture the univariate main effect involving predictors 6 and 7, respectively.

\paragraph{Confounded effects} 
Nonlinear additive effects in the true function share predictor $2$, while predictors $1$ and $2$ also appear as bilinear main effects with predictor $5$. The latter terms are weak in comparison to the other additive effects present in the regression function. The interaction graph in Figure \ref{fig:var_all} shows that AGP successfully recovers the (1,5) main effect, in addition to isolating predictor interaction (2,4) from the three-way (1,2,3) predictor interaction. Weaker edge weights between predictor pairs (2,5), (3,5) and (4,5) indicate such configurations were explored during MCMC, but were less persistent. Consistent with this predictor interaction recovery, AGP has 5 or fewer non-empty components 65\% of the time. AGP settles on such a configuration in 6 / 10 replicate experiments. The remaining runs discover a local configuration mode which is notoriously difficult to escape from. Here, a single AGP component models the bilinear (1,5) main effect, while another models predictor interaction (1,2,3,4). Local neighborhood moves do not subsequently isolate the latter into (1,2,3) and (2,4) constituent interactions (refer to the discussion in Section \ref{sec:discussion}). 

\paragraph{Linear regression} 
AGP and the Lasso both correctly identify predictors 1-5 and 9. AGP also identifies predictor 6 (with inclusion probability 0.33 over a single MCMC chain). The Lasso has a number false positives (i.e., non-zero coefficients for irrelevant predictors), whereas AGP produces but a single false positive (see Figures \ref{fig:var_all} and \ref{fig:simu_competitor_inclusion}). Compared to other test functions, AGP uses a larger number of components to model this fully additive function. The histogram in Figure \ref{fig:var_all} shows that AGP has 6 or fewer non-empty active components roughly 50\% of the time. Since AGP identifies 7 of 10 predictors, components model univariate linear terms in at least half of the MCMC iterations. Edge weights in the interaction graph are relatively consistent with this view. 

\paragraph{Single component} 
This function is difficult to model due to its extreme sparsity, nonlinearity and lack of additive structure. 
With sparsity inducing priors over the model space, along with an ability to adapt to the underlying smoothness of the regression function (via component adaptive scaling), AGP recovers the true support and predictor interaction. All other competitors fail to do so. AGP contains 3 or fewer non-empty components 90\% of the time as depicted in Figure \ref{fig:var_all}. A single component models the (1,2) interaction, while other active components enable occasional exploration of additional joint configurations $\bgamma = \{(1,2), 1, -\},$ and $\bgamma = \{(1,2), 2, -\}$.

\subsection{Predictive performance}
\label{sec:pred_perf}
RMSE is reported on a set of 200 test observations generated for each function described in Section \ref{sec:simulation}. Results are averaged over several independent replications and reported along with standard errors. AGP's improved support and interaction recovery leads to a dramatic improvement in predictive performance and uncertainty quantification over state-of-the-art methods in low and high dimensional settings. This validates the additive-interactive modeling framework for GP regression, and confirms that our adaptive MCMC sampler using paired and inter-component moves provides reliable predictor and interaction recovery in high dimensions.

\begin{figure}[h]
\centering
\includegraphics[width=0.80\linewidth]{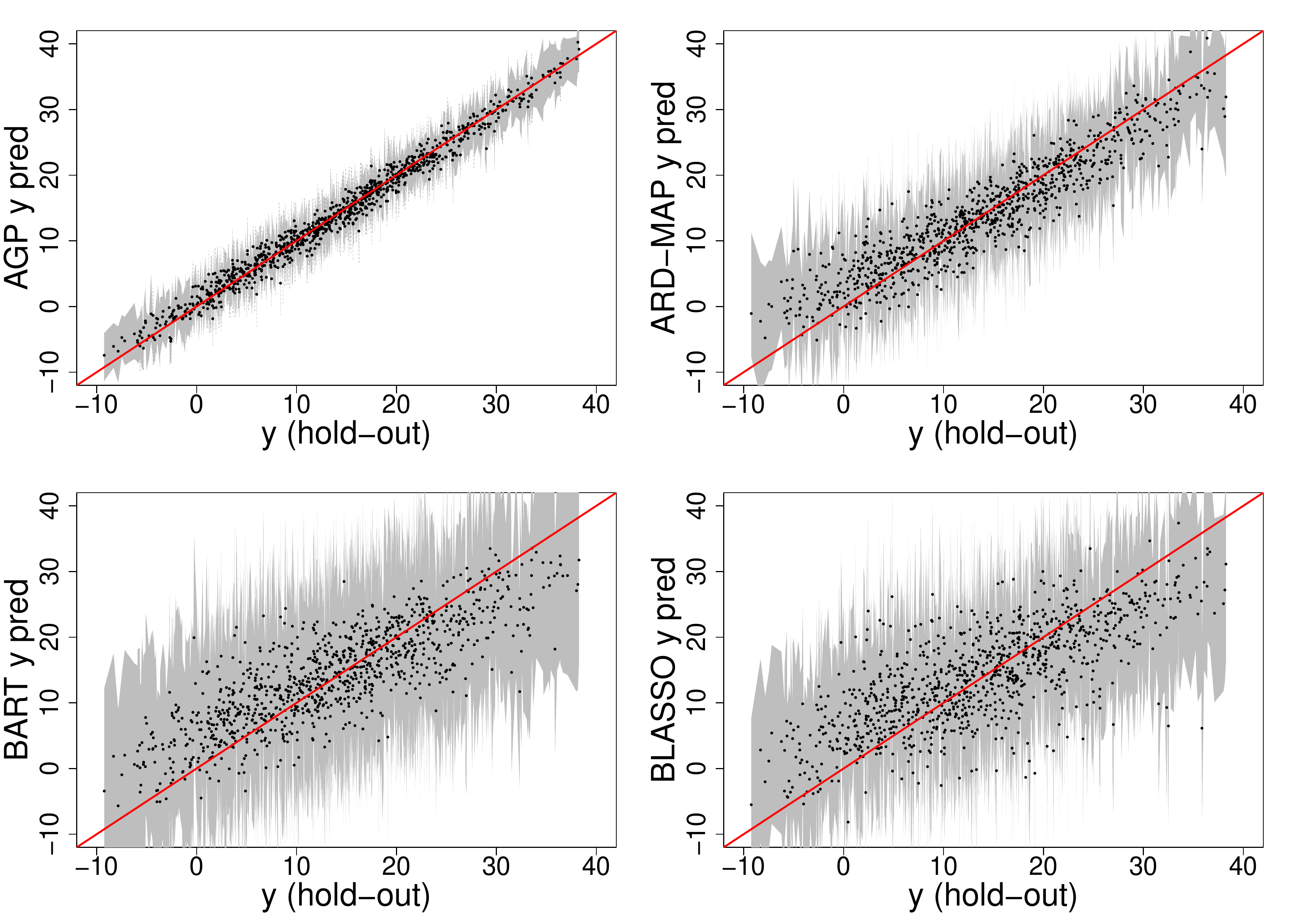}
\caption{Hold-out predictive performance for the modified Friedman function in Section \ref{sec:pred_perf} for experiments with $p = 50$. True response values are plotted along the $x$-axis and predicted values along the $y$-axis, with point-wise 95\% credible intervals overlaid. Here, Bayesian Lasso is used in lieu of the Lasso to enable reporting of credible bands.}
\label{fig:pred_perf}
\end{figure}

Figure \ref{fig:pred_perf} plots predicted response values against generated responses for a simulated test dataset using the modified Friedman function. Consistent with AGP's excellent predictive performance shown in Table \ref{tab:predictive_performance}, values appear close to the ``$y = x$'' line. In addition, 95\% credible interval bands for AGP are significantly narrower as compared to its competitors.
\begin{table}
\caption{\label{tab:predictive_performance} Predictive RMSE for test functions defined in Section \ref{sec:varSelectionInteractionRecovery} with $n = 100$ and $p = 1000$. Standard errors using 10 replicates appear as subscripts on averaged RMSE values.}
\centering
\begin{tabular}{l | c | c | c | c | c }
& AGP & BART & RF & LASSO & NULL\\ 
\hline
Friedman & $\mathbf{1.42}_{0.08}$ & $6.81_{0.50}$ & $6.76_{0.52}$ & $6.18_{0.43}$ & $7.82_{0.39}$\\
Confounded & $\mathbf{2.61}_{0.53}$ & $9.27_{0.41}$ & $9.00_{0.46}$ & $7.87_{1.04}$ & $9.68_{0.28}$\\
Linear regr. & $\mathbf{1.52}_{0.12}$ & $3.38_{0.09}$ & $3.43_{0.10}$ & $1.75_{0.16}$ & $3.70_{0.14}$ \\
Single comp. & $\mathbf{1.97}_{0.27}$ & $7.07_{0.27}$ & $7.08_{0.27}$ & $7.23_{0.53}$ & $7.07_{0.28}$
\end{tabular}
\end{table}

\subsubsection{Predictive comparisons for lower dimensional experiments}
A single GP prior with an ARD squared exponential covariance function\footnote{The ARD covariance function is $C^{\rm \scriptsize ARD}(x, \mt{x}) = \rho^2 \exp\big(-\sum_{j=1}^p \lambda_j^2 (x_j - \mt{x}_j)^2\big)$.} placed on the unknown regression function can identify important predictors and has been the primary workhorse for variable selection in the GP regression setting \citep{neal1997monte, gp-paper}. However, it is well known that MCMC and likelihood based methods for estimation of GP-ARD parameters suffer as the predictor dimension grows. This fact, together with recent theoretical developments for GP regression using an additive-interactive framework, are compelling reasons to pursue the proposed APG model \eqref{eq:agp-model}. It is in instructive, therefore, to compare performance between the two methods. 

Data are generated with $n = 100$ and $p = 50$ for each test function, and predictive performance is summarized in Table \ref{tab:rmse_p50}. The ARD competitor is fit by iteratively computing MAP estimates for discretized covariance parameters until convergence is reached. AGP's improvement over ARD-MAP indicates how modeling of additive structure enables superior estimation of the regression function even when $p$ is small. In addition, as predictor dimension $p$ grows, BART and RF degrade considerably in terms of predictive performance, whereas AGP remains remarkably robust to the 20-fold increase in predictor dimension (compare Tables \ref{tab:predictive_performance} and \ref{tab:rmse_p50}).

\begin{table}
\caption{\label{tab:rmse_p50} Predictive RMSE for test functions defined in Section \ref{sec:varSelectionInteractionRecovery} with $n = 100$ and $p = 50$. Standard errors using 10 replicates appear as subscripts on averaged RMSE values.}
\centering
\begin{tabular}{l | c | c | c | c | c | c}
& AGP & ARD-MAP & BART & RF & LASSO & NULL \\ 
\hline
Friedman & {\bf 1.49$_{0.12}$} & 3.39$_{0.27}$ & 5.46$_{0.26}$ & 5.85$_{0.35}$ & 5.67$_{0.47}$ & 7.71$_{0.22}$ \\
Confounded & {\bf 2.31$_{0.17}$} & 4.50$_{0.24}$ & 7.61$_{0.58}$ & 8.11$_{0.65}$ & 6.89$_{0.43}$ & 9.53$_{0.35}$ \\
Linear regr. & {\bf 1.36$_{0.14}$} & 1.41$_{0.19}$ & 1.88$_{0.17}$ & 2.85$_{0.04}$ & {\bf1.33$_{0.09}$} & 3.67$_{0.07}$ \\
Single comp. & {\bf 1.96$_{0.22}$} & 2.70$_{0.23}$ & 7.28$_{0.25}$ & 7.21$_{0.18}$ & 7.44$_{0.20}$ & 7.31$_{0.30}$
\end{tabular}
\end{table}

\section{Real data illustrations}
\label{sec:real_data_analysis}

We analyze four real datasets which are commonly used in the statistics and machine learning literature, and offer a wide range of predictor effect types, from mostly additive to including high order interaction. In addition, predictor count varies from moderate to very large across these examples. This allows us to validate the AGP model \eqref{eq:agp-model} on data that may satisfy only some of the additive-interaction assumptions. The number of AGP components (and neighborhood budget for PRNS moves) is adaptively tuned using Algorithm \ref{alg2}, with BART, RF, and the Lasso taken to be our competitors as before.

Competing methods vary dramatically in terms of their relative performance (RMSE) across the four datasets as shown in Table \ref{tab:real_data_performance}. For Riboflavin microarray and biscuit-dough spectroscopy (Cookie) data, regularized linear regression (i.e., the Lasso) clearly gives better averaged test prediction than ensemble nonparametric methods, whereas the opposite holds for Crime and Boston housing data (i.e., BART and RF are good competitors). Note that AGP remains competitive (best or second best predictive performance) across the board, owing to its ability to accommodate varying degrees of sparsity, interaction, and additivity. Moreover, this adaptability appears robust across predictor dimension, accommodating even settings with high predictor correlation (e.g., Cookie data).
\begin{table}
\caption{\label{tab:real_data_performance} Performance of AGP against the same by BART, RF, Lasso, and the mean prediction rule (NULL) for well known real datasets. {\em Left}: the fraction of explained variance computed as $R^2 = 1 - (\mathrm{RMSE} / \bar{s}_{y})^2$ for AGP; {\em Right}: predictive (hold-out) RMSE averaged over random test/train splits with standard errors appearing as subscripts.}
\centering
\setlength{\tabcolsep}{2pt}
\begin{tabular}{l | r | r}
& \multicolumn{2}{c}{AGP} \\
\hline
& \multicolumn{1}{c}{$R^2$} & \multicolumn{1}{c}{RMSE}\\
\cline{2-3}
Boston & 0.89$_{0.04}$ & {\bf3.00$_{0.54}$}\\
Crime & 0.64$_{0.15}$ & 421.92$_{70.86}$\\
Riboflavin & 0.74$_{0.13}$ & {\bf0.47$_{0.12}$}\\
Cookie & 0.95 & 0.42
\end{tabular}
\quad
\begin{tabular}{r | r | r | r}
\multicolumn{4}{c}{RMSE}\\
\hline
\multicolumn{1}{c}{BART} & \multicolumn{1}{c}{RF} & \multicolumn{1}{c}{LASSO} & \multicolumn{1}{c}{NULL}\\
\hline
3.48$_{0.49}$ & 3.47$_{0.63}$ & 5.02$_{0.54}$ & 9.15$_{0.47}$ \\ 
{\bf415.54$_{81.54}$} & 430.40$_{78.50}$ & 458.73$_{71.66}$ & 741.92$_{91.63}$ \\
0.54$_{0.08}$ & 0.56$_{0.07}$ & {\bf0.43$_{0.13}$} & 0.80$_{0.10}$ \\
1.55 & 1.44 & {\bf0.25} & 1.98
\end{tabular}
\end{table}

\begin{figure}[h!]
\centering
\includegraphics[width=0.70\linewidth]{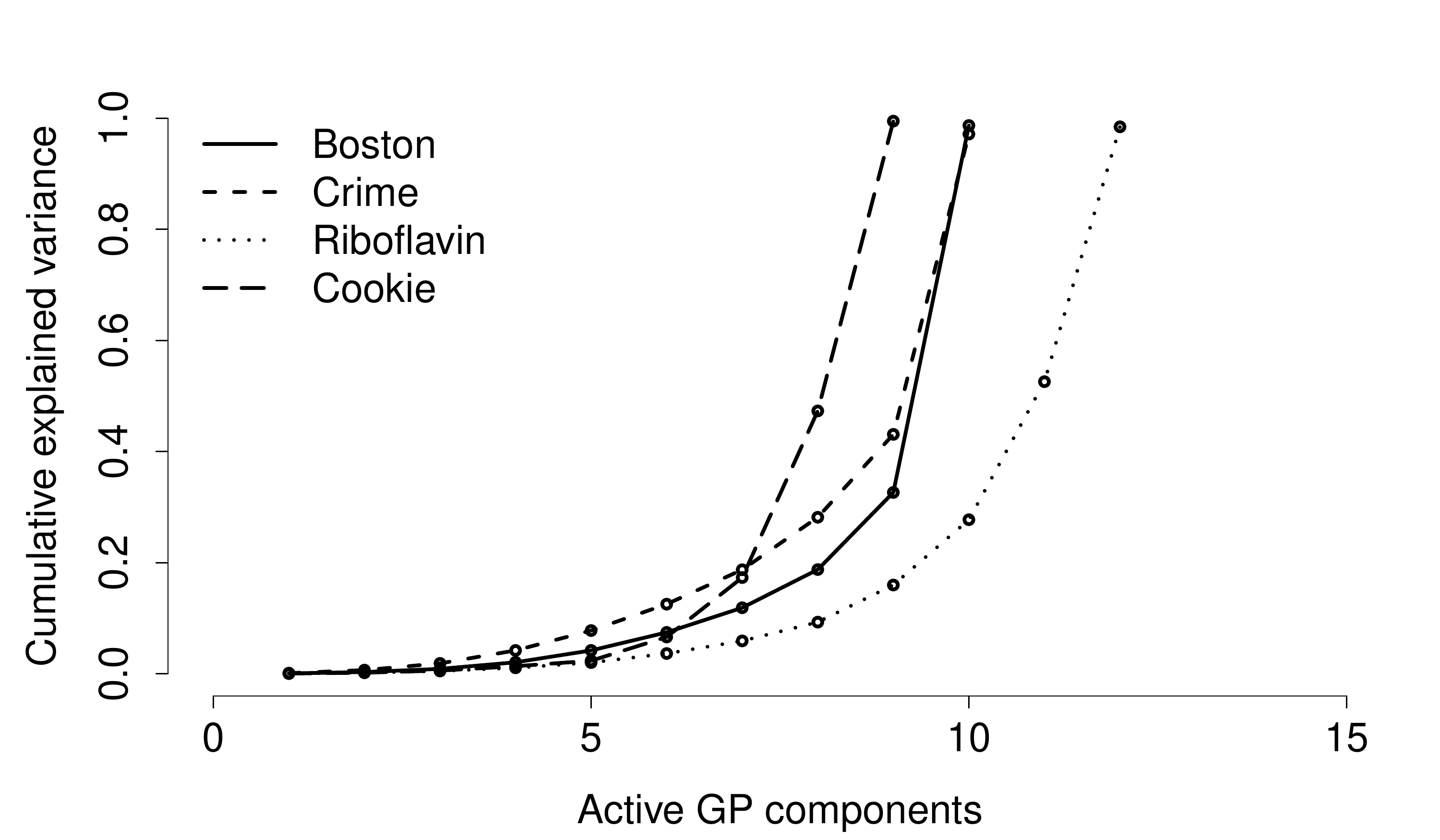}
\caption{A plot of the cumulative variance explained as a function of the median number of active components (sorted by increasing importance). The fraction of the marginal variance explained by an active GP component is $\rho_l^2 / (1+ \sum_{s \in \mathcal{I}_A} \rho_s^2), ~l \in \mathcal{I}_A$. Median active sizes appear as vertical dashed lines in the histograms shown in Figure \ref{fig:real_data_all}.}
\label{fig:explain_var_by_active_comp_real}
\end{figure}

\begin{figure}[!ht]
\caption{\label{fig:real_data_all}Comparison across well known datasets. {\em Left}: number of active vs. non-empty (utilized) components, with median sizes appearing as vertical dashed lines; {\em Middle}: marginal inclusion probabilities for the AGP model (index on log-scale); {\em Right}: an interaction graph among the most important predictors. Vertex sizes are proportional to predictor importance, and edges between predictor pairs are drawn in proportion to their co-appearance across GP components.}
\centering
\arrayrulecolor{gray}
\setlength{\tabcolsep}{2pt}
\begin{tabular}{m{0.22\columnwidth} | C{0.23\columnwidth} C{0.23\columnwidth} 
C{0.27\columnwidth}}
& Components & Marginal inclusion & Interaction graph \\
\hline 
Boston \newline
{\small
\# predictors: 13 \newline
Sample size: 507 \newline
Splits: 6, 70 - 30\%}
& \includegraphics[width=\linewidth]{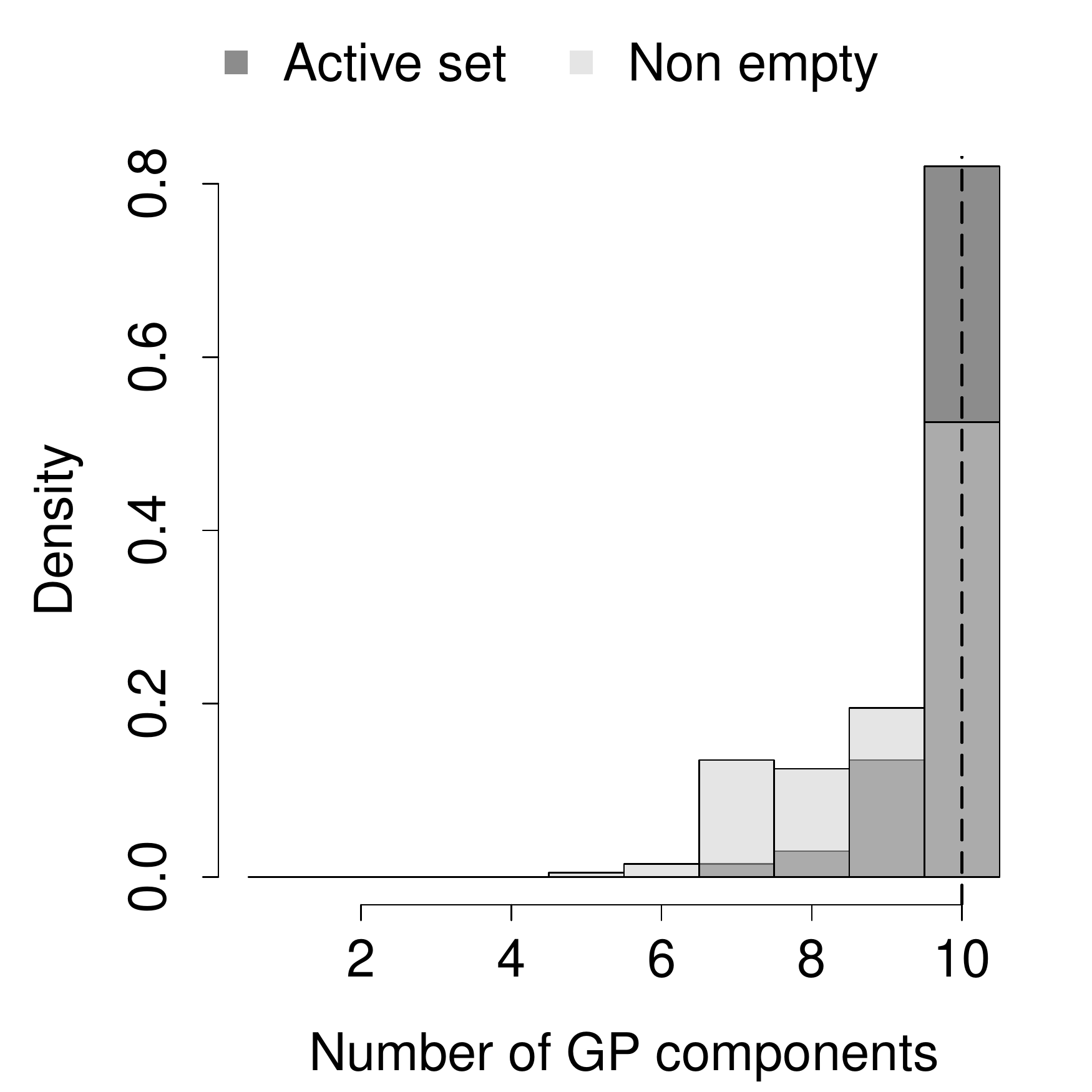}\label{fig:real_boston}
& \includegraphics[width=\linewidth]{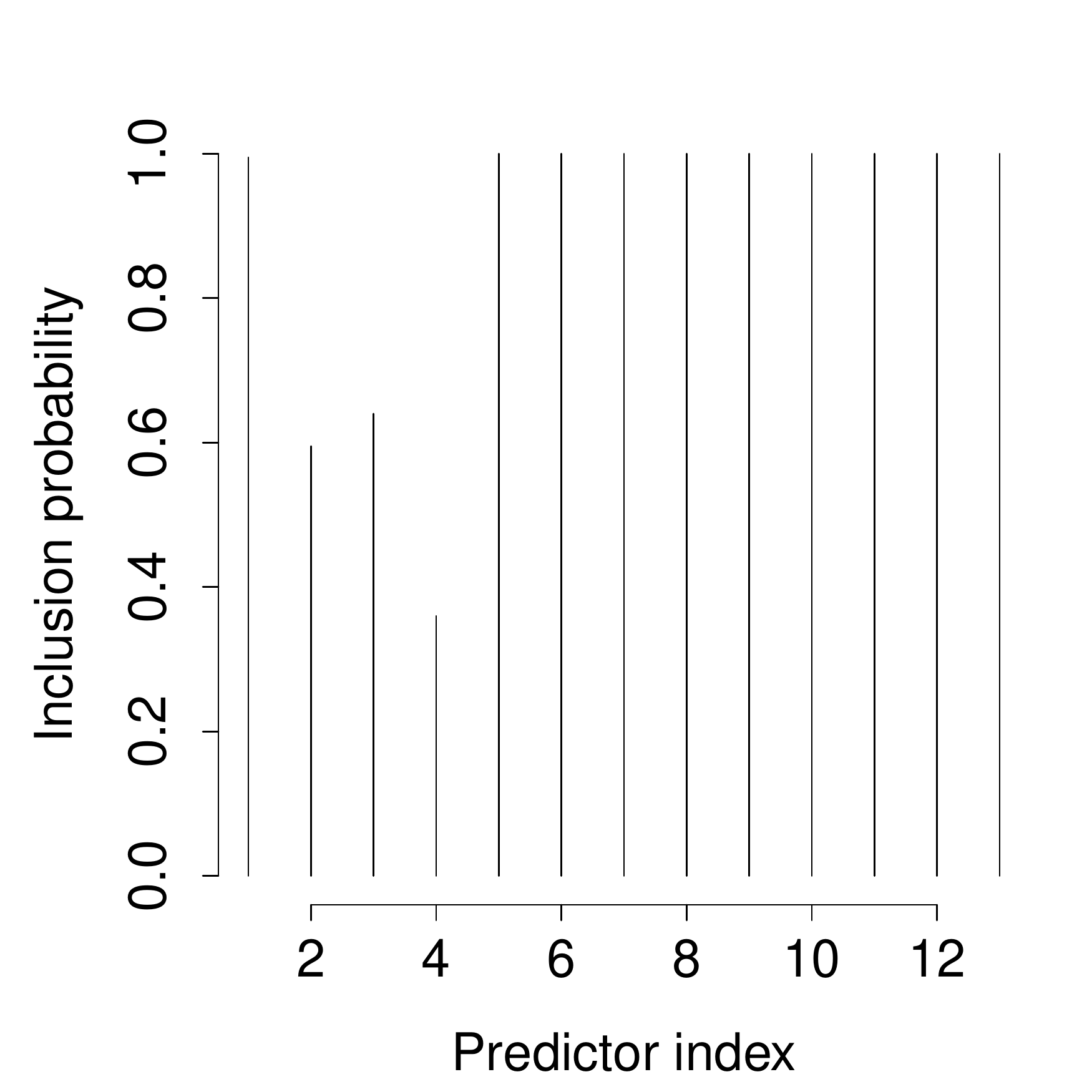}
& \includegraphics[width=\linewidth]{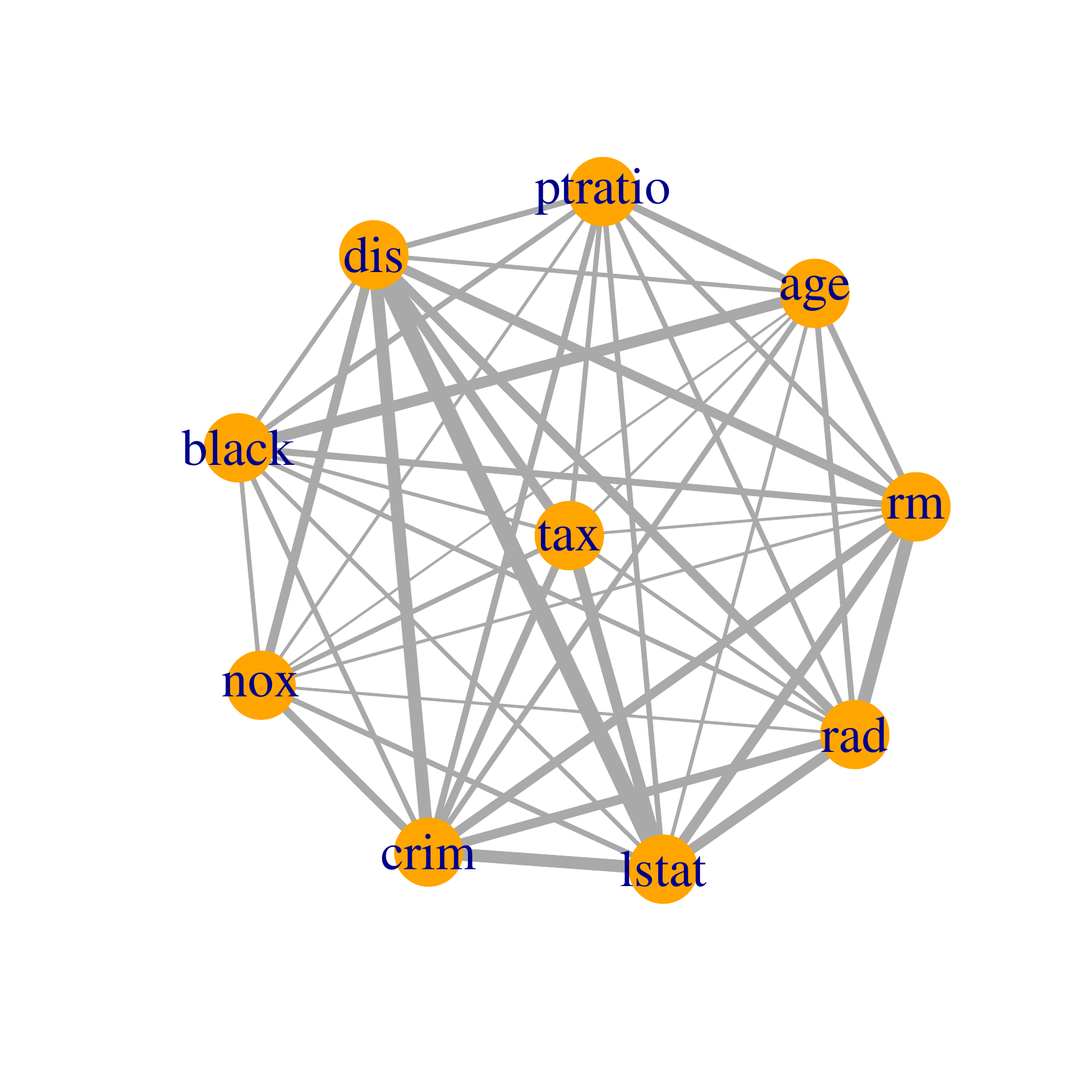} \\
\hline 
Crime \newline
{\small
\# predictors: 102 \newline
Sample size: 416 \newline
Splits: 5, 75 - 25\%}
& \includegraphics[width=\linewidth]{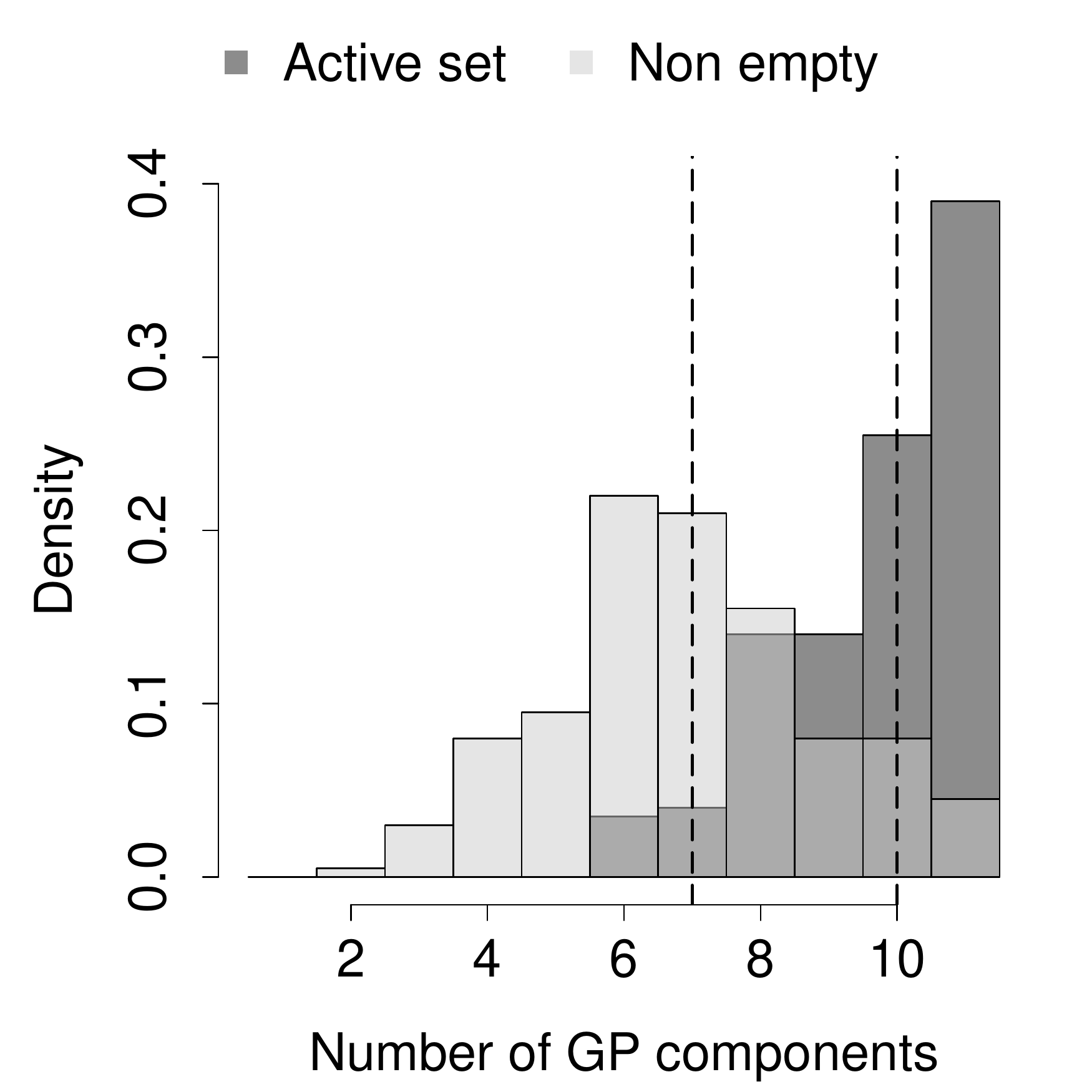}\label{fig:real_crime}
& \includegraphics[width=\linewidth]{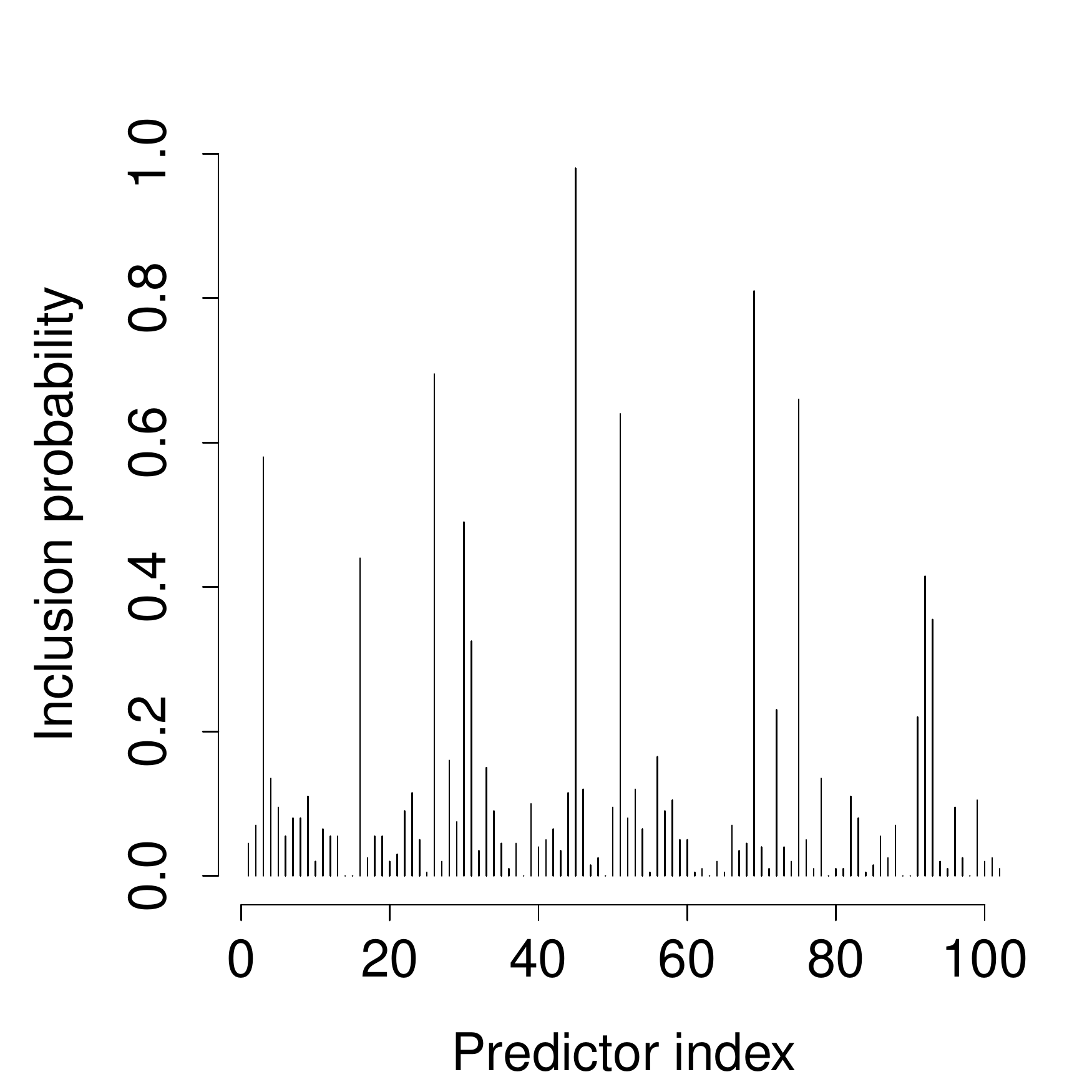}
& \includegraphics[width=\linewidth]{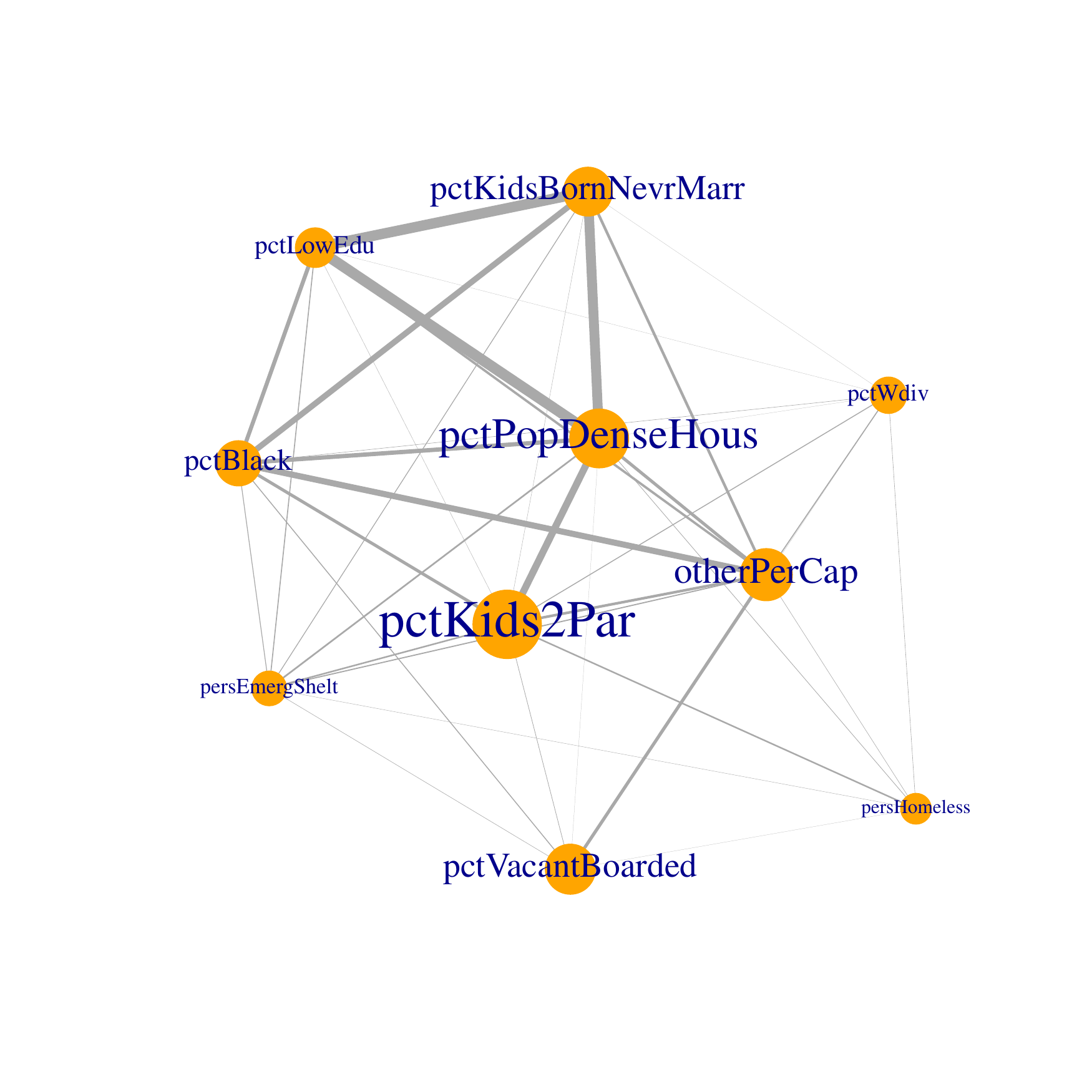} \\
\hline 
Riboflavin \newline
{\small
\# predictors: 4088 \newline
Sample size: 71 \newline
Splits: 5, 80 - 20\%}
& \includegraphics[width=\linewidth]{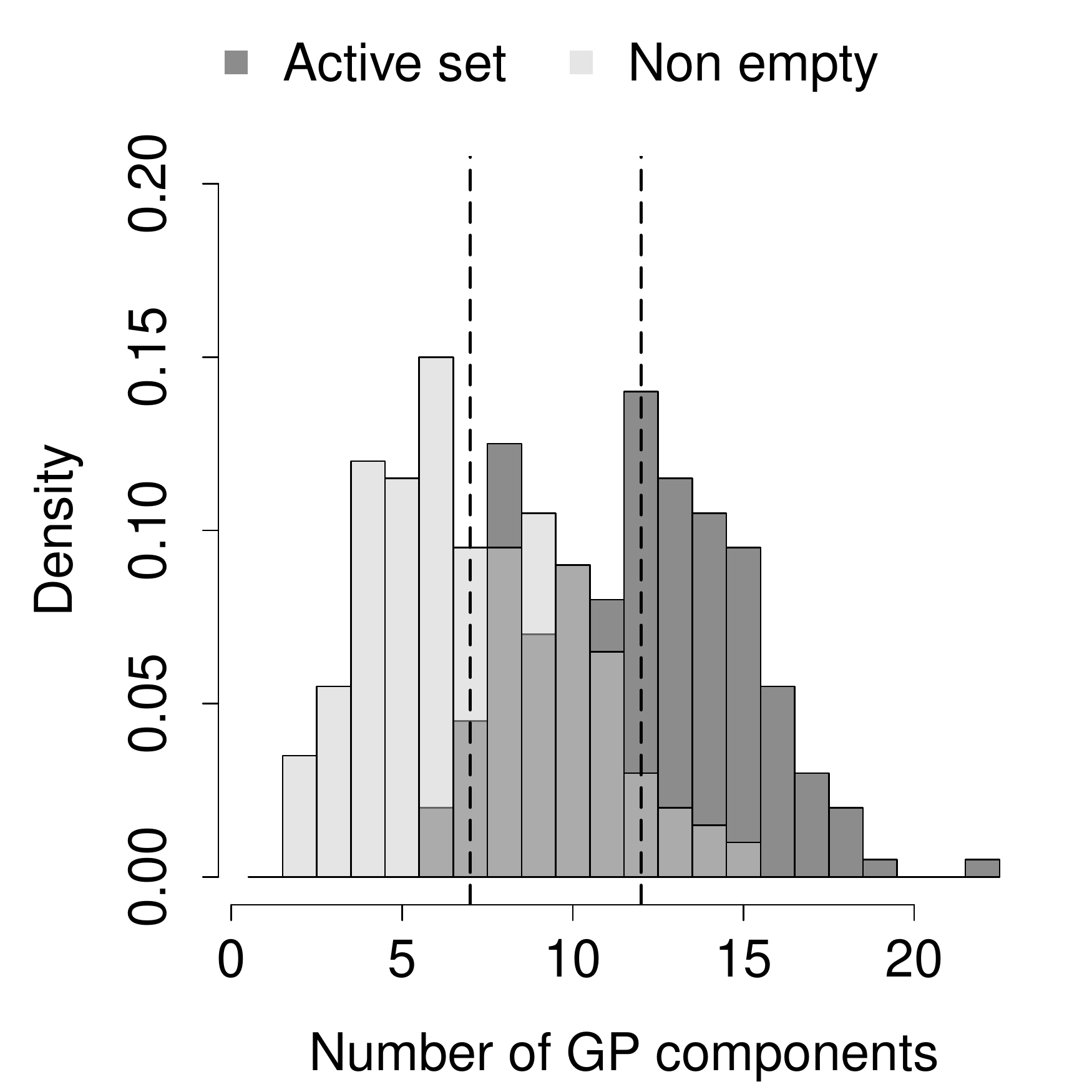}\label{fig:real_ribo}
& \includegraphics[width=\linewidth]{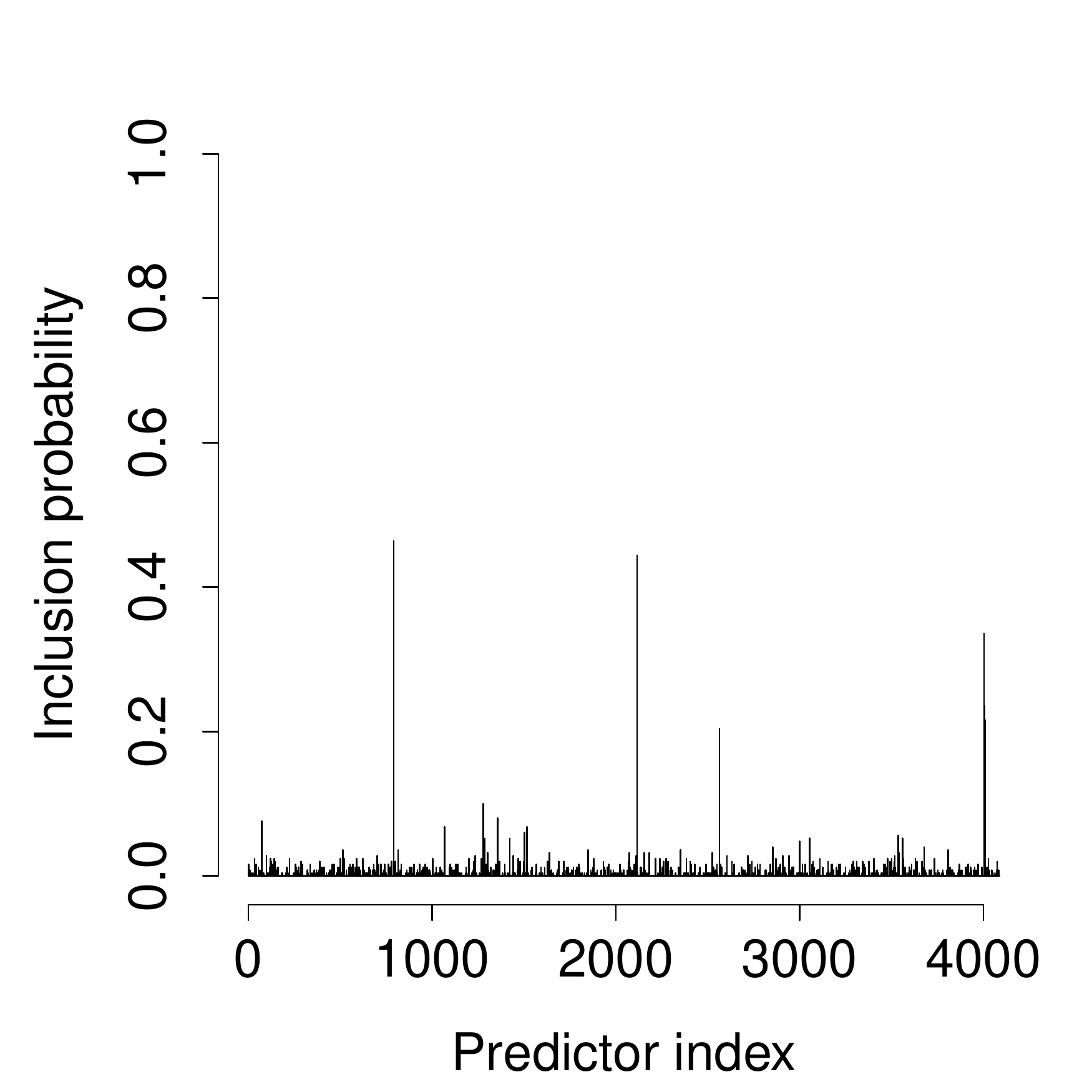}
& \includegraphics[width=\linewidth]{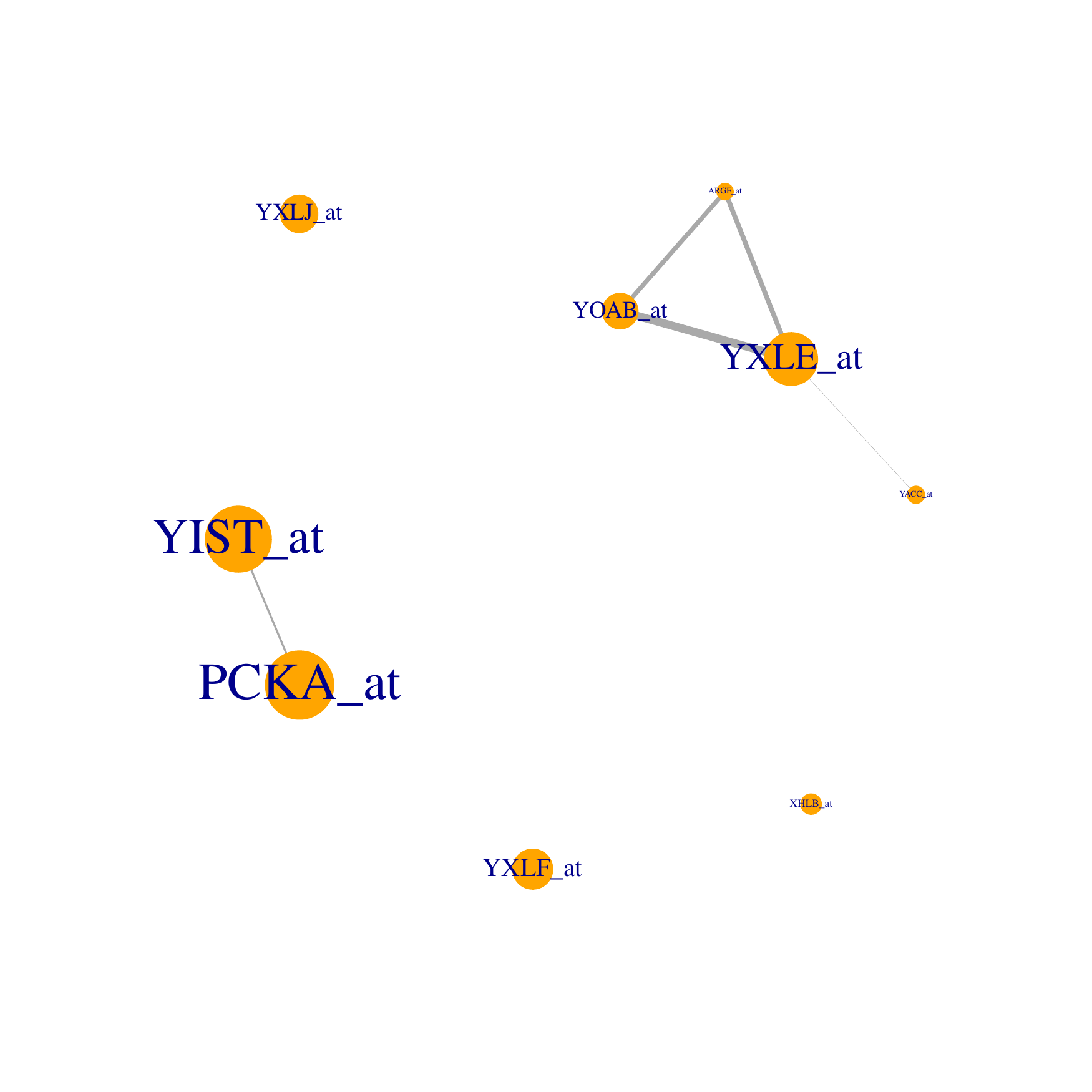} \\
\hline 
Cookie \newline
{\small
\# predictors: 700 \newline
Train size: 40 \newline
Test size: 34}
& \includegraphics[width=\linewidth]{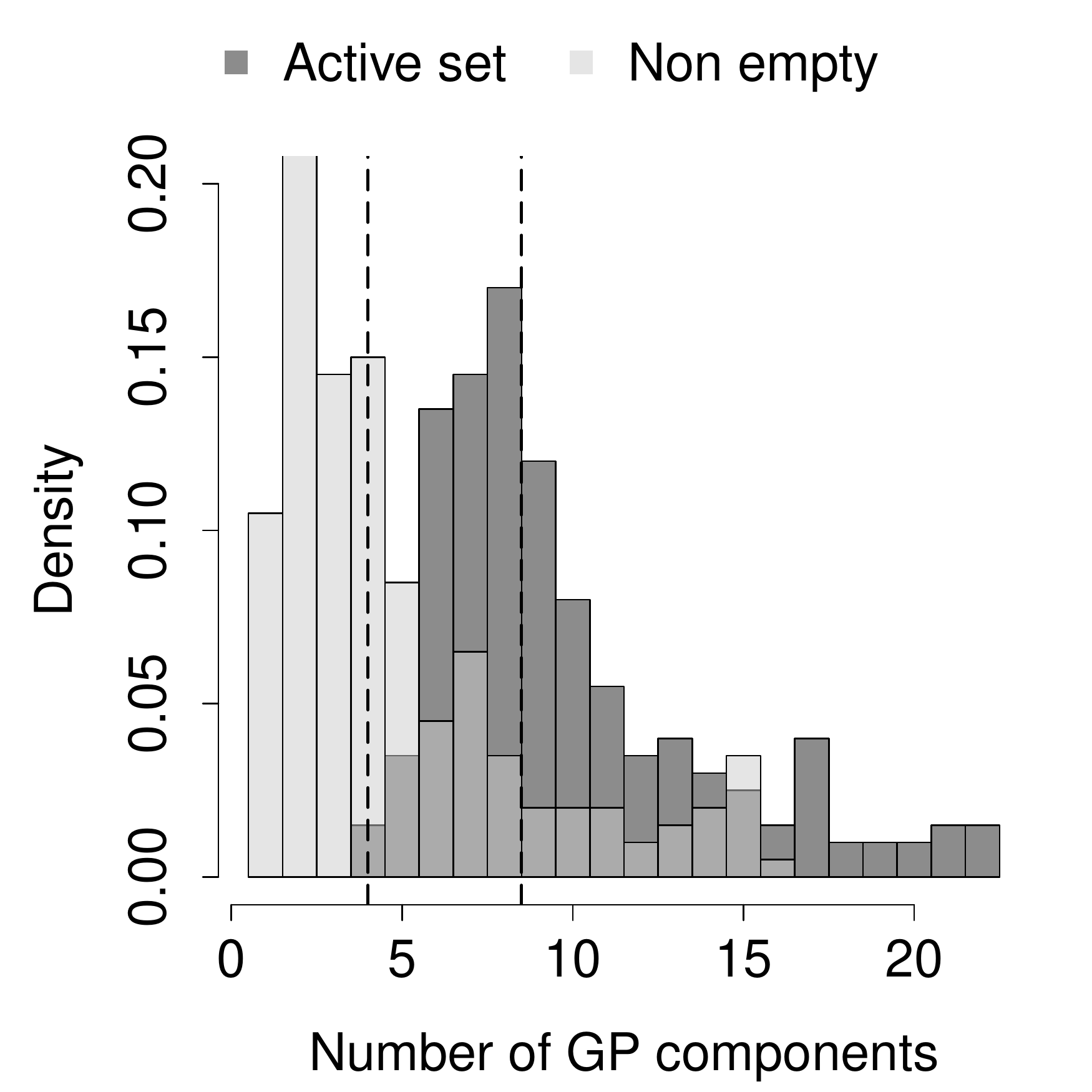}\label{fig:real_cookie}
& \includegraphics[width=\linewidth]{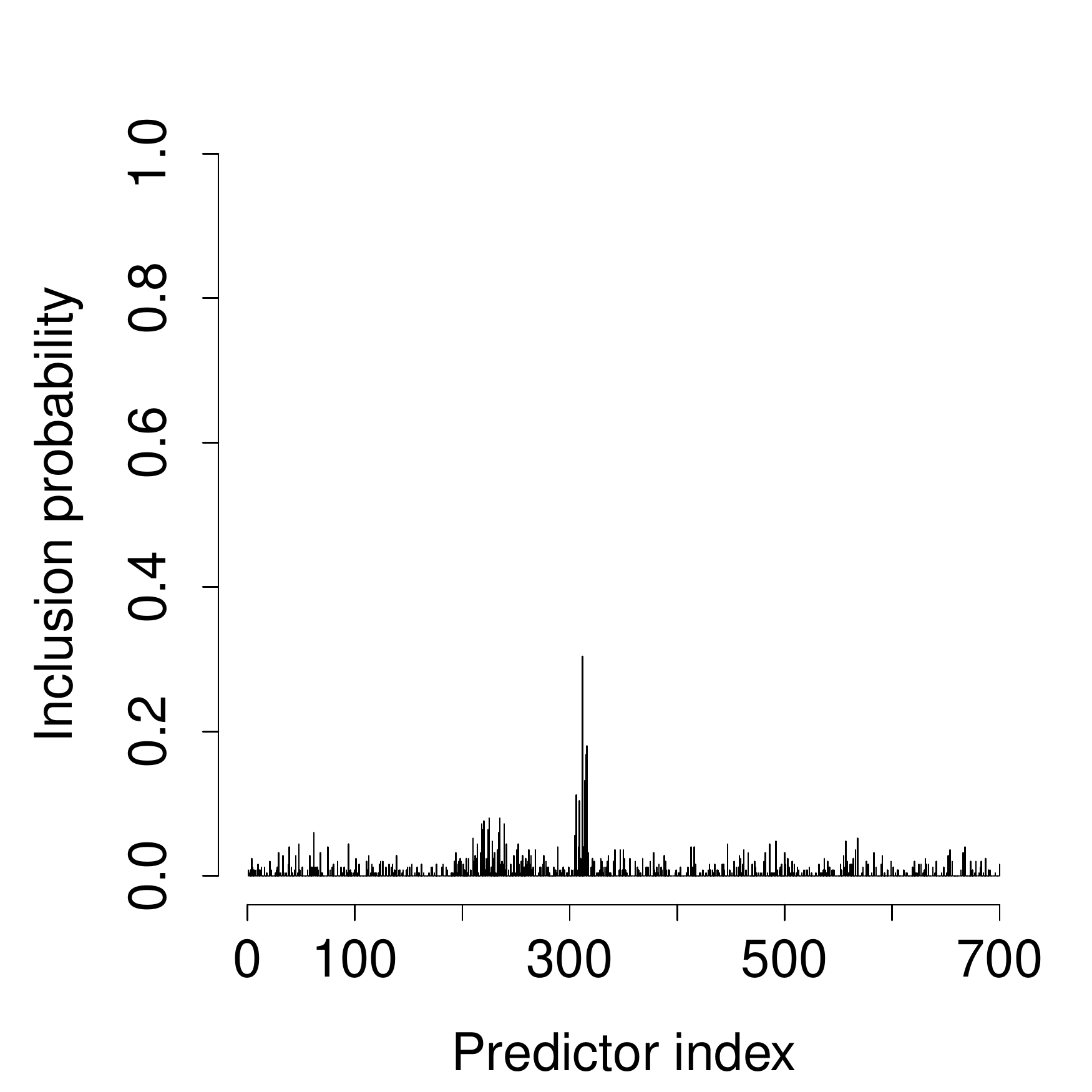}
& \includegraphics[width=\linewidth]{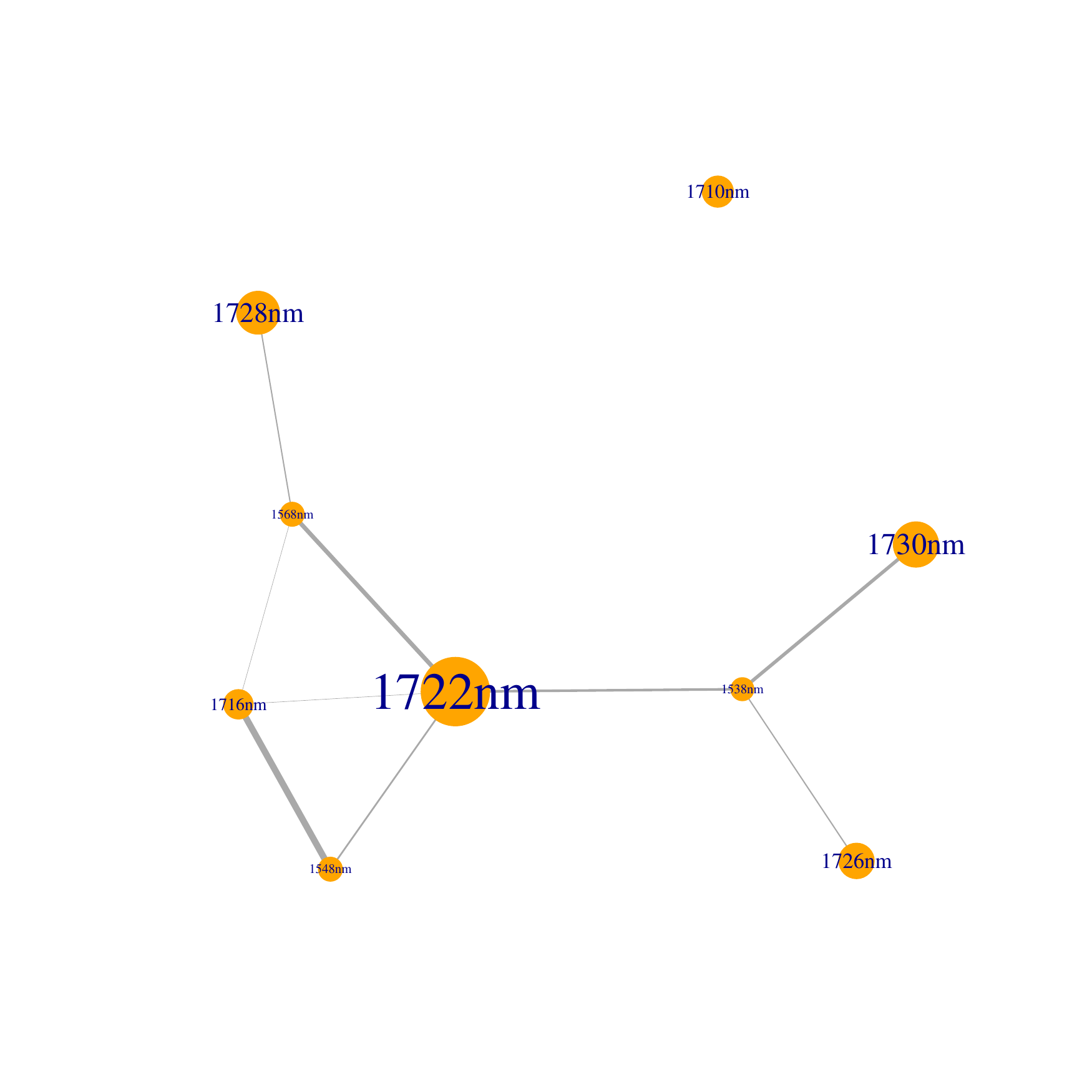}
\end{tabular}
\end{figure}

To understand how learning is distributed across AGP's active components, Figure \ref{fig:explain_var_by_active_comp_real} plots the cumulative variance explained (components sorted by increasing importance) for the median-sized model over all MCMC iterations. These sizes appear as vertical dashed lines in Figure \ref{fig:real_data_all}. For the Riboflavin and Crime data, several active components contribute to AGP's learning, whereas learning in the Boston housing and Cookie data is primarily captured in one or two GP components. The right panel in Figure \ref{fig:real_data_all} plots posterior predictor inclusion probabilities and interactions for the AGP model. As shown, AGP identifies important predictors (only a handful in some, very many in others) and their interactions (strong interaction effects in some; mostly additive, non-interactive effects in others). Table \ref{tab:variance_icm} compares the mean and variance for predictive RMSE (on held out test data) calculated on random train/test partitions of the real datasets. Here, predictive RMSE for MCMC with and without ICM moves are comparable, but estimates from runs using ICM moves generally have smaller variance and are less sensitive to initialization.

For each real dataset, Figure \ref{fig:real_posterior_plots} in Section \ref{sec:additional_figs} of the Appendix displays (a) bar-plots for component smoothness parameters $\rho_l, \lambda_l$, $l \in \mathcal{I}_A$; (b) a trace plot of posterior draws for variance parameter $\sigma^2 \sim \pi(\cdot| \bxi, \by)$ in \eqref{eq:gp_update}; and (c) predictor importance scores $v_j$, $j = 1, \dots, p$ learned using Algorithm \ref{alg2}. The trace plot of the posterior variance serves as a model fit diagnostic and a measure of MCMC mixing and stability.

\begin{table}
\caption{\label{tab:variance_icm} Averaged RMSE (on held out test data) and RMSE variance for real data illustrations for AGP using Algorithm \ref{alg1} for MCMC with and without the use of ICM moves.}
\centering
\begin{tabular}{l|c|c|c|c}
& \multicolumn{2}{c|}{No ICM} & \multicolumn{2}{c}{With ICM} \\
\hline
& Mean & Var & Mean & Var\\
\hline
Boston & 3.16 & 0.55 & 3.00 & 0.54\\
Crime & 424.19 & 74.32 & 421.92 & 70.86\\
Riboflavin & 0.46 & 0.12 & 0.47 & 0.12\\
Cookie & 0.42 & 0.02 & 0.42 & 0.02
\end{tabular}
\end{table}

\section{Discussion} \label{sec:discussion}

Extensions of AGP to accommodate binary outcomes via the probit model \citep{albert1993bayesian} are underway, and future work will extend AGP to survival outcome data as well. An optimized R package for AGP is being written to accommodate larger sample sizes and predictor dimensions (i.e., $n \approx 10^4, \: p \approx 10^6$). Here, developments in GPU and distributed computing, together with low-rank matrix approximation, can enable learning of an association graph between predictors over parallelized MCMC chains. Robust estimates for the predictor correlation matrix \citep{schafer2005shrinkage, bickel2008regularized} may also be used to enhance the efficiency of learning predictor inclusion scores. Finally, additional MCMC strategies for moves in the joint space of component inclusion are required to tackle local modality issues which can arise when additive effects share predictors (e.g., the ``confounded'' test function in Section \ref{sec:varSelectionInteractionRecovery}). There, in some MCMC runs where an AGP component models predictor interaction (1,2,3,4), local neighborhood moves do not subsequently isolate the latter into constituent (1,2,3) and (2,4)  interactions. The ``neighbor of a good neighbor is a better neighbor'' effect was a driving force that lead to the paired-move neighborhood sampler (PRNS), which is effective at overcoming this obstacle when additive components are fairly well separated (e.g., the modified Friedman function in Section \ref{sec:varSelectionInteractionRecovery}). For the ``confounded'' example, this is far from true -- here, a donate ICM move can only overcome this effect in the (low probability) event that predictor 2 is proposed and accepted in another component via an Add move, followed by a donate ICM move from the component in question in a subsequent iteration. In addition, consider the case where (1,2,3) and (4,5) are modeled in separate components. A paired move enables predictor 2 to be added to the second component, and subsequent donate ICM (paired remove) moves allow (2,4,5) to be separated into (2,4) and 5. On the other hand, ICM also enables one to move toward the {\it undesired} local mode, i.e., by considering donate ICM moves from the component with (4,5).

\section*{Acknowledgement} 
SQ is partially supported by grants R01-ES-017436 and R01-ES-017240 from the National Institute of Environmental Health Sciences (NIEHS) of the National Institutes of Health (NIH). ST is partially supported by grant R01-ES-017436 from NIEHS of NIH. Any opinions, findings, and conclusions or recommendations expressed in this material are those of the authors and do not necessarily reflect the views of the NIEHS or NIH.

\clearpage
\appendix
\label{sec:appendix}

\section{Additive Gaussian process sampling}
\subsection{Markov chain sampling for AGP regression}
Section \ref{sec:paired_move_wmtm} introduces a reversible paired-move neighborhood MTM sampler to efficiently explore the space of one-away inclusion vectors. Section \ref{sec:icm_sampling} introduces inter-component moves (ICM) to explore local neighborhoods of joint configuration state $\bgamma = \{\gamma_1, \dots, \gamma_k\}$ that preserves the stationarity distribution of the paired-move sampler. Algorithm \ref{alg1} uses these moves together with predictor importance adaptation (see Section \ref{sec:inclusion_propensity_adapt}) to draw approximate posterior samples for AGP parameters $\bgamma, \brho, \blambda$. 
\begin{algorithm}[ht!]
\begin{algorithmic}[1]
\caption{AGP Markov chain sampling} \label{alg1}
\Require (Defaults)
$k_\mathrm{min} = \lfloor \log(p) \rfloor$, $k_\mathrm{max} = \lceil p^{1/2} \rceil$, $b_0 = 100$, $\Delta = 0.20$.
\Ensure 
Approximate posterior draws for AGP parameters $\{\brho, \blambda, \bgamma\}$
\Function{AGP.SAMPLE}{} 
\Comment{Sampling for AGP model \eqref{eq:agp-model}}

\State Call UPDATE.ACTIVESET($\bxi$, $\mathcal{I}_A$, $B$)
\Comment{see Algorithm \ref{alg2}}
\State Sample $\tau \sim {\rm Beta}\Big(\mu' = \frac{d^* + \sum_{l \in \mathcal{I}_A} d_l}{\nu'}, \nu' = (1 + k_a)p\Big)$
\If{${\rm Unif}(0,1) < \Delta$}
\Comment{Do ICM move}
	\State Select ICM move $m \in \{\rm CD, PD, PS\}$
	\State Propose $\bxi' = (\bgamma', \btheta') \sim p(\cdot | A_m(\bgamma), \btheta, \by)$
	\If{${\rm Unif}(0,1) < \alpha_m(\bxi, \bxi')$}
	\Comment{see Section \ref{sec:icm_sampling} \eqref{eq:icm_acceptance_ratio}}
		\State Set $\bgamma, \brho, \btheta \leftarrow \bgamma', \brho', \btheta'$
		\If{$ m = $ CD}
		\State \textbf{go to} line 13
		\EndIf
		\Else
		\For{$l \in \mathcal{I}_A$}
		\State Sample $\rho_l, \lambda_l \sim \pi(\cdot | \gamma_l, \bxi_{(l)}, \by)$
		\Comment{see Section \ref{sec:backfitting_agp} \eqref{eq:gp_scale_param_update}}
		\EndFor
	\EndIf
	\Else
	\Comment{Do paired-move DMTM}
		\For{$l \in \mathcal{I}_A$}
			\State Select pared-move $m \in \{A, R, S\}$
			\State Propose $\gamma_l' \sim p(\cdot | N_m(\gamma_l), \bxi_{(l)}, \by)$
			\Comment{see Section \ref{sec:paired_move_wmtm} \eqref{eq:paired_move_wmtm_alpha}}
			\If{${\rm Unif}(0,1) < \alpha_m(\gamma_l, \gamma_l')$}
				\State Set $\gamma_l \leftarrow \gamma_l'$
			\EndIf
			\State Sample $\rho_l, \lambda_l \sim \pi(\cdot | \gamma_l, \bxi_{(l)}, \by)$
			\Comment{see Section \ref{sec:backfitting_agp} \eqref{eq:gp_scale_param_update}}
			\\
			\If{$\rho_l > 0, d_l > 0$}
				\State Update $v_j$, $j \in \gamma_l$
				\Comment{see Section \ref{sec:inclusion_propensity_adapt} \eqref{eq:propensity_update}}
			\EndIf
		\EndFor
	\EndIf
\EndFunction
\end{algorithmic}
\end{algorithm}

\subsection{Predictive inference for AGP regression}
Let $\bLambda = \sum_{l = 1}^k \rho_l^2 \bC_l$ represent the scaled covariance of the {\em aggregated} GP $f(x) = \sum_{l = 1}^k f_l(x)$, and $\bC_l, \: 1 \le l \le k$ are the scaled component covariances defined in Section \ref{sec:agp-regression}. Let $\bs_f = (\boldf(\bx_1), \dots, \boldf(\bx_n))'$ stand for the $n$-vector of aggregated GP realizations, $\boldf(\bx_i) = \sum_{l \in \mathcal{I}_A} f_l(x_i)$. For a new observation $(\bx^*, y^*)$, the joint distribution of $\boldf(x^*)$ with aggregated GP realizations over the training data follows from Kolmogorov consistency for Gaussian processes, namely 
\begin{align}
\left(\boldf(\bx^*), \boldf(\bx_1), \dots, \boldf(\bx_n)\right)' \sim \mathrm{N}(\bzero, \sigma^2 \bLambda), \quad
\bLambda = \left( \begin{array}{ll}
\Lambda^* & \bLambda_{\rm cr} \\
\bLambda_{\rm cr}' & \bLambda_{1:n}
\end{array} \right).
\end{align}
This implies $\boldf(\bx^*) | \bs_f, \sigma^2 \sim \mathrm{N}\big(\bLambda_{\rm cr} \bLambda_{1:n}^{-1} \bs_f,\, \sigma^2(\Lambda^* - \bLambda_{\rm cr} \bLambda_{1:n}^{-1} \bLambda_{\rm cr}')\big)$, and the posterior predictive distribution for mean response $\boldf(\bx^*)$ is given by
\begin{align} \label{eq:predictive_distribution}
\pi(\boldf(\bx^*) | \by) = \int \pi(\boldf(\bx^*) | \bs_f, \sigma^2) \: \pi(\bs_f, \sigma^2 | \bxi, \by) \: \pi(\bxi | \by) \: d(\bxi, \bs_f, \sigma^2).
\end{align}

Approximate posterior draws for AGP parameters $\bxi = \{\bgamma, \brho, \blambda\}$ are obtained via Algorithm \ref{alg1}. Sampling  from \eqref{eq:predictive_distribution} proceeds by \begin{inparaenum}[(1)] \item drawing $(\bs_f, \sigma^2)^{(s)} \sim \pi(\cdot, \cdot, | \bxi^{(s)}, \by)$, available in closed-form; and \item drawing $\boldf^{(s)}(x^*) \sim \pi(\cdot | \bs_f, \sigma^2)$ as given. \end{inparaenum} Using the final $S$ draws, the posterior predictive mean is approximated as
\begin{align}
\mathbb{E}(y^* | \bx^*, \by) \approx \frac{1}{S} \sum_{t > T-S} \boldf^{(t)}(\bx^*).
\end{align}
Here, quantiles of predictive estimates $\boldf^{(1)}(\bx^*), \dots, \boldf^{(S)}(\bx^*)$ provide point-wise posterior credible intervals (see Figure \ref{fig:pred_perf}). In addition, promising inclusion configurations may be identified by thresholding $\pi\big(\bgamma^{(t)}, \brho^{(t)}, \blambda^{(t)} | \by \big)$. For a subset of inclusion configurations $\mathcal{M}$, marginal posterior probabilities may be obtained by running an MCMC chain over AGP parameters $\brho, \blambda$ (in parallel for every configuration state $\bgamma \in \mathcal{M}$), and computing
\begin{align} 
\hat{\pi}(\bgamma | \by) = \frac{1}{T} \sum_{t = 1}^T \pi(\bgamma | \brho^{(t)}, \blambda^{(t)}, \by), \quad
\hat{P}(m_\bgamma | \by) = \frac{ \hat{\pi}(\bgamma | \by)}{\sum_{\bgamma \in \mathcal{M}} \hat{\pi}(\bgamma | \by)}. \label{eq:model_score}
\end{align}

\section{Efficient matrix inversion for AGP sampling}

\subsection{Low-rank approximation for AGP} 
\label{sec:scaling_computation}

Evaluating the likelihood function under a GP regression model requires inverting the $n \times n$ (kernel) covariance matrix. The neighborhood sampler for updating component inclusion vectors using back-fitting (see Sections \ref{sec:cond_gp_sampling} and \ref{sec:neighborhood_search}) requires likelihood evaluations for every $\mt{\gamma} \in N(\gamma)$ at each MCMC iteration which grows as $O(p)$. A paired-move neighborhood sampler was introduced in Section \ref{sec:paired_move_wmtm} to probabilistically control the number of neighbors that must be scored. 

When $n$ exceeds several hundred observations the $O(n^3)$ matrix inversion adds significantly to the per-iteration complexity. This arises from Cholesky factorizations for aggregated AGP covariance $\bK = \sum_{l =1}^k \rho_l^2 \bC_l = \bR' \bR$, however we employ a pivoted low-rank matrix approximation which enables matrix inversion at a reduced $O(n r^2)$ cost, $r \ll n$\footnote{By default, Algorithm \ref{alg1} sets pivoted rank $r = \lceil\min\{n, \tfrac{5}{2}(\log n)^2\}\rceil$.}. Here, $\bK \approx \mt{\bR}' \mt{\bR} + \bD$, for an upper triangular matrix $\mt{\bR}_{r \times n}$ and diagonal matrix $\bD$. The marginal likelihood calculation in \eqref{eq:gp_gamma_update} involving $\bSigma^{-1} = (\bI_n + \bK)^{-1} \approx (\bA + \mt{\bR}' \mt{\bR})^{-1}$, $\bA = \bI_n + \bD$, can then be computed efficiently via an application of the Sherman-Morrison Woodburry formula. \cite{harbrecht2012low} provide an overview of pivoted inversion. Other procedures in the predictive-process literature choose a set of ``knot points'' to tackle the cubic-order complexity, where points are adaptively selected in response to changes in the canonical metric induced by uncertainty in the parameters controlling the GP covariance function \citep{banerjee2008gaussian,tokdar.11, banerjee2012efficient}.

\section{Validity of inclusion vector sampling for AGP}
\label{sec:detailed_balance_pfs}

Below, stationary distribution $\pi(\cdot | -)$ is abbreviated as $\pi(\cdot)$. 

\subsection{Reversibility for paired-move neighborhood sampling}
Consider the vanilla paired-move neighborhood sampler having proposal distribution \eqref{eq:paired_move_proposal} and acceptance probability \eqref{eq:prns_mh_alpha}. Below, detailed balance is verified for a paired ``Add-remove'' (``Remove-add'' follows from symmetry, which is precisely the reversibility condition being checked) and ``Swap-swap'' proposals. Given component inclusion vector $\gamma$, one first chooses between Add, Remove and Swap with probabilities $w_A(|\gamma|)$, $w_R(|\gamma|)$ and $w_S(|\gamma|)$, respectively.\\

\noindent \underline{If $m = A$}: construct one-add neighborhood $N_A(\gamma) = \{\tilde \gamma = \gamma + 1_j$ : $j \not\in \gamma \}$. Select $\gamma' = \gamma + 1_k$ with probability $\pi(\gamma') / \sum_{\tilde\gamma \in N_A(\gamma)} \pi(\tilde\gamma)$, namely the RNS proposal distribution \eqref{eq:paired_move_proposal} restricted this set. The paired reverse one-remove neighborhood $N_R(\gamma')$ always contains $\gamma$. Accept with probability $\alpha_m(\gamma, \gamma')$ given by \eqref{eq:prns_mh_alpha}. Then,
\begin{align*}
& \pi(\gamma) T(\gamma, \gamma') = \pi(\gamma) w_A(|\gamma|) \frac{\pi(\gamma')}{\sum_{\mt{\gamma} \in N_A(\gamma)} \pi(\mt{\gamma})} \min\left\{1, \frac{w_R(|\gamma'|)}{w_A(|\gamma|)} \frac{\sum_{\mt{\gamma} \in N_A(\gamma)} \pi(\mt{\gamma} )}{\sum_{\mt{\gamma} \in N_R(\gamma')} \pi(\mt{\gamma})}\right\} \\
&= \pi(\gamma) \pi(\gamma' ) \min\left\{ \frac{w_A(|\gamma|)}{\sum_{\mt{\gamma} \in N_A(\gamma)} \pi(\mt{\gamma})}, \frac{w_R(|\gamma'|)}{\sum_{\mt{\gamma} \in N_R(\gamma')} \pi(\mt{\gamma})} \right\}.
\end{align*}

\noindent \underline{If $m = S$}: construct one-swap neighborhood $N_S(\gamma) = \{\mt{\gamma} = \gamma -1_m + 1_n : (m,n) \in \gamma \times [\gamma]^c\}$. Select $\gamma' = \gamma -1_j + 1_k$ with probability $\pi(\gamma') / \sum_{\tilde\gamma \in N_S(\gamma)} \pi(\tilde\gamma)$, namely the RNS proposal distribution \eqref{eq:paired_move_proposal} restricted this set. The paired reverse swap neighborhood $N_S(\gamma')$ always contains $\gamma$. Accept with probability $\alpha_m(\gamma, \gamma')$ given by \eqref{eq:prns_mh_alpha}. Then,
\begin{align*}
& \pi(\gamma) T(\gamma, \gamma') = \pi(\gamma) w_S(|\gamma|) \frac{\pi(\gamma')}{\sum_{\mt{\gamma} \in N_S(\gamma)} \pi(\mt{\gamma})} \min\left\{1, \frac{\sum_{\mt{\gamma} \in N_S(\gamma)} \pi(\mt{\gamma} )}{\sum_{\mt{\gamma} \in N_S(\gamma')} \pi(\mt{\gamma})}\right\} \\
&= \pi(\gamma) \pi(\gamma' ) \min\left\{ \frac{w_S(|\gamma|)}{\sum_{\mt{\gamma} \in N_S(\gamma)} \pi(\mt{\gamma})}, \frac{w_S(|\gamma|)}{\sum_{\mt{\gamma} \in N_S(\gamma')} \pi(\mt{\gamma})} \right\}, ~ w_S(|\gamma|) = w_S(|\gamma'|).
\end{align*}

The expression $\pi(\gamma)T(\gamma,\gamma')$ is symmetric in $\gamma$ and $\gamma'$ in both cases. This completes the proof.

\subsection{Reversibility for paired-move discrete multiple-try Metropolis sampler}
The proof below makes use of weight function $\omega(\gamma_j, v_j, m)$ defined in \eqref{eq:weight_fn_wmtm}, and acceptance probability $\alpha_m(\gamma, \gamma')$ given by \eqref{eq:paired_move_wmtm_alpha}. Given current inclusion vector $\gamma$, choose between Add, Remove and Swap with probabilities $w_A(|\gamma|)$, $w_R(|\gamma|)$ and $w_S(|\gamma|)$, respectively. Let $m \in \{A, R, S\}$ denote this choice, and $m' \in \{R, A, S\}$ is the corresponding paired reverse neighborhood. In each case, the transition kernel is described in Section \ref{sec:paired_move_wmtm}.\\

\noindent \underline{If $m \in \{A, R\}$}: See steps (2a) and (4a/b) in Section \ref{sec:paired_move_wmtm} for construction of transition kernel $T(\gamma, \gamma')$. Let $\omega_j = \omega(\gamma_j, v_j, m)$ denote the forward bernoulli inclusion probability for predictor $j$, and $\wt{\omega}_j = \omega(\gamma_j', v_j, m')$ denotes its inclusion probability in the reverse paired neighborhood. Then for $\gamma' = \tog(k, \gamma)$,
{\small\begin{align*}
&\pi(\gamma)T(\gamma, \gamma') \\
&= \pi(\gamma) w_m(|\gamma|) \sum_{\substack{\eta, \eta' \in \{0,1\}^p \\ \eta_k = \eta'_k = 1}} \bigg[\omega_k \bigg\{\prod_{j \ne k} \omega_j^{\eta_j} (1 - \omega_j)^{1 - \eta_j} \: \wt{\omega}_j^{\eta_j'} (1-\wt{\omega}_j)^{1-\eta_j'}\bigg\} \\
&\quad\times \frac{\pi(\gamma')}{\sum_{j:\eta_j = 1} \pi(\tog(j, \gamma))} \min\bigg\{1, \frac{w_{m'}(|\gamma'|)}{w_m(|\gamma|)} \frac{\wt{\omega}_k \sum_{j: \eta_{j} = 1} \pi(\tog(j, \gamma))}{\omega_k \sum_{j: \eta'_{j} = 1} \pi(\tog(j, \gamma'))}\bigg\}
\bigg]\\
&= \pi(\gamma)\pi(\gamma')\sum_{\substack{\eta, \eta' \in \{0,1\}^p \\ \eta_k = \eta'_k = 1}} \bigg[ \bigg\{\prod_{j \ne k} \omega_j^{\eta_j} (1 - \omega_j)^{1 - \eta_j} \: \wt{\omega}_j^{\eta_j'} (1-\wt{\omega}_j)^{1-\eta_j'}\bigg\}\\
&\quad\times \min\bigg\{\frac{w_m(|\gamma|) \:\omega_k}{\sum_{j: \eta_{j} = 1} \pi(\tog(j, \gamma))}, \frac{w_{m'}(|\gamma'|) \:\wt{\omega}_k}{\sum_{j: \eta'_{j} = 1} \pi(\tog(j, \gamma'))}\bigg\}\bigg].
\end{align*}}

\noindent \underline{If $m = S$}: See steps (2b) and (4c) in Section \ref{sec:paired_move_wmtm} for construction of transition kernel $T(\gamma, \gamma')$. Denote forward bernoulli inclusion probabilities for a pair of predictors $(r,a) \in \gamma \times [\gamma]^c$ as $\omega_r = \omega(\gamma_r, v_r, m = R)$, $\omega_a = \omega(\gamma_a, v_a, m = A)$. Likewise for reverse inclusion probabilities, let $\omega_r' = \omega(\gamma_r', v_r, m = R)$, $\omega_a' = \omega(\gamma_a', v_a, m = A)$ and $(r,a) \in \gamma' \times [\gamma']^c$. For $\gamma' = \tog((r^*,a^*), \gamma)$, the following facts are noted: \begin{inparaenum}[(i)] \item $\omega_r = \omega'_r$, $r \in \gamma \setminus \{r^*\}$; \item $\omega_a = \omega'_a$, $a \in [\gamma]^c \setminus \{a^*\}$; \item $\omega_{r=r^*} = \omega'_{r=a^*} = 1$; \item $\omega_{a=a^*} = f(v_{a^*})$ and $\omega'_{a=r^*}=f(v_{r^*})$, see \eqref{eq:predictor_inclusion_pr}; and \item $w_S(|\gamma|) = w_S(|\gamma'|)$.~\end{inparaenum}
Then,
{\small\begin{align*}
&\pi(\gamma)T(\gamma, \gamma') \\
&= \pi(\gamma) w_S(|\gamma|) \\
&\quad\times\sum_{\substack{(r,a) \in \gamma \times [\gamma]^c\hfill \\ \eta_{r^*} = \eta_{a^*} = 1\hfill \\ \eta_{a^*}' = \eta_{r^*}' = 1\hfill}} \bigg[\omega_{r^*} \omega_{a^*} \bigg\{\prod_{\substack{(r,a) \\ \ne (r^*,a^*)}} \omega_r^{\eta_r + \eta'_r}(1 - \omega_r)^{2 - \eta_r - \eta'_r} ~ \omega_a^{\eta_a + \eta'_a}(1 - \omega_a)^{2 - \eta_a - \eta'_a}\bigg\} \\
&\quad\times \frac{\pi(\gamma')}{\sum_{(r,a) : \eta_r = \eta_a = 1} \pi(\tog((r,a), \gamma))} \\
&\quad\times \min\bigg\{1, \frac{w_S(|\gamma'|)}{w_S(|\gamma|)} \frac{\omega'_{r^*} \omega'_{a^*} \sum_{(r,a) : \eta_r = \eta_a = 1} \pi(\tog((r,a), \gamma))}{\omega_{r^*} \omega_{a^*} \sum_{(r,a): \eta_r' = \eta_a' = 1} \pi(\tog((r,a), \gamma'))}\bigg\}
\bigg]\\
&= \pi(\gamma)\pi(\gamma') \sum_{[\cdots]} \bigg[\bigg\{\prod_{(r,a) \ne (r^*,a^*)} \omega_r^{\eta_r + \eta'_r}(1 - \omega_r)^{2 - \eta_r - \eta'_r} ~ \omega_a^{\eta_a + \eta'_a}(1 - \omega_a)^{2 - \eta_a - \eta'_a}\bigg\} \\
&\quad\times \min\bigg\{\frac{w_S(|\gamma|) \: \omega_{r^*} \omega_{a^*}}{\sum_{(r,a) : \eta_r = \eta_a = 1} \pi(\tog((r,a), \gamma))}, \frac{w_S(|\gamma'|) \: \omega'_{r^*} \omega'_{a^*}}{\sum_{(r,a): \eta_r' = \eta_a' = 1} \pi(\tog((r,a), \gamma'))}\bigg\}\bigg].
\end{align*}}

In all cases, i.e. for paired-moves $m \in \{A, R, S\}$, the expression $\pi(\gamma)T(\gamma,\gamma')$ is symmetric in $\gamma$ and $\gamma'$. This completes the proof.

\subsection{Proof of Lemma \ref{lem:icm_stationarity}} 

Given state vector $\bxi = (\bgamma, \btheta)$, $\xi_l = (\gamma_l, \rho_l, \lambda_l)$ and $l \in \mathcal{I}_A$, an ICM move proceeds by choosing between ``cross donate'', ``paired donate'', or ``paired swap'' moves with probabilities $w_{\rm CD}$, $w_{\rm PD}$ and $w_{\rm PS}$, respectively. For move $m \in \{\rm CD, PD, PS\}$, the corresponding ICM neighborhood $A_m(\bgamma)$ is constructed (see Section \ref{sec:icm_sampling}). Select $\bxi'$ with probability $\pi(\bxi' | -) / \sum_{\tilde\bxi \in A_m(\bgamma)} \pi(\tilde\bxi | -)$ and accept it with probability $\alpha_m(\bxi, \bxi')$ given by \eqref{eq:icm_acceptance_ratio}. Below, $k_{ne} \le k_a$ denotes the number of non-empty active components.\\

\noindent \underline{If $m = $ CD}: For $\bgamma' = \{\dots, \gamma_u + 1_k, \gamma_v - 1_k, \dots\}$, where component $\gamma_v$ donates predictor $k$ to component $\gamma_u$, and
\begin{align*}
&\pi(\bgamma | \btheta, -) T(\bgamma, \bgamma') \\
&= \pi(\bgamma | \btheta, -) \frac{w_{\rm CD}}{k_{ne}} \frac{\pi(\bgamma' | \btheta, -) }{\sum_{\tilde\bgamma \in N_{\rm CD}(\bgamma;v)} \pi(\tilde\bgamma | \btheta, -)} \min\left\{1, \frac{\sum_{\tilde\bgamma \in N_{\rm CD}(\bgamma;v)} \pi(\tilde\bgamma | \btheta, -)}{\sum_{\tilde\bgamma \in N_{\rm CD}(\bgamma';u)} \pi(\tilde\bgamma | \btheta, -)}\right\} \\
&= \pi(\bgamma | \btheta, -) \pi(\bgamma' | \btheta, -) \min\left\{\frac{w_{\rm CD} / k_{ne}}{\sum_{\tilde\bgamma \in N_{\rm CD}(\bgamma;v)} \pi(\tilde\bgamma | \btheta, -)}, \frac{w_{\rm CD} / k_{ne}}{\sum_{\tilde\bgamma \in N_{\rm CD}(\bgamma';u)} \pi(\tilde\bgamma | \btheta, -)}\right\}.
\end{align*}

\noindent \underline{If $m \in \{\rm PD, PS\}$}: Choose component indices $(u, v) \in \mathcal{I}_A$ with $d_u, d_v > 0$. If $m = {\rm PD}$, then $\bgamma' = \{\dots, \gamma_u + 1_k, \gamma_v - 1_k, \dots\}$, where component $\gamma_v$ donates predictor $k$ to component $\gamma_u$; otherwise $m = {\rm PS}$, and for $j \in \gamma_u (\not\in \gamma_v)$ and $k \in \gamma_v(\not\in \gamma_u)$, $\bgamma' = \{\dots, \gamma_u -1_j + 1_k, \gamma_v -1_k + 1_j, \dots\}$. In addition, here $A_m(\bgamma; (u,v)) = N_m(\bgamma; (u,v)) \times \mathcal{G}^2$ and
\begin{align*}
&\pi(\bxi) T(\bxi, \bxi') \\
&= \pi(\bxi) \frac{w_m}{{k_{ne} \choose 2}} \frac{\pi(\bxi') }{\sum_{\tilde\bxi \in A_m(\bgamma;(u,v))} \pi(\tilde\bxi)} \min\left\{1, \frac{\sum_{\tilde\bxi \in A_m(\bgamma;(u,v))} \pi(\tilde\bxi)}{\sum_{\tilde\bxi \in A_m(\bgamma';(u,v))} \pi(\tilde\bxi)}\right\} \\
&= \pi(\bxi) \pi(\bxi') \frac{w_m}{{k_{ne} \choose 2}} \min\left\{\frac{w_m / k_{ne}}{\sum_{\tilde\bxi \in A_m(\bgamma;(u,v))} \pi(\tilde\bxi)}, \frac{w_m / k_{ne}}{\sum_{\tilde\bxi \in A_m(\bgamma';(u,v))} \pi(\tilde\bxi)}\right\}.
\end{align*}

In all cases, i.e. $m \in \{\rm CD, PD, PS\}$, the expression $\pi(\bxi) T(\bxi, \bxi')$ is symmetric in its arguments. Hence the mixture ICM proposal distribution comprising ``cross donate'', ``paired donate'', and ``paired swap'' moves {\em preserves} stationarity of the paired-move DMTM sampler.

\clearpage
\section{Additional figures} \label{sec:additional_figs}

\begin{figure}[!ht]
\caption{\label{fig:simu_competitor_inclusion} Marginal predictor importance (inclusion probability) for BART, RF, and Lasso on each simulated test function considered in Section \ref{sec:varSelectionInteractionRecovery}. Plots generated below use the same data which used to generate Figure \ref{fig:var_all} $(n = 100, p = 1000$).}
\centering
\arrayrulecolor{gray}
\setlength{\tabcolsep}{2pt}
\begin{tabular}{l | C{0.27\columnwidth} C{0.27\columnwidth} C{0.27\columnwidth} }
& BART & RF & LASSO \\
\hline 
Friedman
& \includegraphics[width=\linewidth]{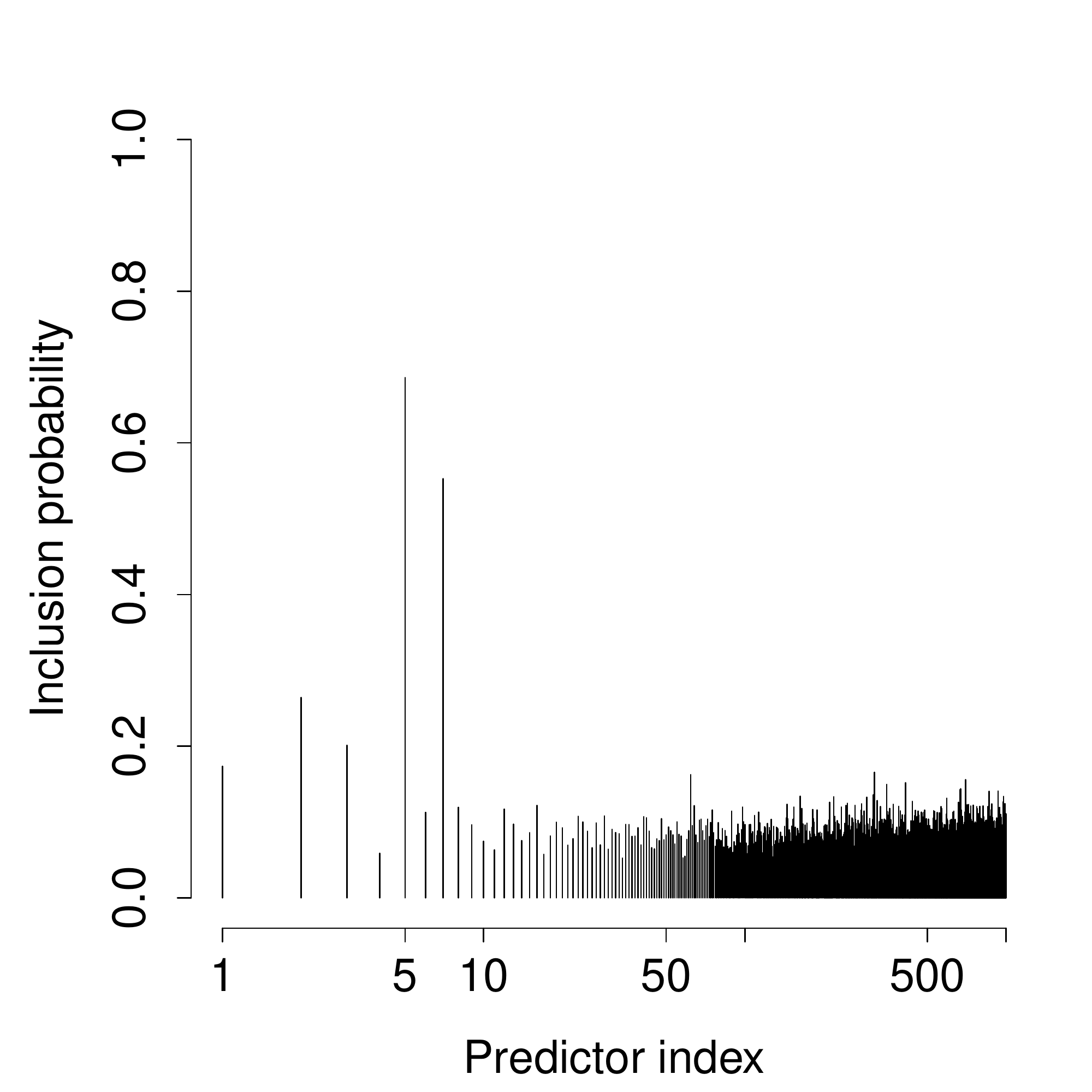}
& \includegraphics[width=\linewidth]{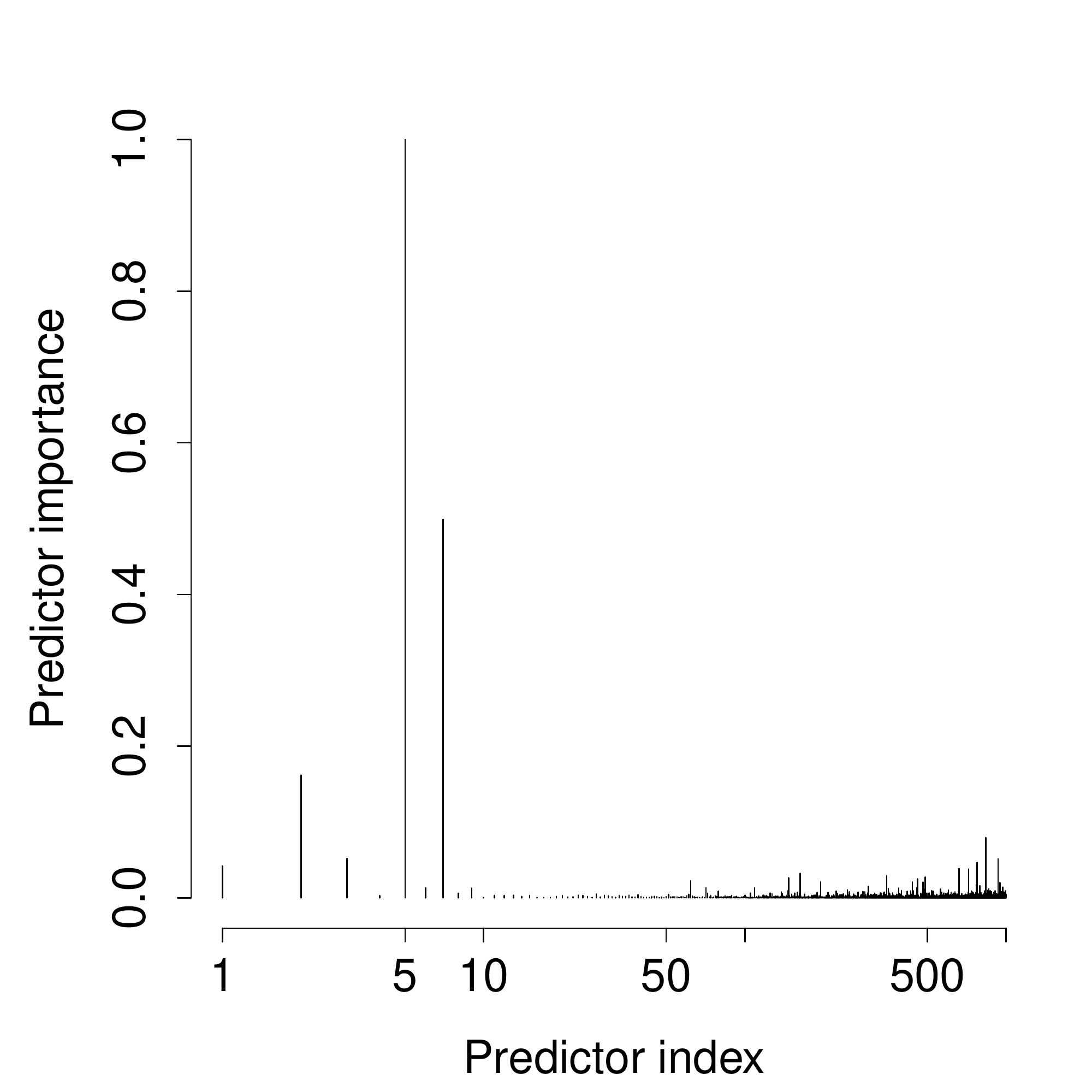}
& \includegraphics[width=\linewidth]{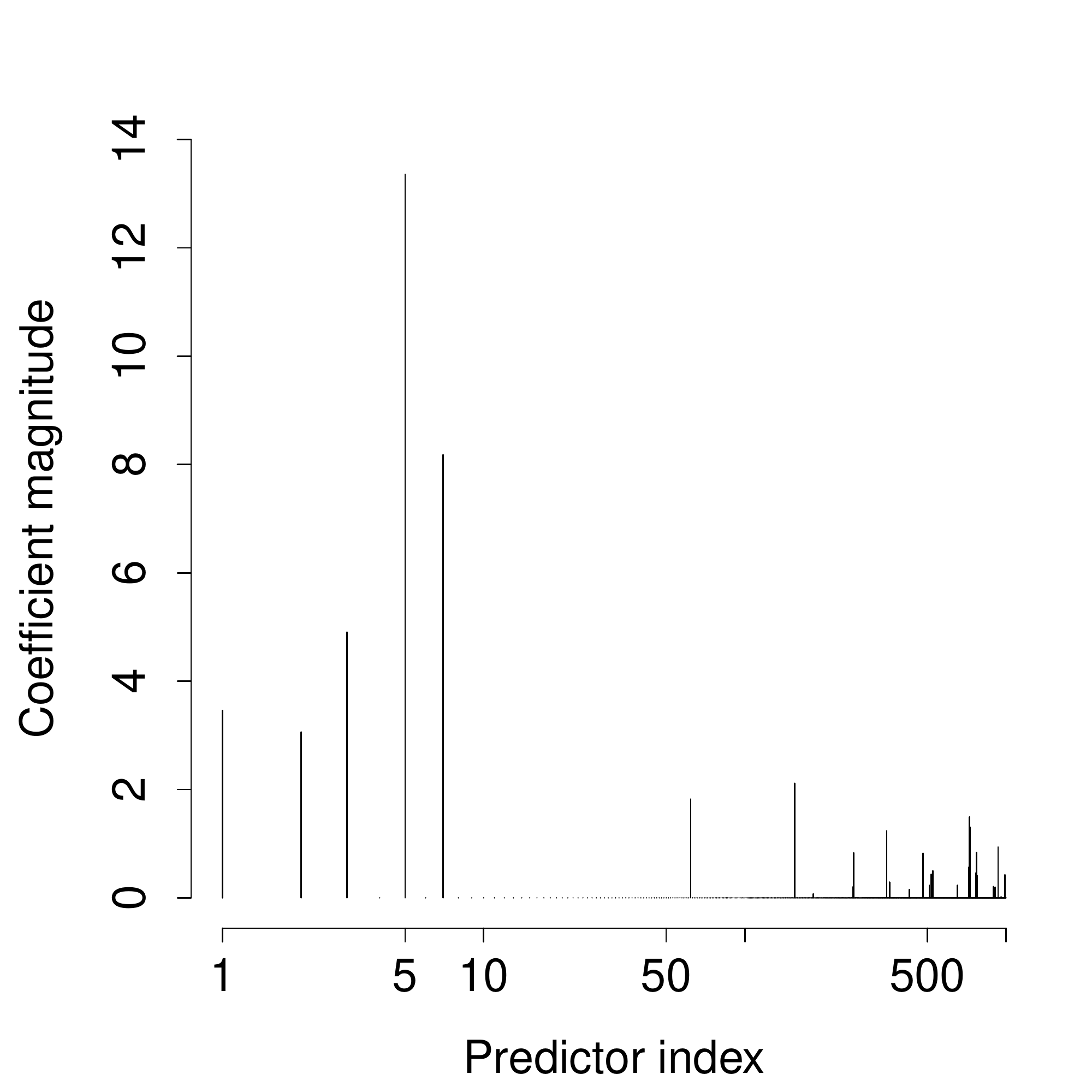} \\
\hline
Confounded
& \includegraphics[width=\linewidth]{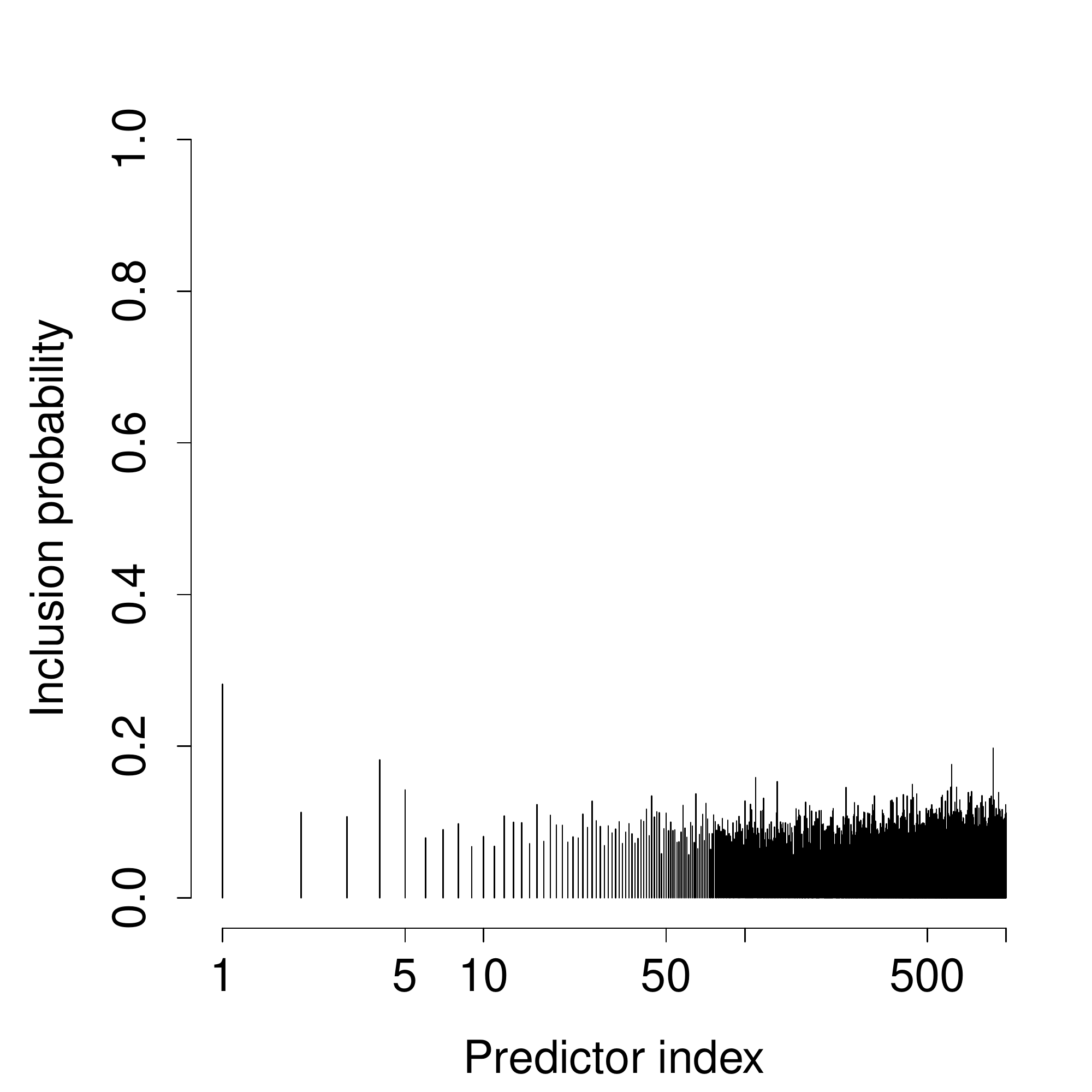}
& \includegraphics[width=\linewidth]{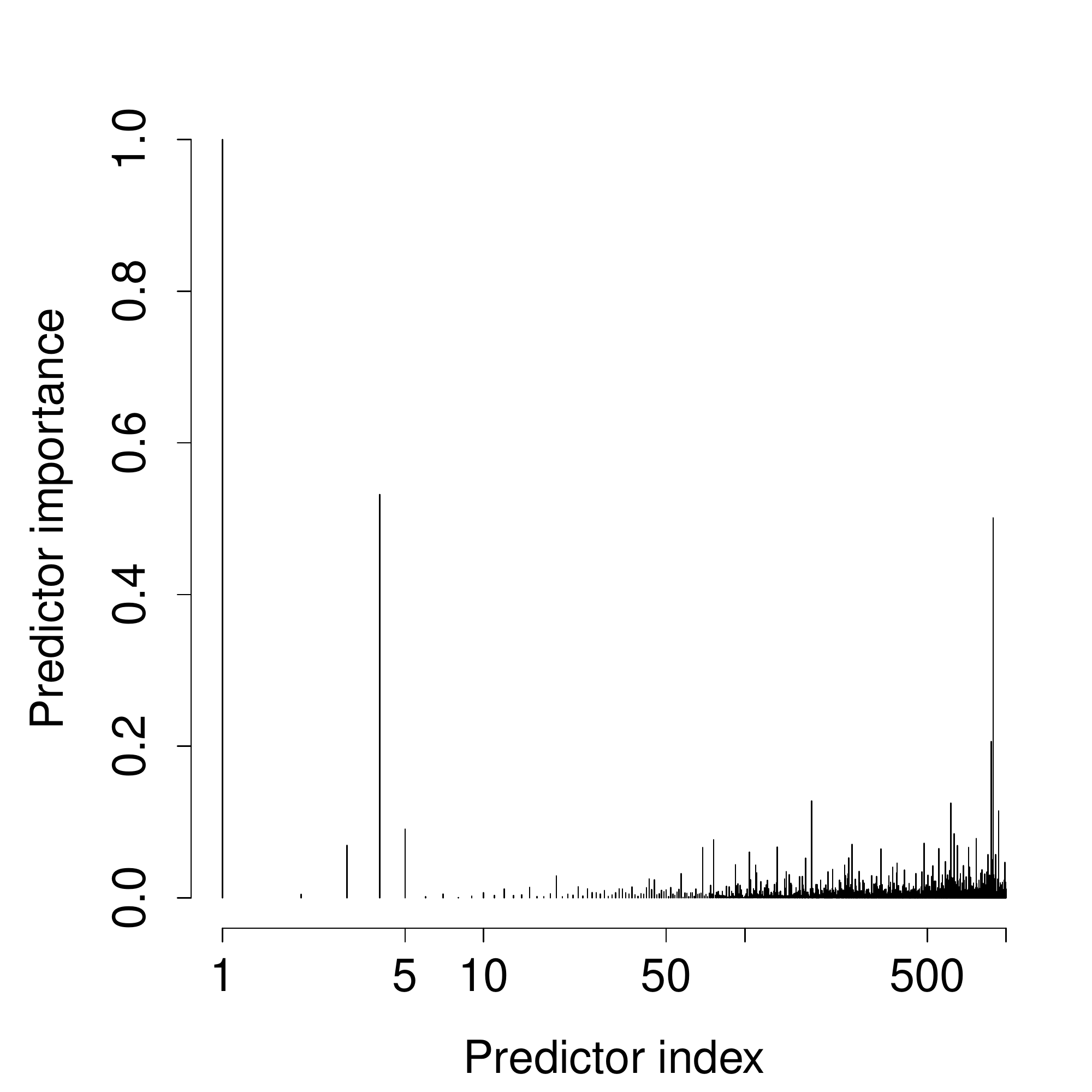}
& \includegraphics[width=\linewidth]{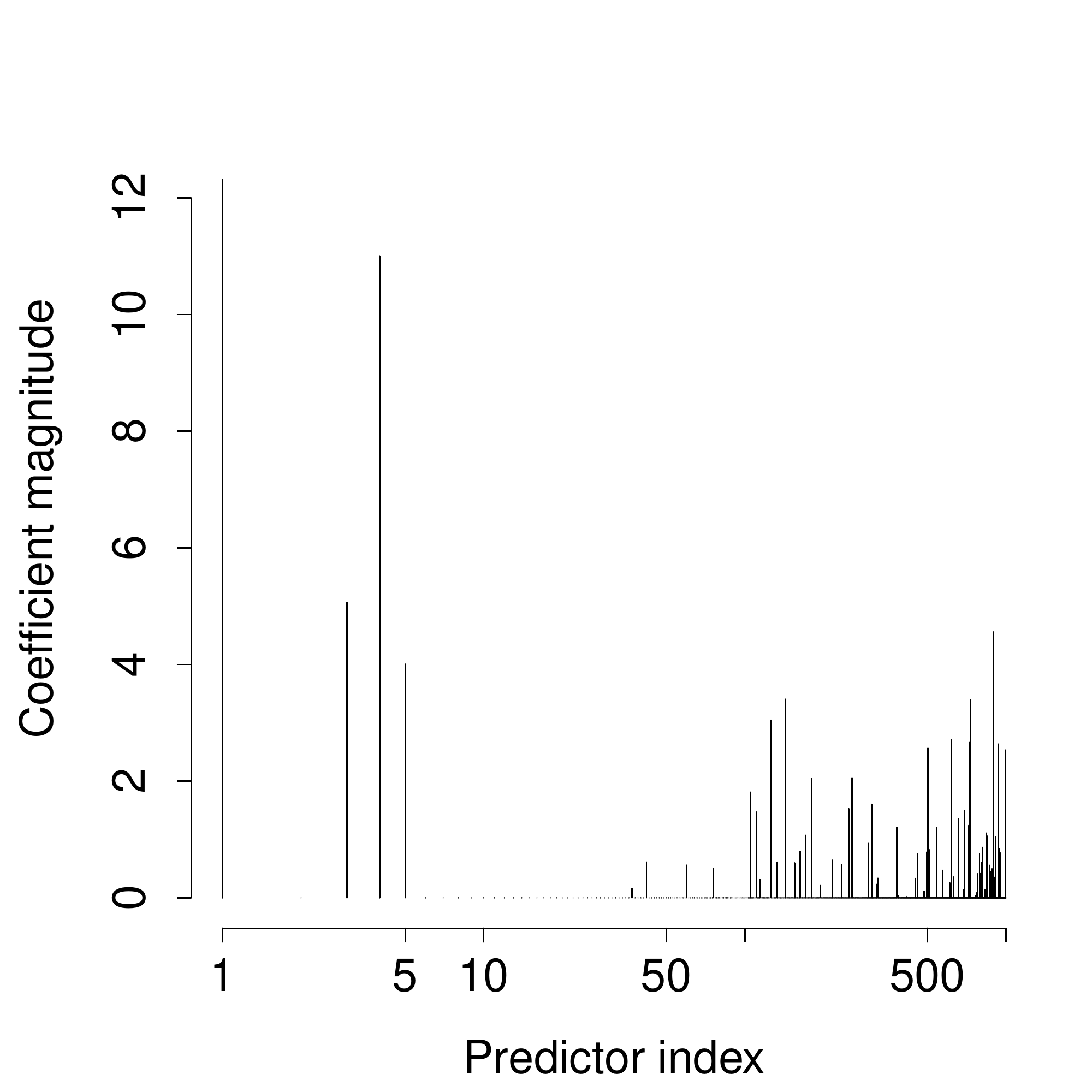} \\
\hline
Linear
& \includegraphics[width=\linewidth]{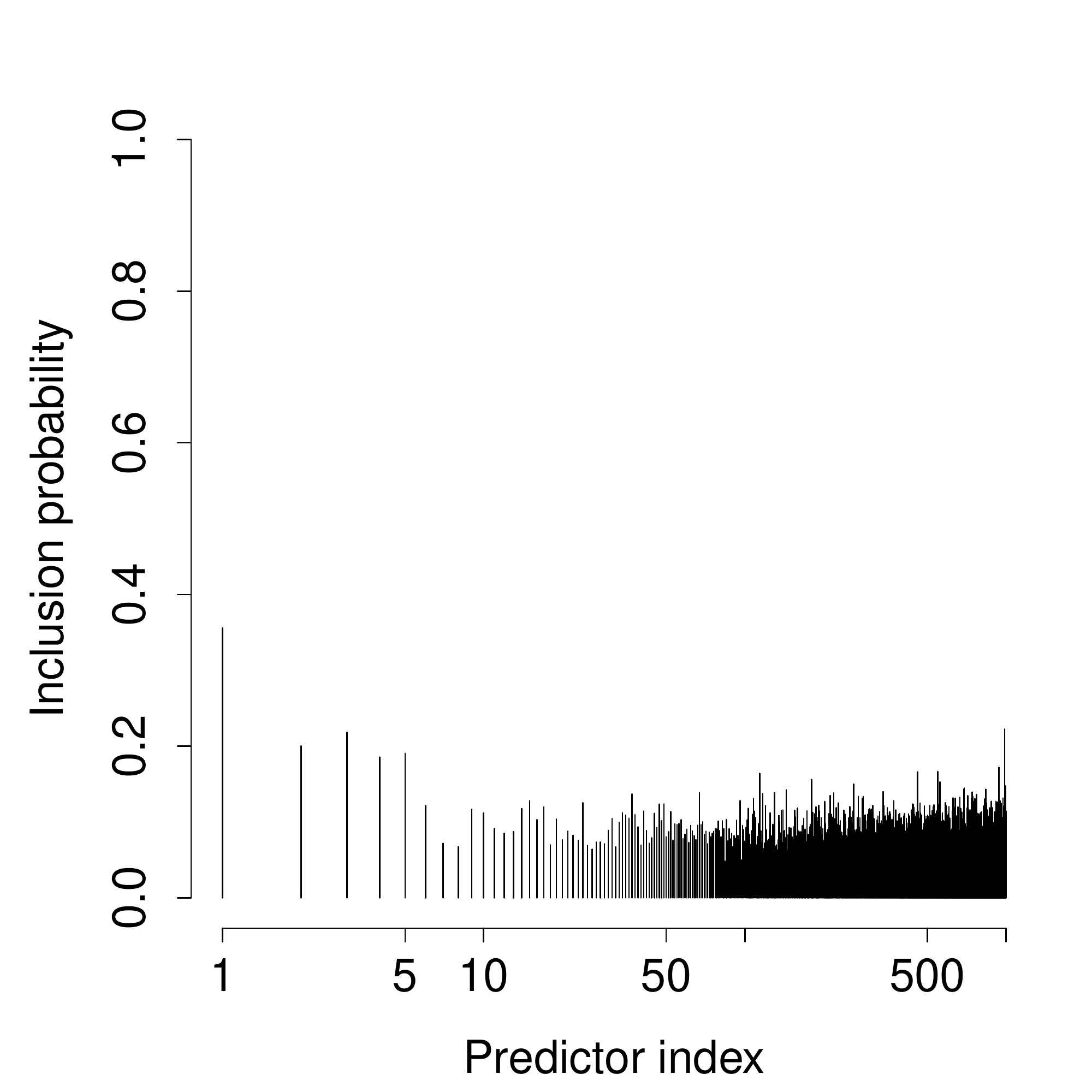}
& \includegraphics[width=\linewidth]{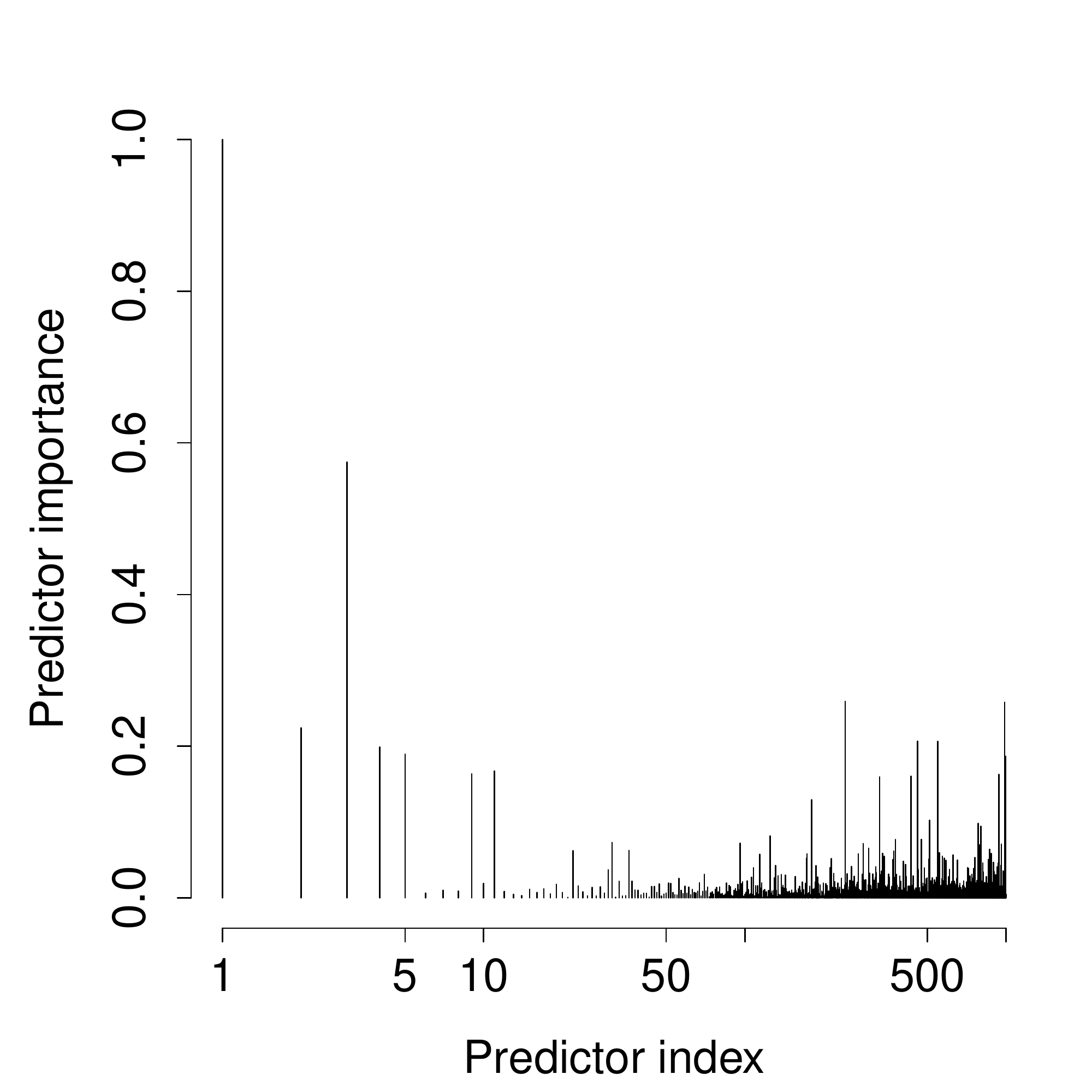}
& \includegraphics[width=\linewidth]{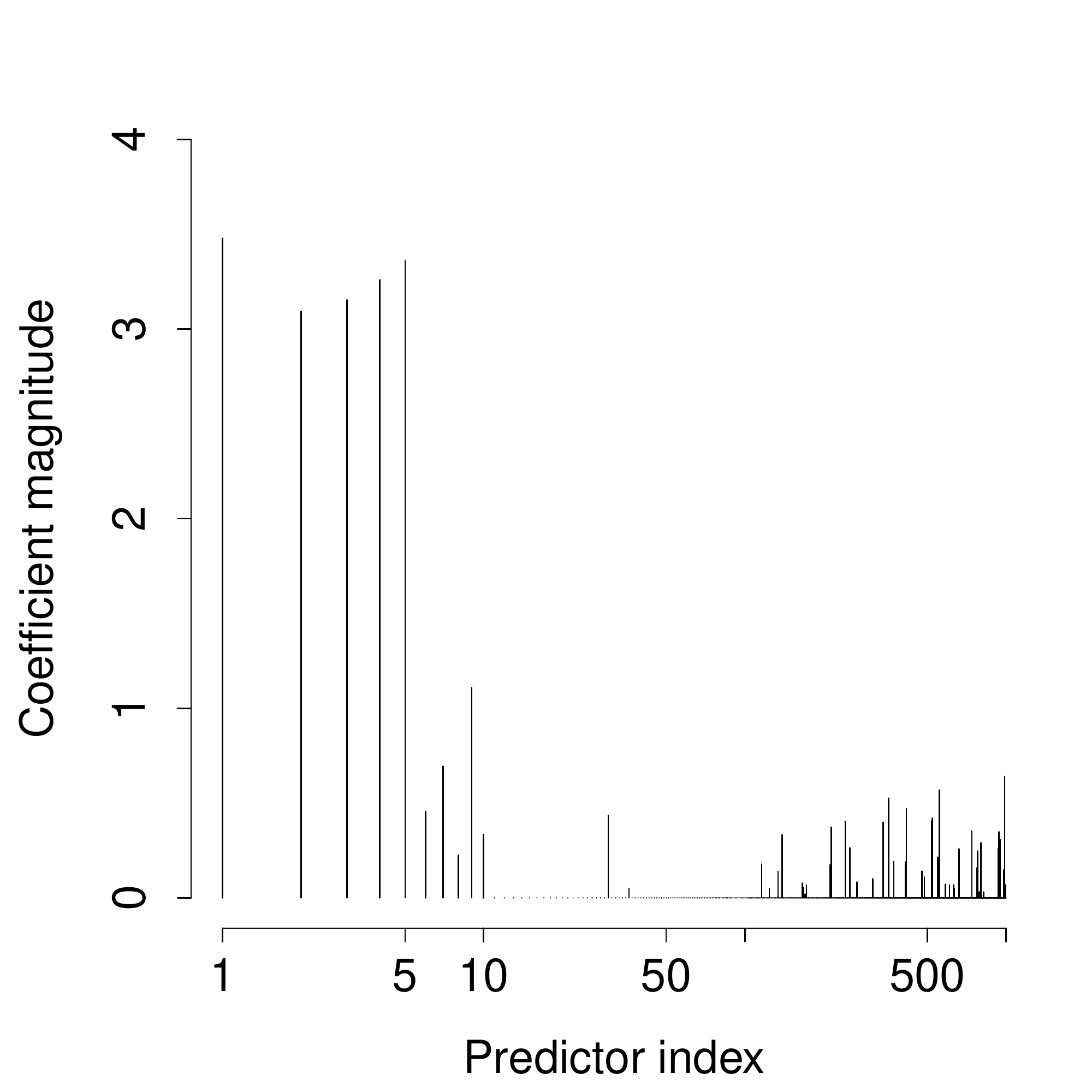} \\
\hline
Single
& \includegraphics[width=\linewidth]{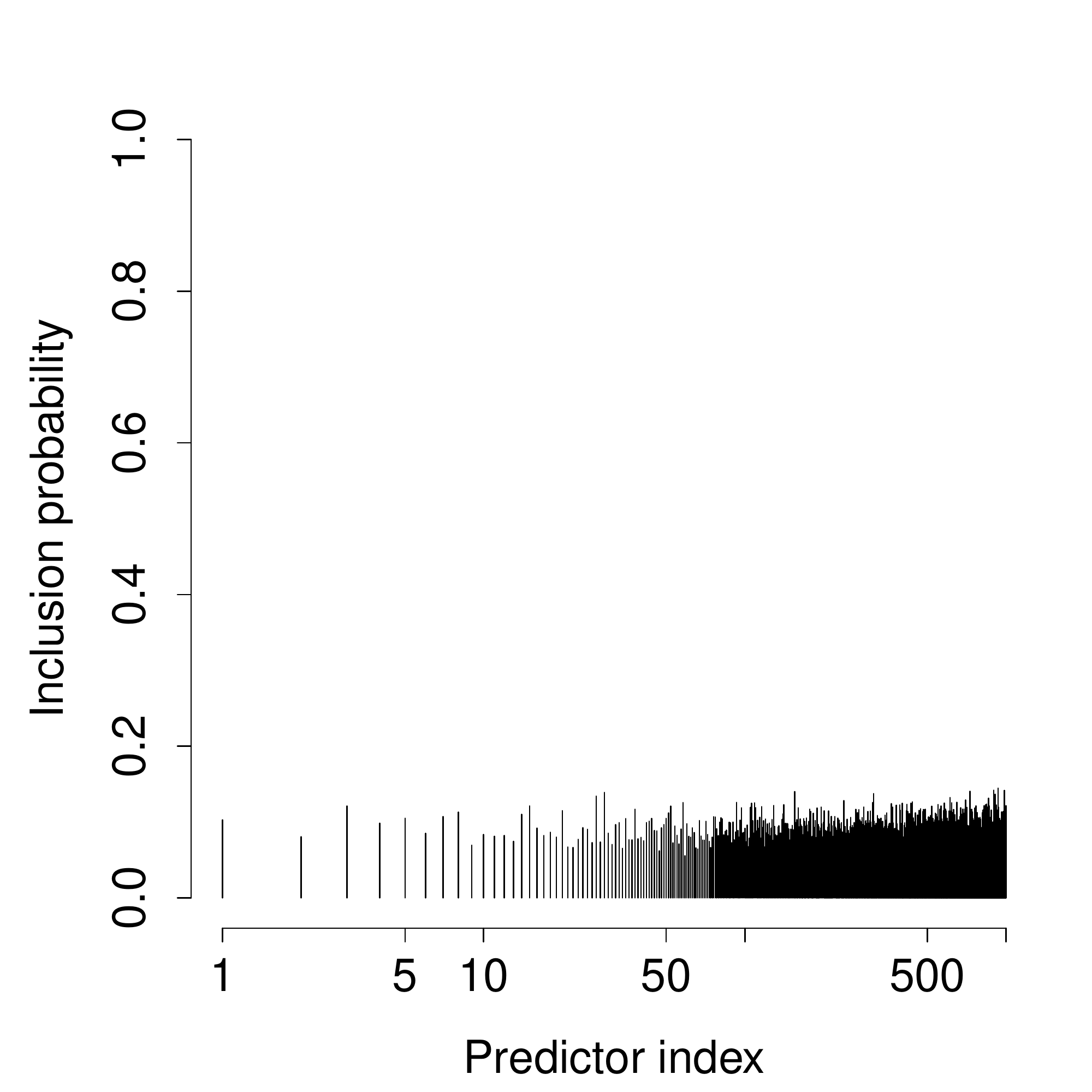}
& \includegraphics[width=\linewidth]{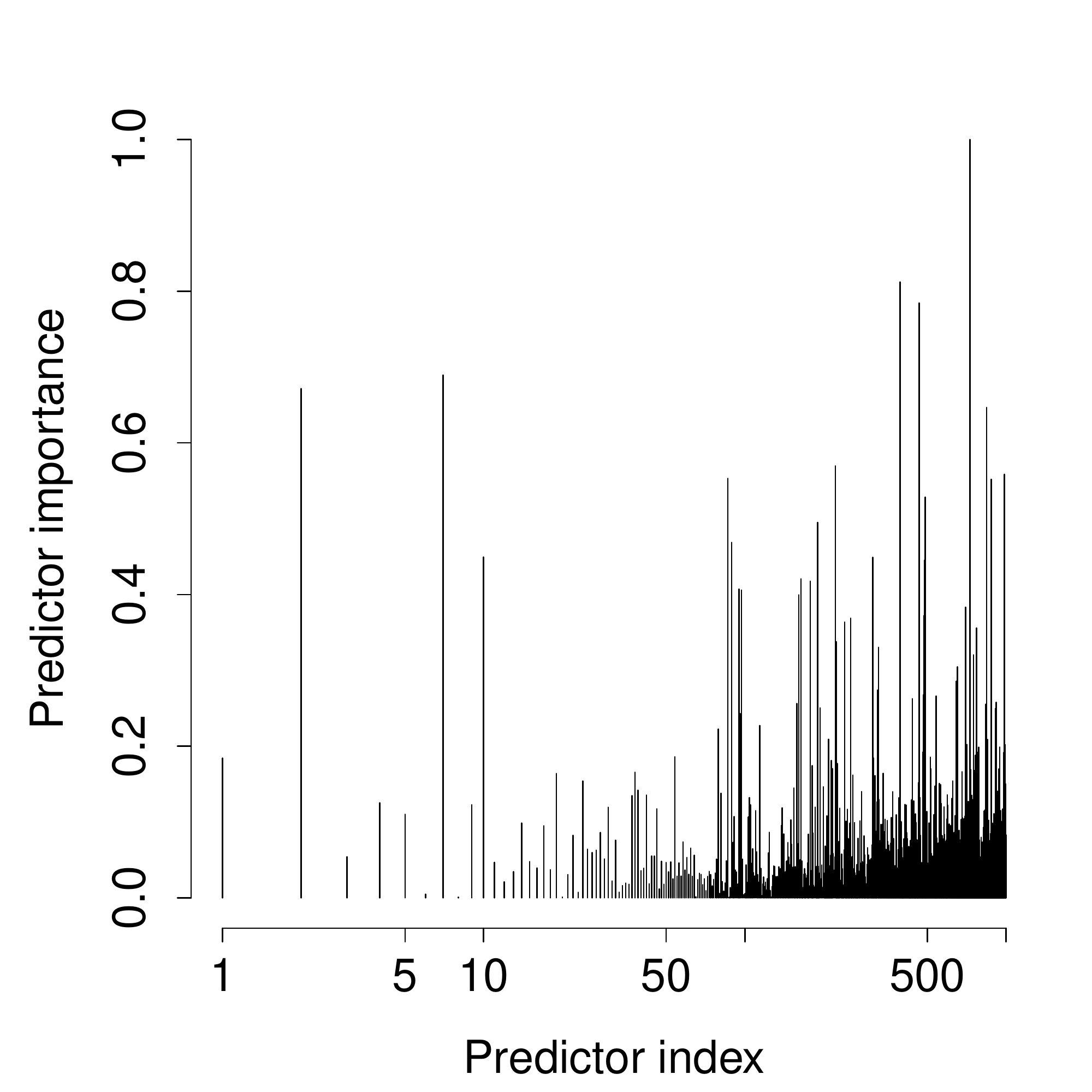}
& \includegraphics[width=\linewidth]{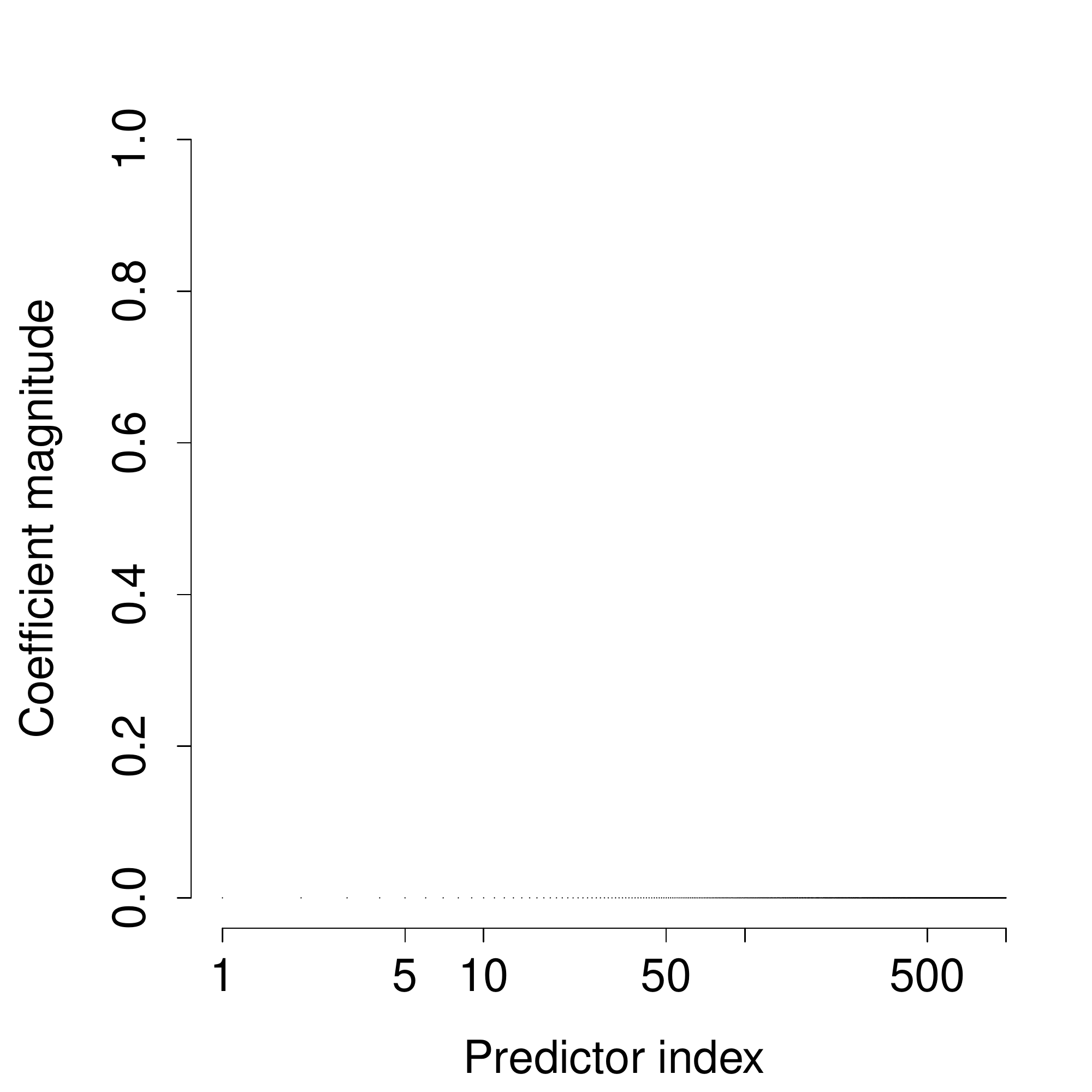}
\end{tabular}
\end{figure}

\begin{figure}[!ht]
\caption{\label{fig:real_posterior_plots} Posterior plots and comparisons for the real datasets considered in Section \ref{sec:real_data_analysis}. {\em Left}: plot of predictor importance scores, $v_j, ~ j = 1,\dots, p$. {\em Middle}: bar-plots of GP scale parameters $\rho, \lambda$ for contributing components (i.e., having 75\% quantile over posterior draws for $\rho_l > 0$). {\em Right}: trace plot of posterior draws for variance parameter, $\sigma^2$.}
\centering
\arrayrulecolor{gray}
\setlength{\tabcolsep}{2pt}
\begin{tabular}{l | C{0.27\columnwidth} C{0.27\columnwidth} C{0.27\columnwidth} }
& Importance scores $\{v_j, j = 1, \dots, p\}$ & GP parameters \{$\rho_l, \lambda_l, l \in \mathcal{I}_A\}$ & Trace plot ($\sigma^2$) \\
\hline 
Boston
& \includegraphics[width=\linewidth]{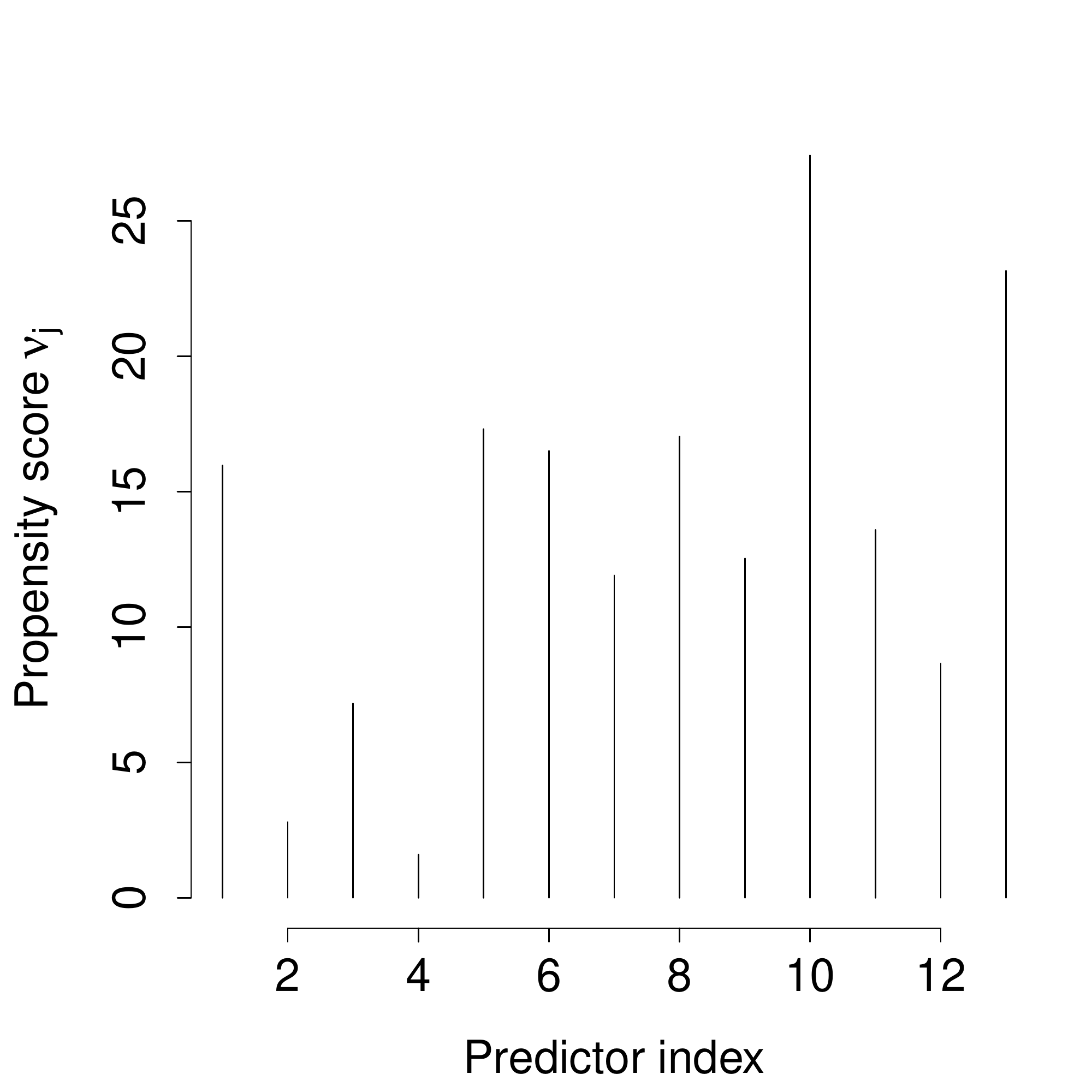}
& \includegraphics[width=\linewidth]{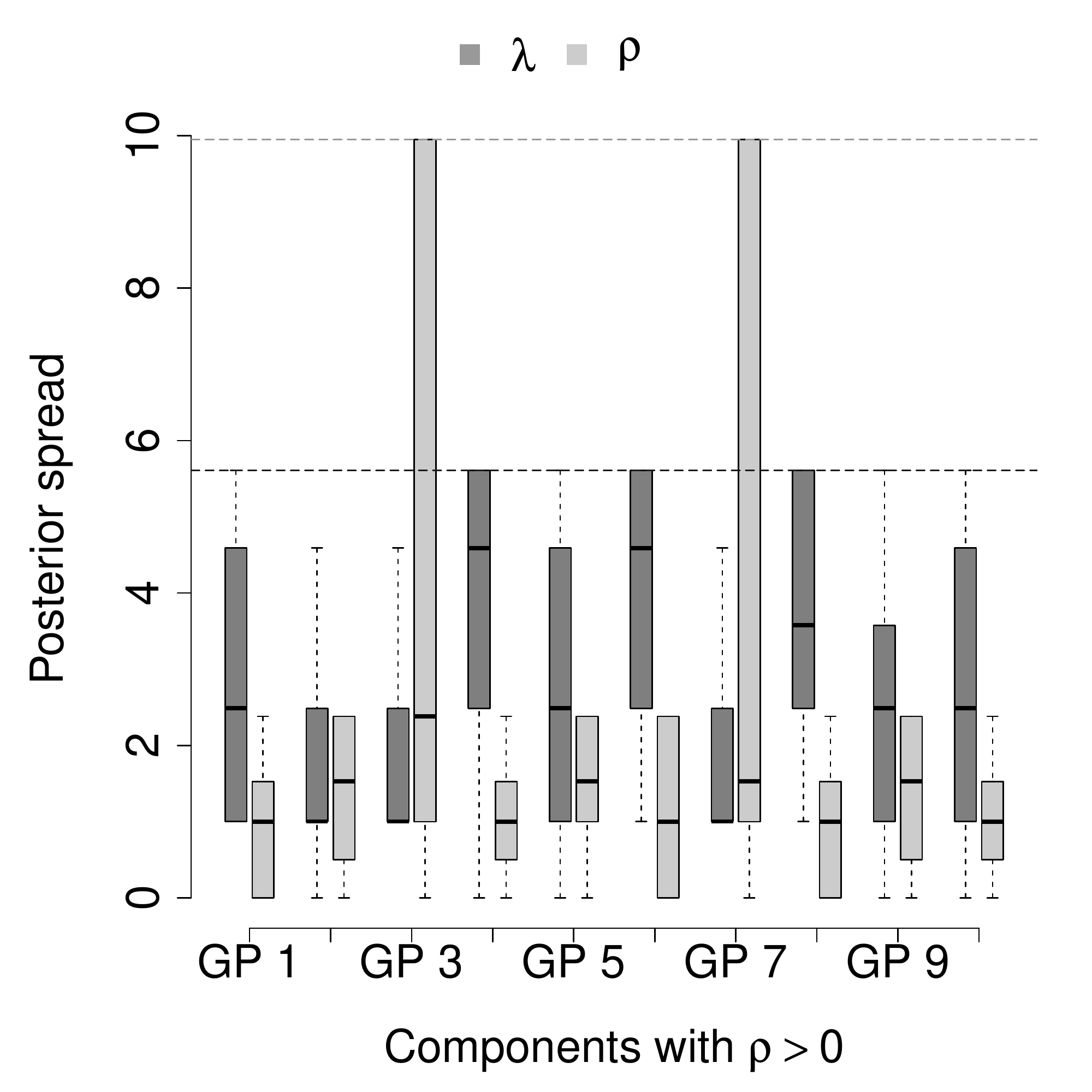}
& \includegraphics[width=\linewidth]{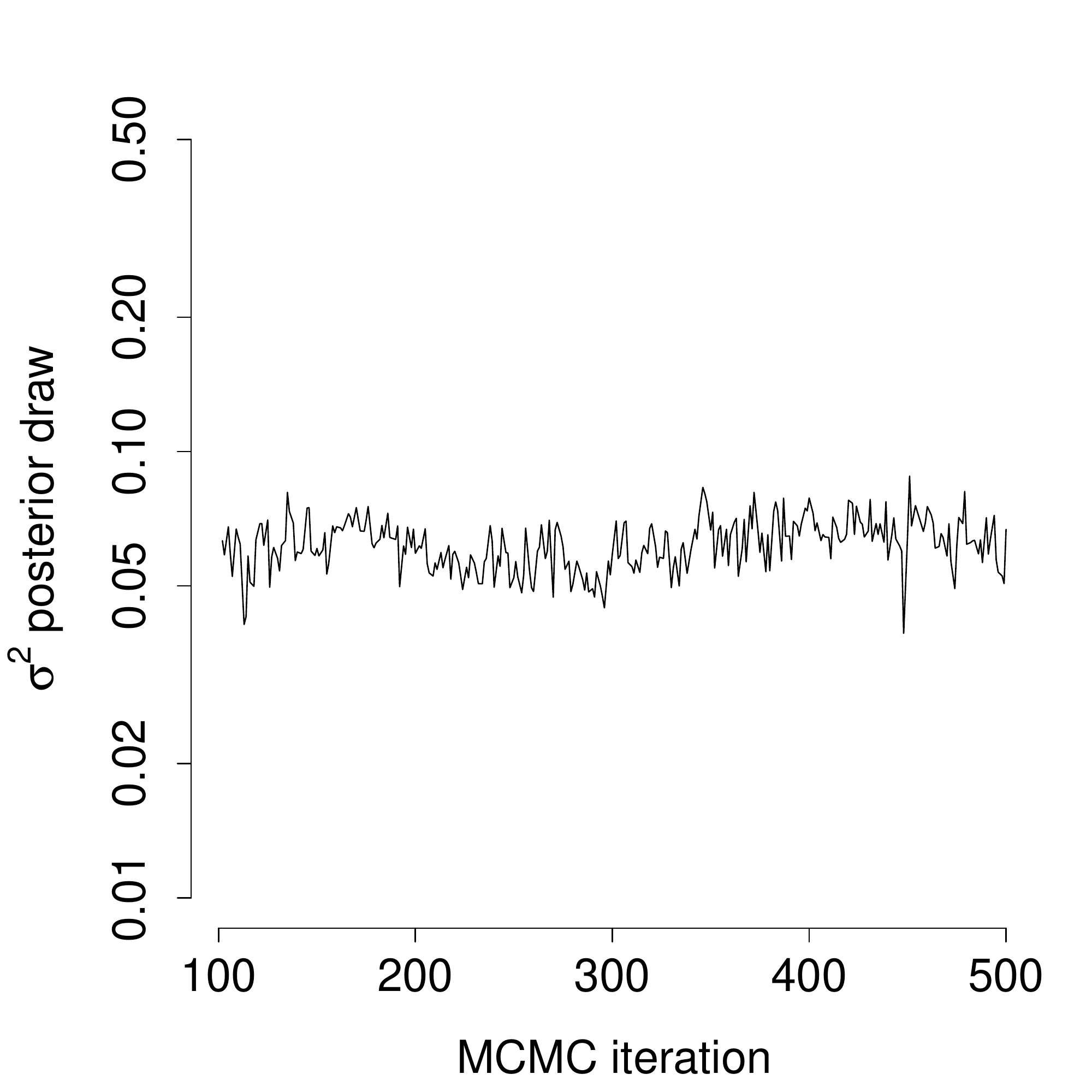} \\
\hline
Crime
& \includegraphics[width=\linewidth]{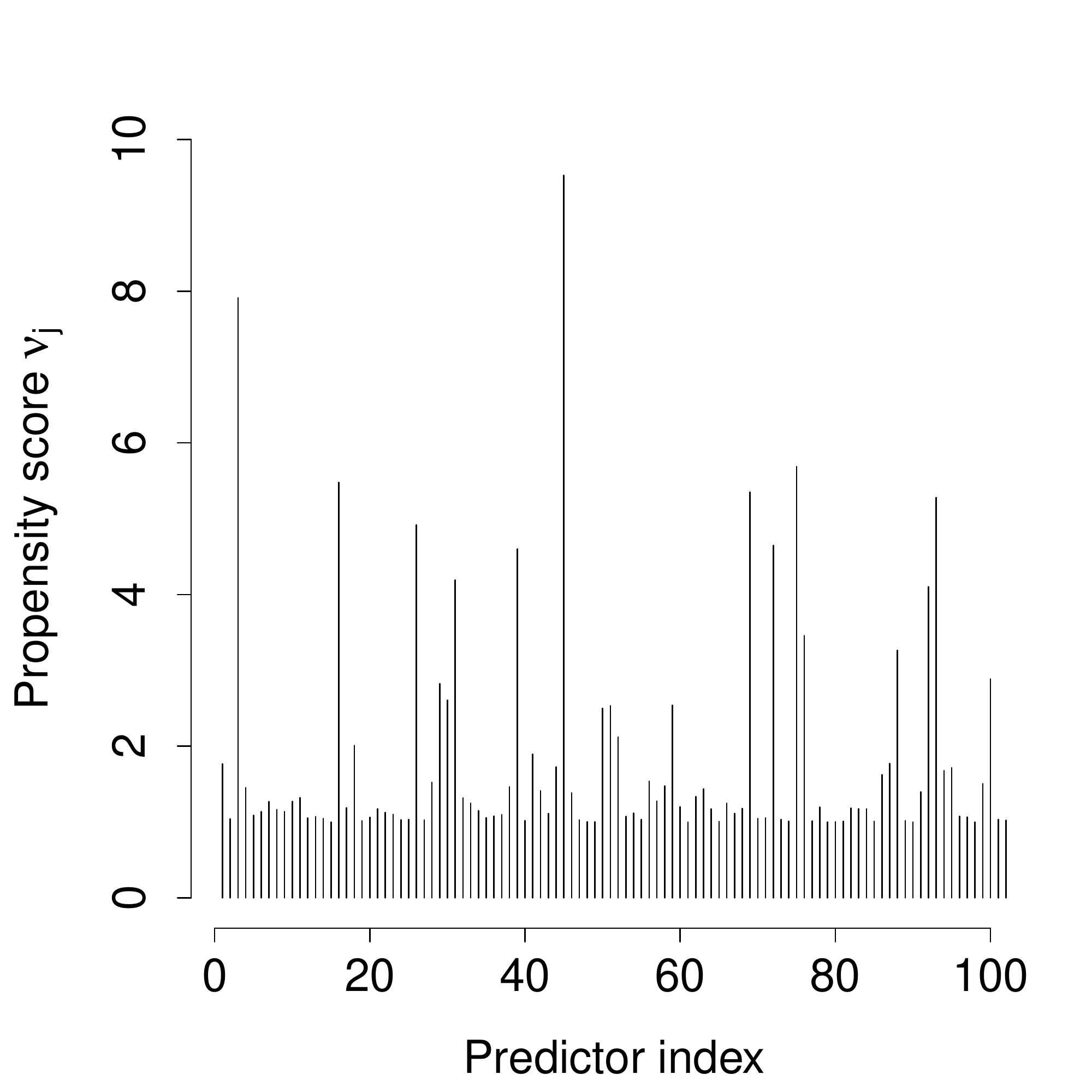}
& \includegraphics[width=\linewidth]{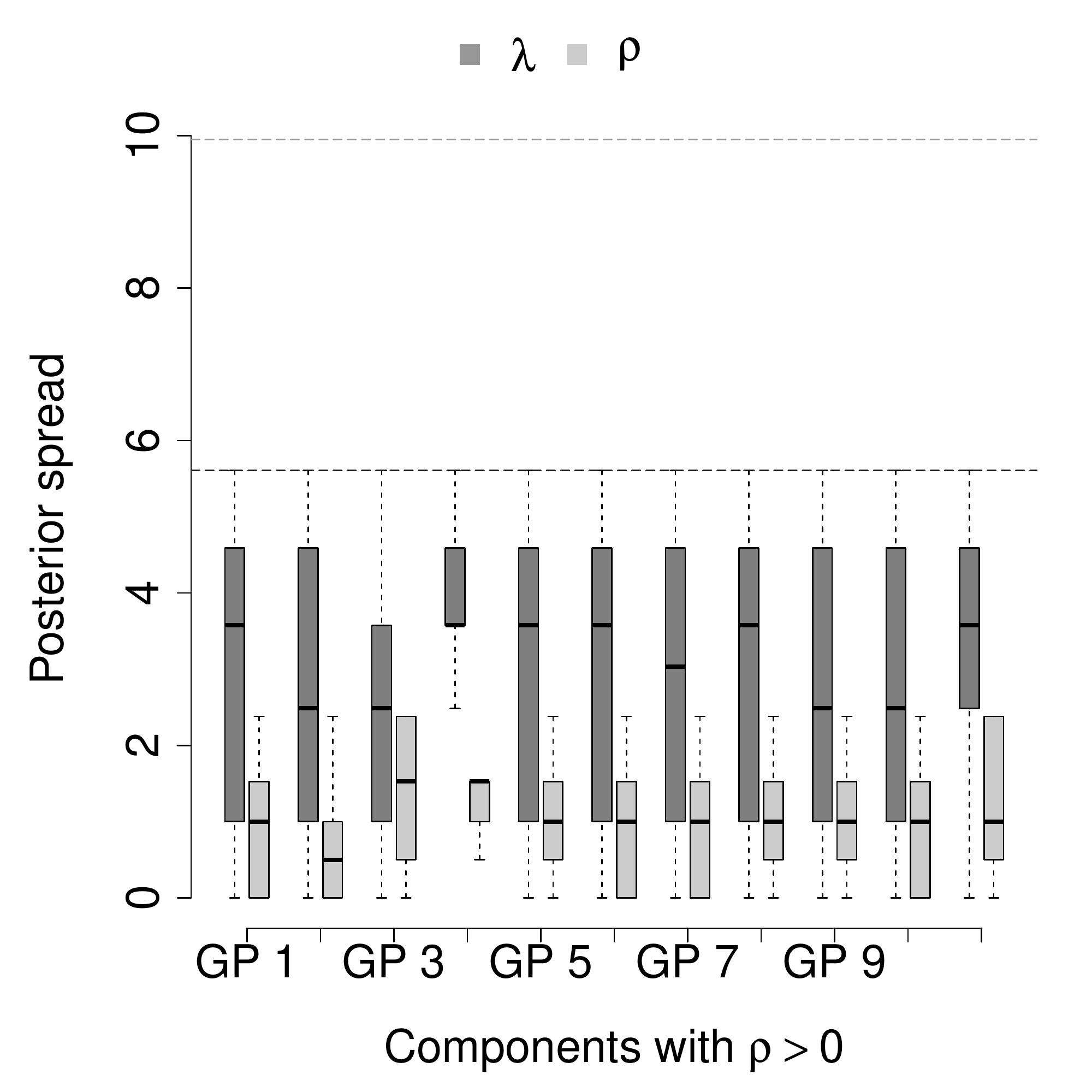}
& \includegraphics[width=\linewidth]{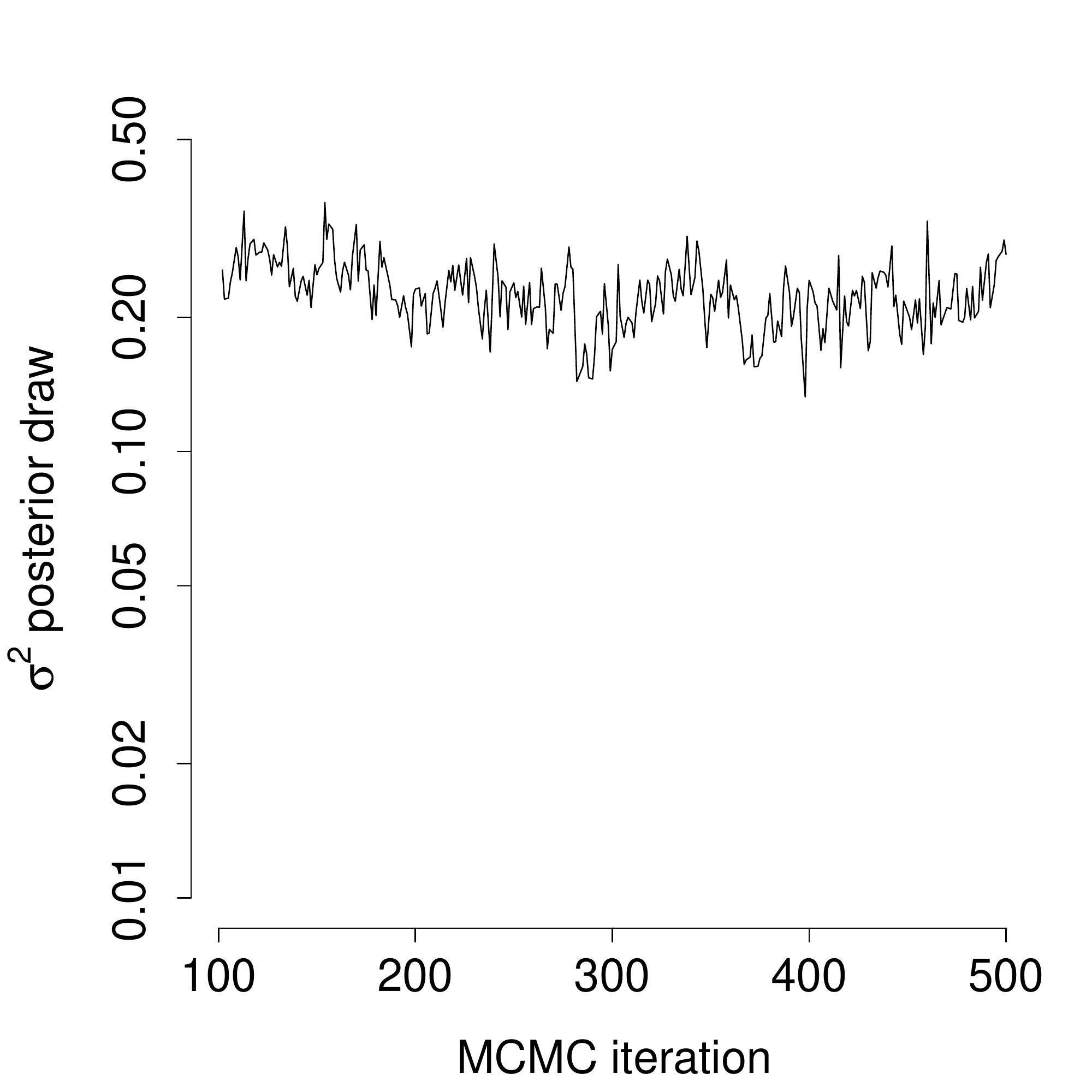} \\
\hline
Riboflavin
& \includegraphics[width=\linewidth]{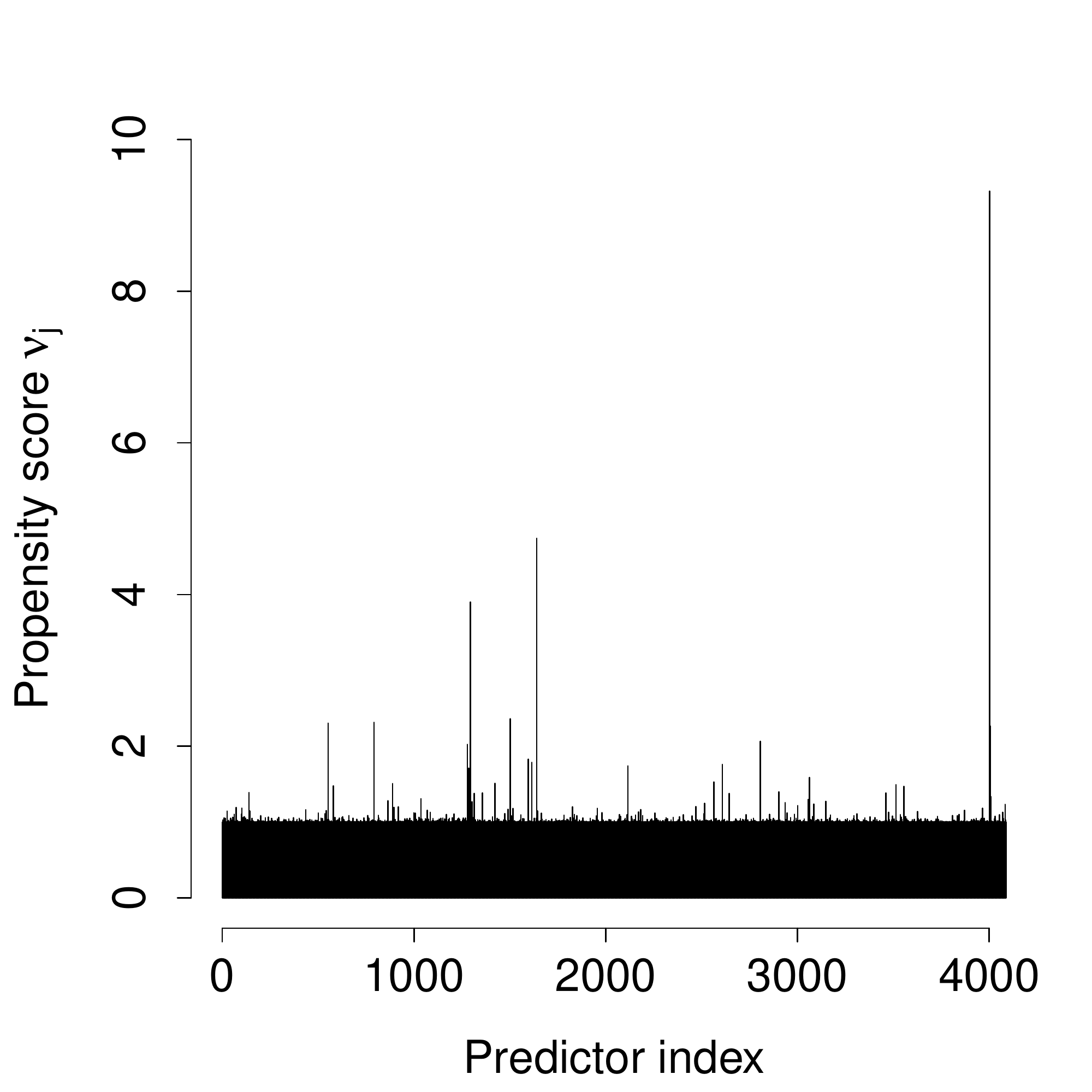}
& \includegraphics[width=\linewidth]{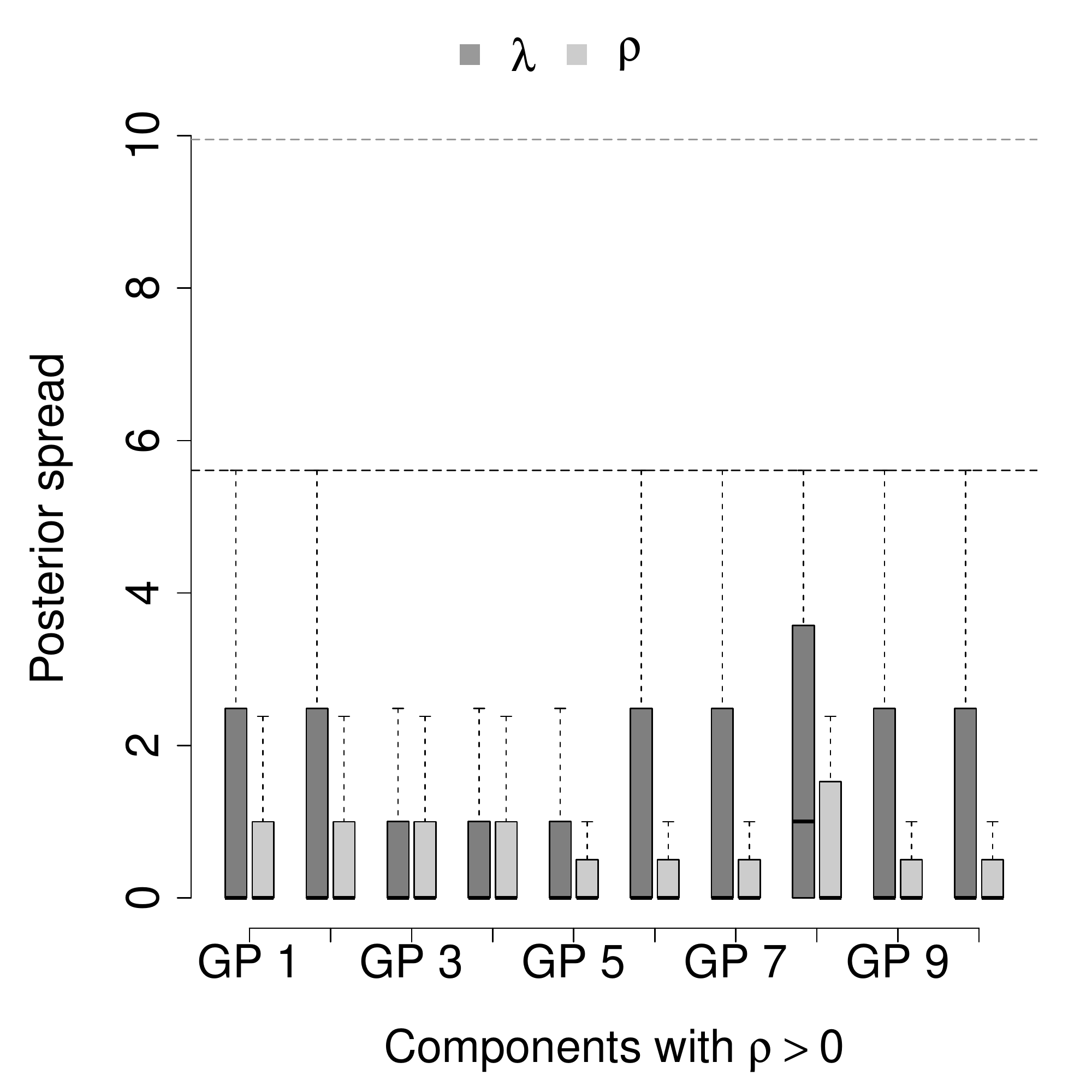}
& \includegraphics[width=\linewidth]{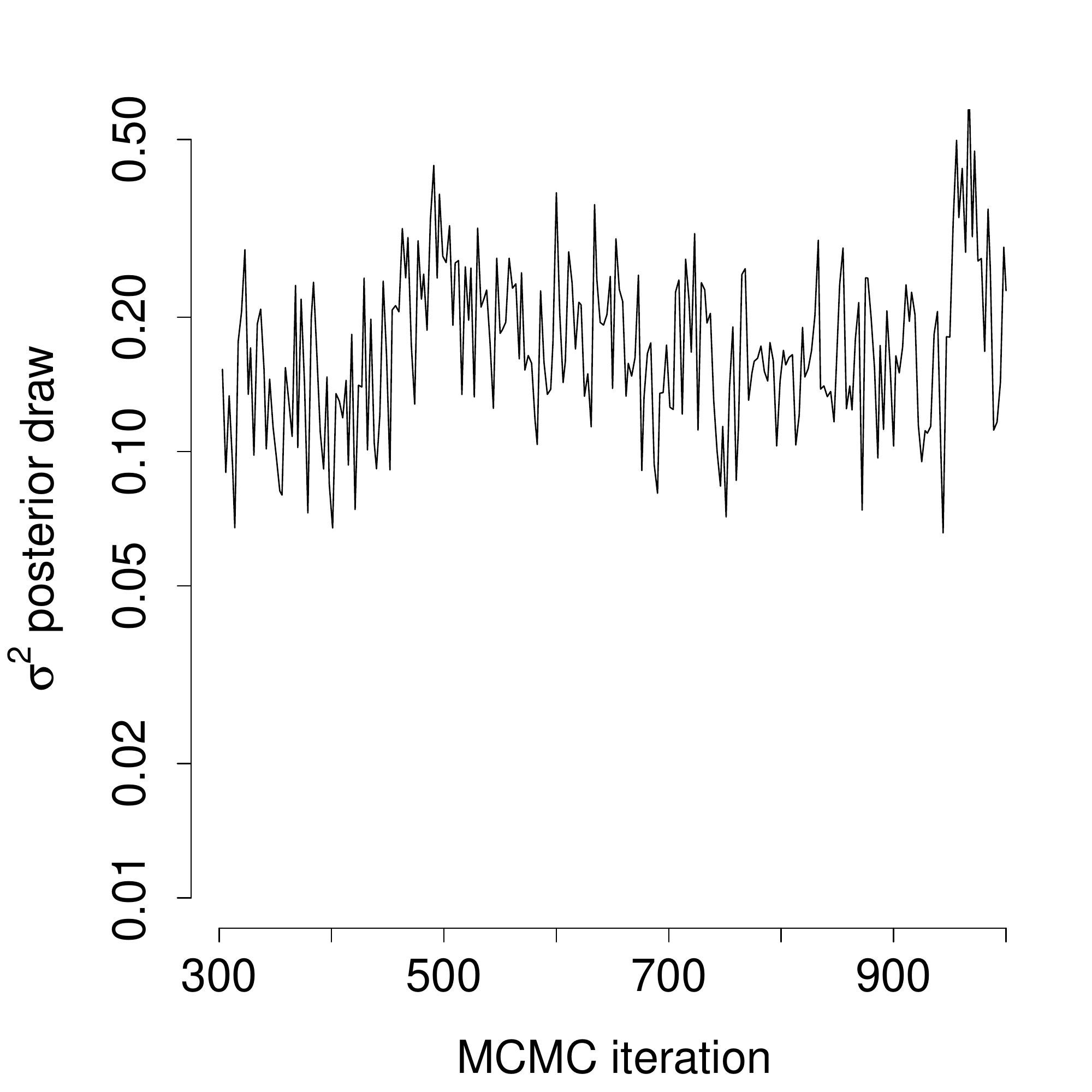} \\
\hline
Cookie
& \includegraphics[width=\linewidth]{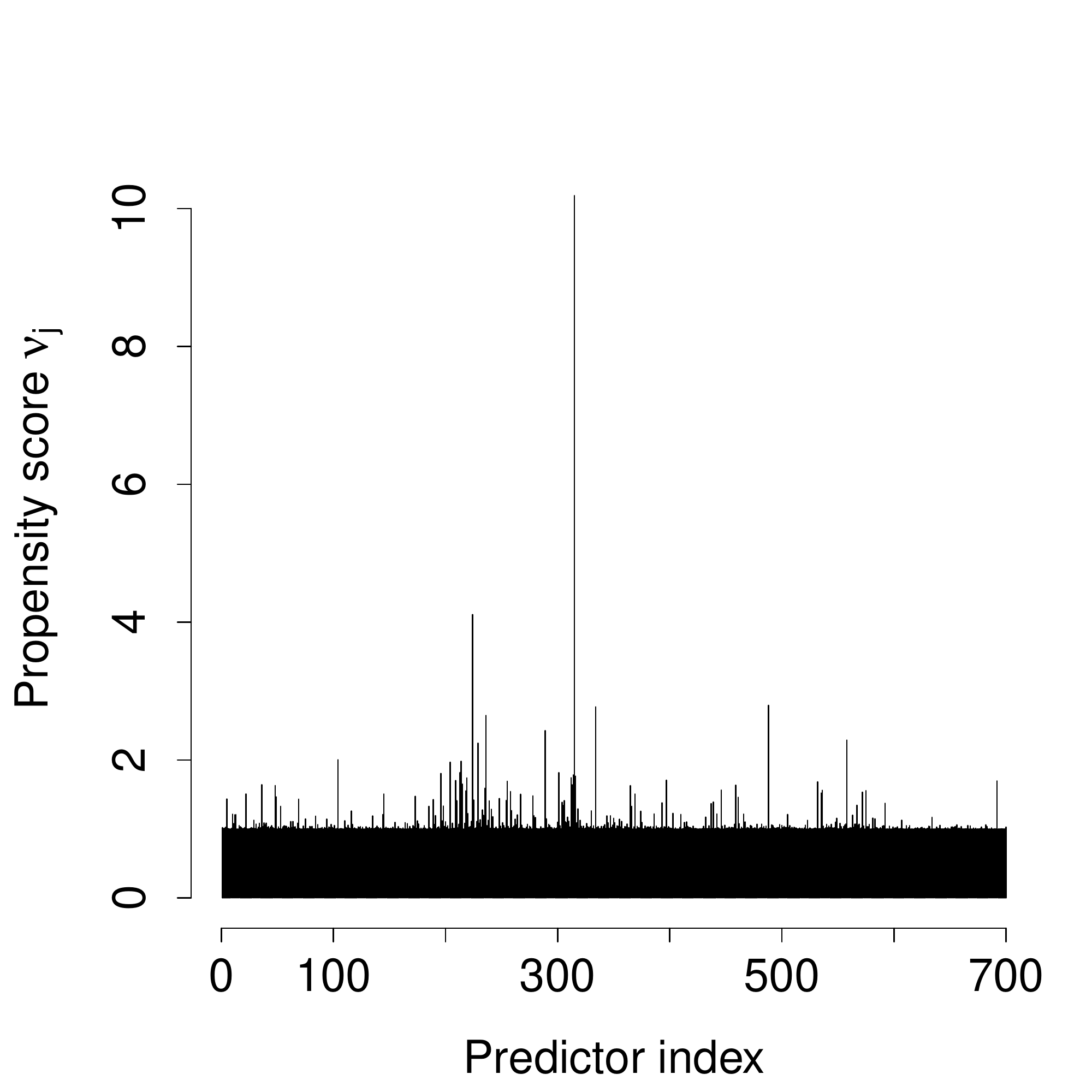}
& \includegraphics[width=\linewidth]{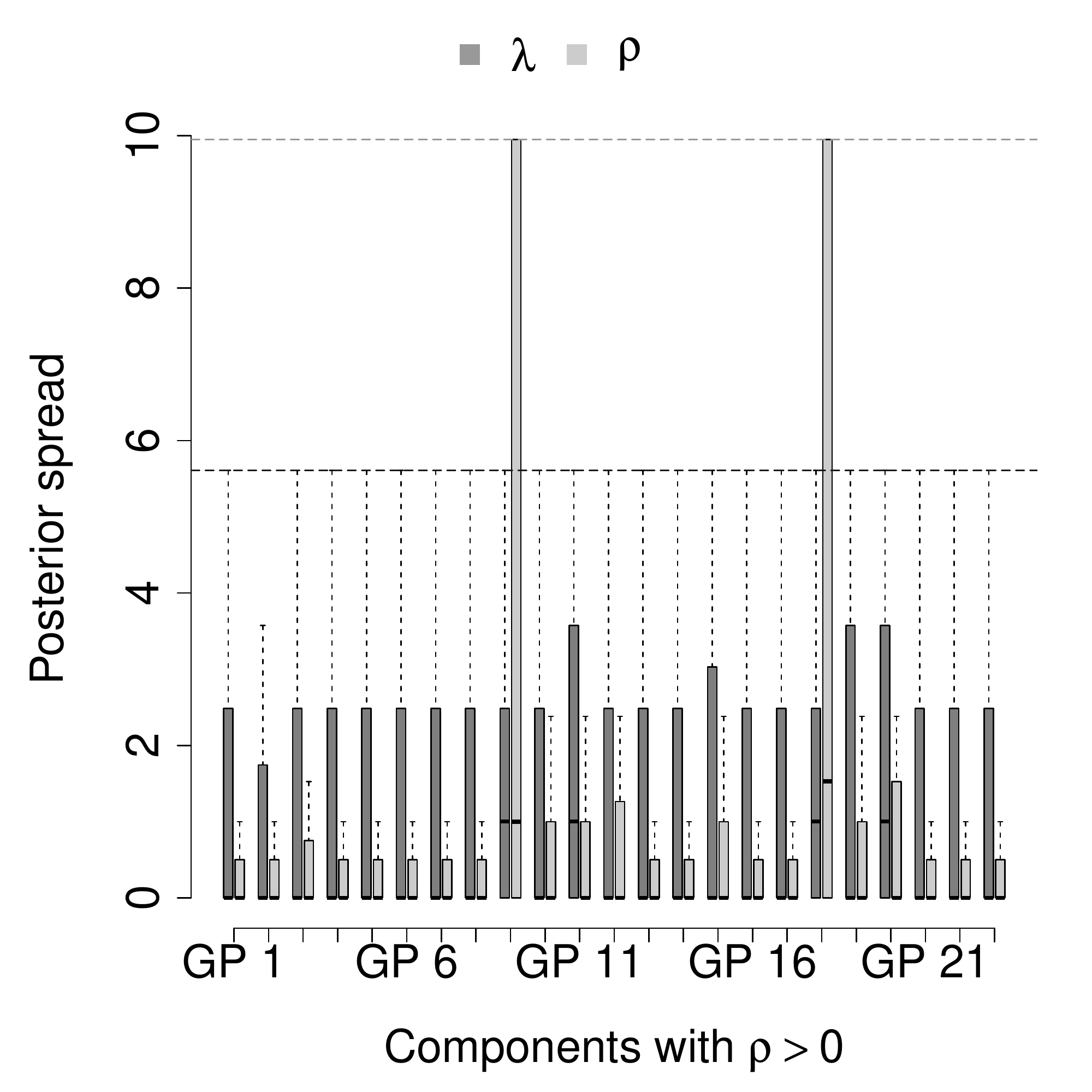}
& \includegraphics[width=\linewidth]{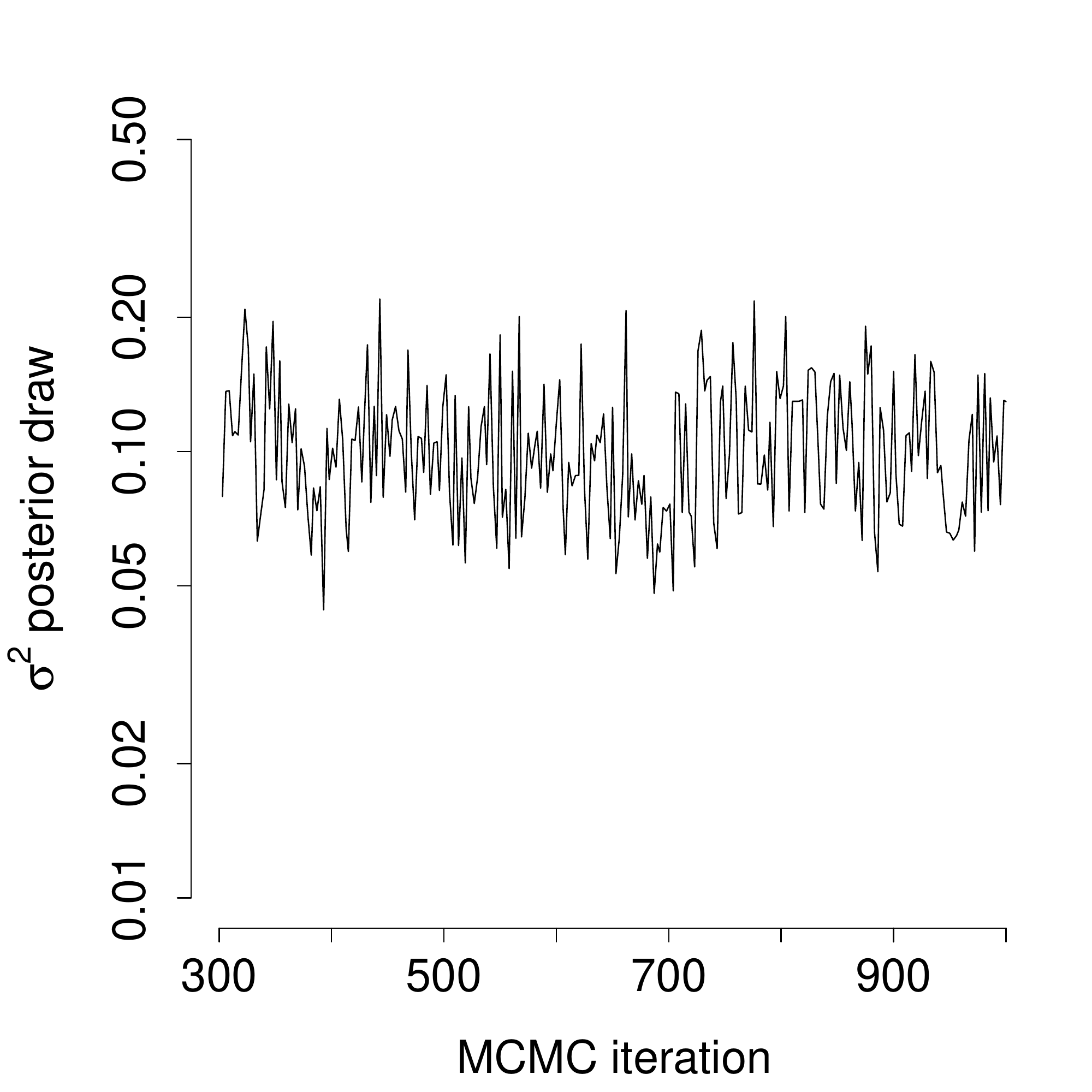}
\end{tabular}
\end{figure}

\clearpage 

\bibliographystyle{chicago}
\bibliography{references,minimax,mybib}

\end{document}